\documentclass[aps,prx,twocolumn,preprintnumbers,superscriptaddress,10pt,showkeys]{revtex4-1}
\usepackage{graphicx}
\usepackage{dcolumn}
\usepackage{bm}
\usepackage{hyperref}
\usepackage{colortbl}

\newcommand{\eqref}[1]{(\ref{#1})}

\begin{document}

\title{Dense and Hot QCD at Strong Coupling}

\author{Tuna Demircik}
\email{tuna.demircik@apctp.org}
\affiliation{Asia Pacific Center for Theoretical Physics, Pohang, 37673, Korea}
\author{Christian Ecker}
\email{ecker@itp.uni-franfurt.de}
\affiliation{Institut f\"ur Theoretische Physik, Goethe Universit\"at, Max-von-Laue-Str. 1, 60438 Frankfurt am Main, Germany}
\author{Matti J\"arvinen}
\email{matti.jarvinen@apctp.org}
\affiliation{Asia Pacific Center for Theoretical Physics, Pohang, 37673, Korea}

\date{\today}
 
\begin{abstract}
We present a novel framework for the equation of state of dense and hot Quantum Chromodynamics (QCD),
which focuses on the region of the phase diagram relevant for 
neutron star mergers and core-collapse supernovae.
The model combines predictions from the gauge/gravity duality with input from lattice field theory, QCD perturbation theory, chiral effective theory and statistical modeling. It is therefore, by construction, in good agreement with theoretical constraints both at low and high densities and temperatures. The main ingredients of our setup are the non-perturbative V-QCD model based on the gauge/gravity duality, a van der Waals model for nucleon liquid, and the DD2 version of the Hempel-Schaffner-Bielich statistical model of nuclear matter. 
By consistently combining these models, we also obtain a description for the nuclear to quark matter phase transition and its critical endpoint.
The parameter dependence of the model is represented by three (soft, intermediate and stiff) variants of the equation of state, all of which 
agree with 
observational constraints from neutron stars and their mergers. We discuss resulting constraints for the equation of state, 
predictions for neutron stars and the location of the critical point.
\end{abstract}

\preprint{APCTP Pre2021 - 034}

\maketitle


\section{Introduction\label{sec:Intro}}

Solving QCD at intermediate density and temperature is a long-standing open problem.
Recent and upcoming developments in relativistic heavy ion collision experiments and astrophysical observations of compact stars 
as well as 
their mergers urgently demand progress in the theoretic modelling of QCD at few times nuclear saturation density $n_s=0.16\,\rm{fm}^{-3}$ and temperatures up to around $100$\,MeV.
First principle approaches such as lattice QCD and perturbation theory are not applicable in the relevant regime,
and effective field theory only works up to densities around the nuclear saturation density.
Despite recent progress in finite temperature chiral effective theory computations at high loop-order \cite{Keller:2020qhx}, typical densities and temperatures such as estimated by realistic binary neutron star merger simulations \cite{Most:2018eaw,Perego:2019adq,Endrizzi:2019trv,Hammond:2021vtv,Figura:2021bcn,Raithel:2021hye} remain currently out of reach.

A central quantity that is absolutely crucial in the modelling of compact stars is the equation of state (EoS).
Due to the aforementioned lack of first-principles results, the QCD EoS at intermediate densities
has currently large uncertainties 
at zero temperature, and even less is known about the temperature dependence. 
These uncertainties motivate us to formulate a novel framework,
which combines predictions from various different approaches in different temperature and density regions of the QCD phase diagram
where they are expected to work best.
The main idea is to use gauge/gravity duality to model the physics at large and intermediate densities, and combine this with effective theory at low densities. This combination allows us to establish a unified  description of QCD matter for a wide range of densities and temperatures including, but not limited to, the ranges 
relevant for core-collapse supernovae and neutron star mergers.

The QCD phase diagram is conjectured to include a critical point where the nuclear to quark matter transition ends. Ongoing experiments at RHIC (the beam energy scan) already probe the region of the phase diagram where the critical point may lie~\cite{STAR:2020tga}. Future experiments at FAIR \cite{CBM:2016kpk,Durante:2019hzd} and NICA \cite{Sissakian:2009zza} will provide more detailed information about this region, and will reach substantially higher densities, i.e., densities well above the nuclear saturation density.
Consequently, it is  timely to improve theoretical predictions for the location of the critical point and the EoS in its vicinity. 
Our approach leads to EoSs which are in good agreement, among other things, with lattice data at small density as well as with ab initio calculations and
observational
data from neutron star measurements at finite density and small temperature. Therefore we are able to obtain controlled interpolations of the EoS to the theoretically challenging region of intermediate densities and temperatures, and sound estimates for the location of the critical point.

\begin{figure}
    \includegraphics[height=0.33\textwidth]{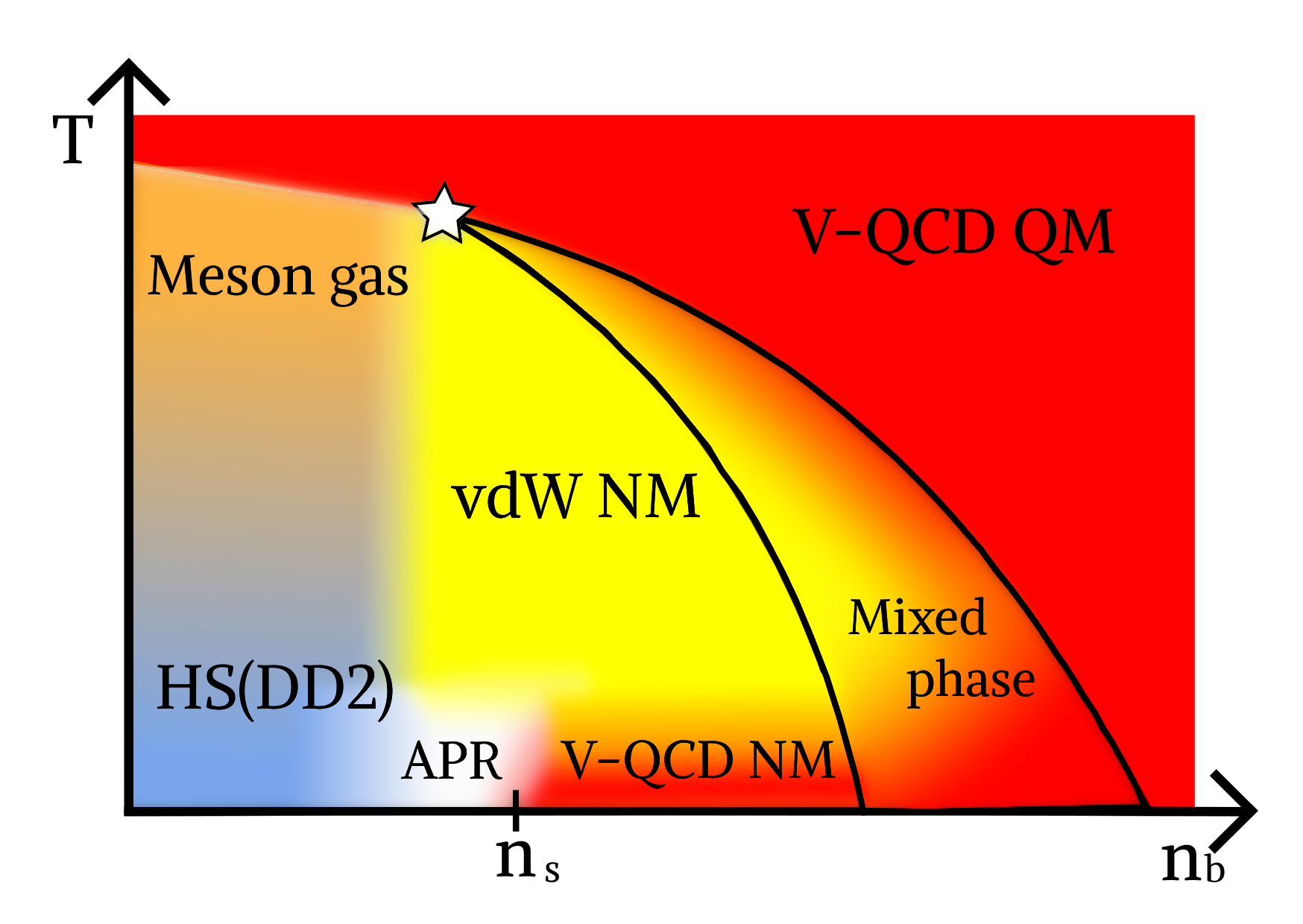}
    \caption{
    \small A schematic diagram that shows the construction for the temperature dependence in the model. See text for details.
    }
    \label{fig:phaseDiagram}
\end{figure}

The model, which we construct in this article, is a thermodynamically consistent combination of three main approaches: the holographic V-QCD model~\cite{Jarvinen:2011qe,Ishii:2019gta}, an adjusted van der Waals model of nuclear matter~\cite{Vovchenko:2020lju}, and the nuclear theory model Hempel-Schaffner-Bielich (HS) EoS~\cite{Hempel:2009mc} with DD2 relativistic mean field theory interactions~\cite{Typel:2009sy}.
We also use the Akmal-Pandharipande-Ravenhall (APR) model~\cite{Akmal:1998cf} of cold nuclear matter~\footnote{We use APR rather than HS(DD2) for cold nuclear matter near the saturation density because if we used HS(DD2) only, the resulting EoSs would be so stiff that they would conflict with the LIGO/Virgo measurement of the tidal deformability for GW170817.} and a meson gas model near the QCD crossover region (see Fig.~\ref{fig:phaseDiagram}), but they play a smaller role in the final EoS.

We use V-QCD \cite{Jarvinen:2011qe,Ishii:2019gta}, which is a non-perturbative model for QCD that is based on the gauge/gravity duality, at intermediate and high densities, i.e., densities above the nuclear saturation density.
This model allows us to describe both the nuclear and quark matter phases and therefore the phase transition in a single framework.
It has been shown~\cite{Ecker:2019xrw,Jokela:2020piw} to lead to a feasible EoS for cold QCD matter which is in good agreement with all available data.
Its extension to finite temperature is however problematic 
due to a generic limitation in gauge/gravity duality that the EoS in confined phases, including the nuclear matter phase, is independent of temperature and therefore not fully realistic.
In the absence of other reliable ways to estimate the temperature dependence of the EoS for dense nuclear matter we use essentially the simplest approach: a van der Waals type model of nuclear matter, i.e., a 
gas of nucleons and electrons with excluded volume corrections and an effective potential.
The effective potential is tailored for the model to agree with V-QCD at zero temperature, and therefore the van der Waals model 
gives an extrapolation of the V-QCD nuclear matter result to finite temperatures.
The third and final main constituent of the model is the nuclear theory model at low densities (below and around nuclear saturation density). 
In this article, we use the HS(DD2) model.

Fig.~\ref{fig:phaseDiagram} shows a sketch of the building blocks and the resulting phase structure of our model and their dependence on temperature and density, which is the main topic of this article.
This dependence is completed by the dependence on the charge fraction $Y_q$, i.e., allowing deviation from beta-equilibrium, which is required for realistic simulations of neutron star mergers and core-collapse supernovae. 
For this dependence we use the prediction of the HS(DD2) model in the NM phase and a simple model, arising from the pressure of free electrons, in the QM phase. 
We will provide three tabulated variants of the density, temperature and charge fraction dependent EoSs of this article in the CompOSE database \cite{Typel:2013rza,composeurl}.

Finally, let us comment on how the model of this article is related to other recent research on the EoS of QCD. 
The cold versions of the model, i.e., hybrid EoSs using V-QCD for dense nuclear and quark matter at zero temperature and at beta equilibrium, and various models for the low density nuclear matter, were established and studied in~\cite{Ecker:2019xrw,Jokela:2020piw}. They have been successfully applied in simulations of binary neutron star mergers \cite{Ecker:2019xrw} and the study of rapidly rotating isolated stars \cite{Demircik:2020jkc}. A related work~\cite{Chesler:2019osn} combined the HS(DD2) model, as well as two other general purpose models for the EoS, directly with the quark matter EoS from V-QCD, producing models for the EoS with dependence on temperature and charge fraction outside beta equilibrium. In the current article, we extend this work, among other things, by including the description of nuclear matter from gauge/gravity duality and its temperature dependence, which is our main focus, and also allows a consistent description of the mixed nuclear-quark matter phase. A recent article~\cite{Fujimoto:2021dvn} also considered models in the vdW class for hot and dense nuclear matter. Their focus was matching the predictions of chiral effective theory with the vdW models at low density, and use the vdW models to extend the results to higher densities and temperatures, whereas in the current work we use the vdW model to extrapolate the holographic cold hybrid EoS (which is feasible for all densities) to finite temperatures.

The article is organized as follows. In Sec.~\ref{sec:Model}, we review the various building blocks of our approach. In Sec.~\ref{sec:comb}, we show how the results for the EoS from these building blocks can be combined into a single thermodynamically consistent EoS. In Sec.~\ref{sec:Result} we analyze the EoS, predictions for the critical point, and predictions for nonrotating neutron stars. Our conclusions and future directions are given in Sec.~\ref{sec:Conclusion}.
In three appendices we provide technical details on the construction of our EoSs (App.~\ref{app:eosdetails}), compare the vdW model to the nuclear matter part of V-QCD (App.~\ref{app:vdWcomparison}) and explain how we determine our predictions for the QCD critical point (App.~\ref{app:CP}).

Throughout this article we use Planck units where $c=\hbar=k_B=1$.

\section{Building blocks of the model}\label{sec:Model}

We start the discussion of our model by briefly reviewing the various building blocks of the model.

\subsection{Holographic V-QCD}

A central ingredient in our setup is the use of the gauge/gravity duality.
It maps, in general, 
strongly coupled four dimensional field theory  
to classical 
five dimensional gravity. 
Therefore challenging questions on the field theory side can be solved by carrying out a simple classical analysis on the dual gravity side.
In this work we use the V-QCD model in regions which do not admit a weakly coupled description in terms of quarks, gluons or hadrons.
V-QCD 
is an effective gauge/gravity model in the sense that it contains a relatively large number of parameters that are tuned to match with QCD data from experiments, lattice analysis, and perturbation theory. 
This is particularly useful in the context of hot and dense QCD as there is plenty of lattice data available at low density, and gauge/gravity models can be used to extrapolate this data to higher densities 
at which first-principles methods are not available~\cite{DeWolfe:2010he,Knaute:2017opk,Critelli:2017oub,Grefa:2021qvt}.
Applying these ideas to V-QCD has been shown to lead to feasible and well constrained EoSs for both dense quark~\cite{Jokela:2018ers} and nuclear matter~\cite{Ishii:2019gta,Ecker:2019xrw,Jokela:2020piw}. 
This means the model is also able to describe the nuclear to quark matter phase transition in a single framework.

Let us then explain briefly how V-QCD is constructed. 
We discuss here only the main features of the model and refer the reader to~\cite{Jarvinen:2021jbd} for a complete review with precise definitions.
The model contains both a gluon sector, given by the improved holographic QCD model (five-dimensional dilaton gravity)~\cite{Gursoy:2007cb,Gursoy:2007er}, and a flavor sector arising through a pair of space filling flavor branes~\cite{Bigazzi:2005md,Casero:2007ae}. 
The flavors are dynamical: full backreaction of the branes to the geometry
is included formally by working in the Veneziano limit~\cite{Veneziano:1979ec} where both the number of colors $N_c$ and flavors $N_f$ is large but their ratio is $\mathcal{O}(1)$~\cite{Jarvinen:2011qe}.

A basic feature of the gauge/gravity duality is that various phases in the field theory map to different geometries in the five dimensional gravity theory. 
In case of V-QCD,
there are two possible geometries. 
The first is a horizonless geometry ending at a ``good" kind of IR singularity \cite{Gubser:2000nd}  
and the second is ``planar" black hole solution.  The black holes can be charged, which is interpreted as a nonzero baryon number arising from deconfined quark matter~\cite{Alho:2012mh,Alho:2013hsa}.
The gravitational solution includes a scalar condensate in the confined phase, which is dual to the chiral condensate of QCD therefore implementing chiral symmetry breaking~\cite{Bigazzi:2005md,Casero:2007ae,Jarvinen:2011qe}.
These geometries are therefore dual to a chirally broken confined hadron gas phase and a chirally symmetric deconfined quark-gluon plasma 
phase, respectively. 
For the plain confined geometry, the thermodynamics is trivial in the sense that for example the pressure is independent of temperature and chemical potential;
we will use other methods in this region as we will discuss below.
For the deconfined phase, the temperature and entropy density are calculated through black hole thermodynamics~\cite{Gursoy:2008bu,Gursoy:2008za,Alho:2012mh,Alho:2013hsa}.  
Apart from hadron gas and quark matter, we consider nuclear matter by employing the approach of~\cite{Ishii:2019gta}. 
The nuclear matter phase is obtained in an approximation 
that is based on
a homogeneous five-dimensional bulk field in the confined horizonless geometry. The five-dimensional action of the model is then obtained schematically as a sum of three terms:
\begin{eqnarray}
S_\mathrm{V-QCD}=S_\mathrm{g}+S_\mathrm{f}+S_\mathrm{nm}\,, \label{action}
\end{eqnarray}
where the first one ($S_\mathrm{g}$) is the action for improved holographic QCD, i.e., the gluon sector of the model, the second one ($S_\mathrm{f}$) is the flavor brane action important in the quark gluon plasma and quark matter phases, and  ($S_\mathrm{nm}$) is the action for homogeneous nuclear matter derived in~\cite{Ishii:2019gta}. 

We do not present the details of the actions here, but as we pointed out above, they contain a relatively high number of parameters that need to be tuned to match the model with QCD data. First, the model must agree with known features of QCD such as asymptotic freedom, confinement, linear glueball and meson trajectories, and chiral symmetry breaking. Second, those parameters that are left free after considering such qualitative constraints, are fitted to lattice data for thermodynamics of QCD~\cite{Alho:2015zua,Jokela:2018ers}. Specifically, we use lattice data for the equation of state of large-$N_c$ pure Yang-Mills~\cite{Panero:2009tv} and data for $N_c=3$ QCD with $N_f=2+1$ flavors at physical quark masses~\cite{Borsanyi:2013bia,Borsanyi:2011sw} at small baryon number density. Interestingly, the fit is stiff in the sense that the dependence of the various model parameters is mild, but despite this a fit of high quality 
is 
possible with all the parameter values remaining in a natural range.
At the moment, the fit uses flavor independent quarks with zero mass.
See the review~\cite{Jarvinen:2021jbd} for a detailed discussion of this fit and on the comparison of the model with QCD in general.

After determining the model parameters,
V-QCD has a natural phase diagram that includes both nuclear and quark matter~\cite{Ishii:2019gta,Jarvinen:2021jbd}.
The EoS for quark matter (QM) agrees with QCD lattice data at zero density and finite temperature, with perturbative QCD for high values of the baryon number chemical potential and/or temperature,
and with constraints for the EoS of cold quark matter EoS at low density ~\cite{Jokela:2018ers,Chesler:2019osn}. 
Notice however that we have not included electrons or photons, and as the quark flavors are identical, there is no dependence on charge fraction.
For the full model EoS in the quark matter phase, we also need to model these features, i.e., the electron pressure and the dependence on charge fraction. We will discuss this in Sec.~\ref{sec:Model}.

In the nuclear matter phase, the model is feasible at zero temperature and can be used to construct phenomenologically viable EoSs
for cold QCD matter as was done in~\cite{Ecker:2019xrw,Jokela:2020piw,Jokela:2021vwy}.
However, there is an issue with the extension to finite temperature:
the thermodynamics  
is temperature independent in the confined phases, as we pointed out above.  
This is a rather generic feature of gauge/gravity duality and arises due to taking the limit of large $N_c$: 
the pressure of confined color singlet hadron states, which are the constituents of the confined matter, is suppressed by $1/N_c^2$ with respect to the pressure of the deconfined matter \footnote{In principle working in the Veneziano limit, where $N_f \sim N_c$ should enhance the pressure of the confined matter. However it turns out that such contributions are not captured only by backreacting the flavor action to the geometry, but one should include nontrivial string loop diagrams~\cite{Alho:2015zua}.}. 
Due to this issue, the thermodynamics in the confined ``hadron gas'' phase without nuclear matter is trivial, i.e., that of empty space. But also in the nuclear matter phase, 
the temperature dependence is absent. 
While this result may be a rather good zeroth-order approximation, it 
prevents us from building a fully realistic EoS directly based on V-QCD.
In order to cure this issue, we will use 
a vdW model instead
for the temperature dependence, which we shall discuss next.

\subsection{van der Waals Hadron Gas model (vdW-HG)}

The vdW model consists of a bosonic and a fermionic sector.
The ideal gas pressure of the whole system is:
\begin{eqnarray}
p_\mathrm{id}(T,\{\mu_k\})&=&\sum_i p_\mathrm{FD}^{(i)}(T,\mu_i,m_i)\nonumber\\
&&+\sum_j p_\mathrm{BE}^{(j)}(T,m_j)+p_{\gamma}(T).\qquad
\end{eqnarray}
The first term, $p_\mathrm{FD}^{(i)}$ is the fermionic contribution that 
describes
nucleons, antinucleons, electrons and positrons
denoted by the index $i\in\{n,\bar{n},p,\bar{p},e,\bar{e}\}$.
The bosonic sector is given by 
the last two terms, where $p_\mathrm{BE}^{(j)}$ and $p_{\gamma}$ are the contributions of mesons and photons, respectively.
We include all mesons with $m_j\leq1$ GeV from the particle data group 
listings~\cite{ParticleDataGroup:2020ssz}. 
The photon contribution $p_{\gamma}(T)=\pi^2T^4/45$ is simply that of a black-body photon gas. 
The pressure of a relativistic ideal Fermi (Bose) gas is given by
\begin{eqnarray}
p_\mathrm{FD/BE}^{(k)}(T,\mu_k,m_k)=\frac{g_k}{6\pi^2}\int_0^\infty \frac{p^4}{E_{k}}\frac{dp}{e^{(E_{k}-\mu_k)/T}\pm1},
\end{eqnarray}
wherein the index $k$ denotes the fermion (boson) species with relativistic dispersion relation $E_k=\sqrt{p^2+m_k^2}$, $+$ $(-)$ is for fermions (bosons) and $g_k$ is the spin degeneracy factor.
We also note that different fermion and antifermion species 
have 
different chemical potentials while  
$\mu_k=0$ for the bosonic sector.  

It is well known that the ideal hadron gas (IHG) picture has  
vital shortcomings. 
Perhaps most importantly, 
it fails in describing the ground state of NM, viz., saturation of NM at $T=0$ and $n=n_s$. 
The reason for this shortcoming is the fact that the IHG does not capture repulsive and attractive van der Waals (vdW) interactions of nucleons.
Repulsive vdW interactions are often implemented via excluded volume (EV) corrections describing a hard-core repulsion of nucleons.
The thermodynamically consistent formulation of an IHG with EV correction was developed in \cite{Rischke:1991ke} (see also \cite{Hagedorn:1980kb}).
A more natural picture of vdW interactions that realizes both the short range repulsive and the intermediate range attractive interactions was formulated in   \cite{Vovchenko:2016rkn,Vovchenko:2017zpj} (see \cite{Vovchenko:2020lju} for a more complete list of references). 
The parameters 
of the vdW interactions are typically 
fixed by requiring consistency with  the saturation density $n_s$ and  binding energy per nucleon of the ground state  $\epsilon_B/n_b=-16$MeV \cite{Vovchenko:2015pya,Redlich:2016dpb}.  

In our construction, we follow a different strategy and use holography as guidance to model vdW interactions. 
In the rest of this subsection, we present our implementation of repulsive vdW interactions via EV corrections by employing the formulation in \cite{Rischke:1991ke}.
Attractive vdW interactions are incorporated by a direct matching to the cold V-QCD hybrid EoSs at beta-equilibrium described in section~\ref{vdw-holo-sec}.  

The excluded volume corrected pressure is defined implicitly by introducing a shifted chemical potential $\tilde{\mu}_i$:
\begin{eqnarray}
&&p_\mathrm{ex}(T,\{\mu_k\})=p_\mathrm{id}(T,\{\tilde{\mu}_k\})\,;\label{pexc} \\&&\tilde{\mu}_i=\mu_i-v_ip_\mathrm{ex}(T,\{\mu_k\})\, .
\end{eqnarray}
We choose $v_p=v_{\bar{p}}=v_n=v_{\bar{n}}=v_0$, $v_e=v_{\bar{e}}=0$ and set $v_0=0.56\,\text{fm}^3$. 
The choice of $v_0$ is motivated by comparison with V-QCD EoSs at $T=0$. 
More details about our choice and a comparison of different values for $v_0$ are given in Appendix \ref{app:vdWcomparison}.

The number density 
is found by differentiating (\ref{pexc}): 
\begin{eqnarray}
n_\mathrm{ex}^{(i)}(T,\{\mu_k\})&=&\frac{\partial p_\mathrm{ex}(T,\{\mu_k\})}{\partial\mu_i}\nonumber\\
&=&\frac{n_\mathrm{id}^{(i)}(T,\{\tilde\mu_k\})}{1+v_0\sum_l n_\mathrm{id}^{(l)}(T,\{\tilde\mu_k\})},\;\;\label{nex}
\end{eqnarray}
where $n_\mathrm{id}^{(i)}=\partial p_\mathrm{id}(T,\{\tilde\mu_k\})/\partial \mu_i$ and $i$ denotes the fermion species, while the index $l$ runs over only nucleons and antinucleons.
At this point, we fix the chemical potentials of antifermions as $\mu_{\bar i} = -\mu_{i}$.
The total number density for each fermion species is then given by the difference between the corresponding particle and antiparticle number densities:
\begin{equation}
\tilde{n}_\mathrm{ex}^{(i)}(T,\{\tilde\mu_k\})=n_\mathrm{ex}^{(i)}(T,\{\tilde\mu_k\})-n_\mathrm{ex}^{(\bar{i})}(T,\{\tilde\mu_k\})\label{nex2} \,,
\end{equation}
where $i\in \{p,n,e\}$.
Requiring charge neutrality in addition,
\begin{equation}
\tilde{n}_\mathrm{ex}^{(e)}(T,\{\mu_k\})=\tilde{n}_\mathrm{ex}^{(p)}(T,\{\tilde\mu_k\}) \,,\label{chn}
\end{equation}
leaves only two free chemical potentials which we choose to be $\mu_p$ and $\mu_n$.
By using (\ref{nex}) and (\ref{nex2}), the definitions for the baryon number density $n_b$, the charge fractions $Y_q$  and charge neutrality condition are  
obtained as
\begin{eqnarray}
n_b(T,\mu_p,\mu_n)&=&\tilde{n}_\mathrm{ex}^{(p)}(T,\mu_p,\mu_n)+\tilde{n}_\mathrm{ex}^{(n)}(T,\mu_p,\mu_n)\,,\\
Y_q(T,\mu_p,\mu_n)
&=&\tilde{n}_\mathrm{ex}^{(p)}(T,\mu_p,\mu_n)/n_b(T,\mu_p,\mu_n)\,. 
\end{eqnarray}

For future convenience, we conclude this section by giving the definition of the free energy which is the natural thermodynamic potential in the 
canonical ensemble:
\begin{equation}
f_\mathrm{ex}(T,n_b,Y_q)=\sum_i n_\mathrm{ex}^{(i)} \mu_i(T,n_b,Y_q) -p_\mathrm{ex}(T,n_b,Y_q),
\end{equation}
where the thermodynamics is expressed in terms of variables $n_b$ and $Y_q$ instead of $\mu_p$ and $\mu_n$.

\subsection{HS(DD2) model} 
HS(DD2) is a commonly used EoS that was originally developed to simulate core-collapse of supernovae \cite{Hempel:2009mc,Typel:2009sy}.
It provides reliable modelling of NM below and around the saturation density, with a consistent description for the transition from nonuniform to uniform nuclear matter. 

HS(DD2) consists of two different sectors. 
The first sector models light and heavy nuclear clusters in nuclear statistical equilibrium with EV correction.
The second sector describes unbound nucleons in relativistic mean-field (RMF) theory. 
HS(DD2) includes all possible light nuclei (e.g. deuterium, tritium, etc.) in addition to $\alpha$-particles and heavy nuclei up to mass number $A\sim330$. 

The RMF approach is used to model interactions of unbound nucleons with the exchange of $\sigma$, $\omega$ and $\rho$ mesons. 
At low densities, the interactions become negligible and the system reduces to an ideal Fermi-Dirac gas of nucleons.
Photons are added separately as a free Bose gas and the Wigner-Seitz approximation is employed for the Coulomb interaction between electrons and nuclei. 
EV effects are implemented in a way to treat unbound nucleons and nuclei in a different manner: while the volume of all baryons is excluded for nuclei, unbound nucleons only feel the volume of nuclei since the interaction among them is already modelled by the RMF model.   

The parameters of the RMF model are the masses of the nucleons and the mesons, and the coupling constants.
In HS(DD2), the so called TMA 
parameter set is used \cite{TOKI1995c357}.
For the masses of nuclei, the experimental data \cite{AUDI2003337} is used when it is available, otherwise the data is taken from nuclear calculations \cite{Geng:2005yu}. 

By construction, HS(DD2) is a thermodynamically consistent model for both nonuniform matter of light and heavy clusters within the gas of unbound nucleons at low density, uniform matter described by the RMF model above densities higher than $n_s$, and the transition between them.
The resulting EoS are provided in three dimensional ($T$,$n_b$,$Y_q$) tabular form covering a wide range.
The EoS tables are publicly available on the CompOSE database \cite{Typel:2013rza,composeurl}.
Different variants of the model were constructed by using different sets of RMF parameters~\cite{Hempel:2011mk,Steiner:2012rk}.

\section{Combining the building blocks}\label{sec:comb}

In this section we discuss in detail how the various building blocks of the model are combined into a unified model (see Fig.~\ref{fig:phaseDiagram}).

\subsection{Construction of the cold EoS}\label{sec:coldEoS}

We start from the construction of the nuclear matter EoS at zero temperature, which mostly follows Refs.~\cite{Ecker:2019xrw} and~\cite{Jokela:2020piw}. That is, we first construct cold hybrid EoSs by combining the predictions of V-QCD with nuclear theory models 
at beta equilibrium. 

Our approach for nuclear matter in V-QCD, which is based on a homogeneous bulk field, 
is natural and realistic for densities well above $n_s$. 
At such high densities the average distance between neighboring nucleons are comparable or smaller than their diameters, so that their wave functions overlap, and approximating the system as homogeneous matter is expected to work. 
However, this is not case at densities below $n_s$, wherein homogeneous approximations will break down. 
This is not a problem, since in this region traditional NM models have proven to be reliable and feasible.
In this sense traditional NM models and homogeneous NM in V-QCD complement each other.
In \cite{Ecker:2019xrw,Jokela:2020piw}, this idea was implemented to 
construct  
hybrid EoSs: low density  EoSs from various traditional nuclear theory models were combined with the high density V-QCD NM EoSs, matching them continuously at an intermediate density around $1.5n_s$ to $2n_s$. 

A potential weakness of the construction of~\cite{Ecker:2019xrw,Jokela:2020piw} is that the matching point between the low and high density models introduces, in effect, a second order phase transition, where for example the speed of sound is discontinuous. 
While such discontinuities are ubiquitous in constructions of EoSs in the literature and appear in particular in the commonly used polytropic EoSs, we find that for the extension to finite temperature, it is better if the cold EoS has continuous speed of sound. 
In order to smooth it out, we consider an improved matching setup with two separate matching densities in the same region of densities slightly above the saturation density. The EoS in the intermediate region between the two matching densities is chosen to have a speed of sound which is linear in the baryon number density, whereas the low and high density regions are treated as before. The slope of the speed of sound in the intermediate region is determined by requiring continuity of the speed of sound, i.e., third order phase transitions at both the transition densities. See Appendix~\ref{app:coldEoS} for details. The resulting EoS (free energy as a function of baryon number density) is denoted below as $f_\mathrm{cold}(n_b)$.

In this work, we will use three variants of hybrid EoSs for nuclear matter at zero temperature. We select the Akmal-Pandharipande-Ravenhall (APR) model \cite{PhysRevC.58.1804} for the low density regime and three variants of V-QCD at  high density: these variants are defined by the data fits 5b, 7a, and 8b of~\cite{Jokela:2018ers,Jokela:2020piw}. These choices represent the leftover parameter dependence of V-QCD after the comparison with lattice data and taking into account observational constraints. If we used the approach of~\cite{Ecker:2019xrw,Jokela:2020piw} with a single transition density $1.6n_s$ the hybrids with these three choices would be exactly the soft, intermediate and stiff variants of the V-QCD(APR) EoS published in the CompOSE database \cite{composeurlVQCDsoft,composeurlVQCDinter,composeurlVQCDstiff}. 
Here the stiffness refers to a property of dense nuclear matter: the stiff EoS reaches a noticeably higher speed of sound than the soft one.
In the improved approach of this article, we choose the two transition densities to be $1.4n_s$ and $1.8n_s$. That is, the choices of the cold nuclear matter EoS in this article are practically the same as the published V-QCD(APR) variants, but the kink (say, in pressure as a function of density) at the transition density has been smoothed out.

\subsection{Matching the vdW model with holography }\label{vdw-holo-sec}

We then discuss how the EoS for cold nuclear matter is extrapolated to finite 
temperatures, and also outside beta equilibrium. As pointed out above, we use the vdW model for the temperature dependence in the dense nuclear matter phase. This approach is motivated by the fact that, for an appropriate choice of the excluded volume $v_0$, the EoSs of the vdW model and the nuclear matter of the V-QCD model are relatively close (see Appendix~\ref{app:vdWcomparison}). 
Therefore adding only a small, mostly attractive potential is required to match the vdW model with the cold hybrid EoS. Moreover, we add the dependence on the charge fraction outside beta equilibrium by using the HS(DD2) model. In principle, one could use the vdW model also for the charge fraction. The vdW model is however too simple to satisfy the experimental constraints for nuclear matter below and around saturation density. 
In particular, the symmetry energy is too low -- this is known to happen for a free gas, and excluded volume effects alone are not sufficient to improve the result~\cite{Baldo:2016jhp}. We could in principle use the vdW model at higher densities, were experimental constraints do not apply, but for simplicity we adopt the $Y_q$ dependence from HS(DD2) everywhere in the nuclear matter regime.  

The above adjustments are taken into account by redefining the free energy of the dense NM phase as
 \begin{eqnarray}
 \label{eq:fvdWdef}
 f_\mathrm{vdW}(T,n_b,Y_q)=f_\mathrm{ex}(T,n_b,Y_q)+\Delta f(n_b,Y_q),
 \end{eqnarray}
where $\Delta f$ models the mean contribution from an attractive potential.  
It is 
simpler to specify the free energy difference $\Delta f$ directly rather than start from the definition of the potential.
Because the contribution from the potential separates (see, e.g.,~\cite{Vovchenko:2020lju}),
the two ways of formulating this contribution are practically equivalent.
That is, we take~\footnote{We use $T_\mathrm{min}=0.1$~MeV instead of $T=0$ when evaluating these definitions numerically because this value is the lowest available in the data grid of the HS(DD2) EoS.}
\begin{eqnarray}
\label{eq:Deltaf}
\Delta f(n_b,Y_q)&=
&f_\mathrm{cold}(n_b)\nonumber\\
&&-f_\mathrm{ex}(T=0,n_b,Y_q)\nonumber\\
&&+f_\mathrm{HS(DD2)}(T=0,n_b,Y_q)\nonumber\\&&-f_\mathrm{HS(DD2)}(T=0,n_b,Y_q^\mathrm{eq}(n_b)),
\end{eqnarray}
where $f_\mathrm{cold}$ is the free energy of the one dimensional cold hybrid EoS constructed as discussed above,
$f_\mathrm{HS(DD2)}$ is the free energy of the HS(DD2) model, and
$Y_q^\mathrm{eq}$ is the value at beta-equilibrium  
 for the HS(DD2) EoS. 

The first two lines on the right hand side of~\eqref{eq:Deltaf} adjust the dependence of the EoS on $n_b$ such that it matches with the cold hybrid EoS at zero temperature. 
The last two lines in~\eqref{eq:Deltaf} adjust the dependence on $Y_q$ such that it agrees with that of the HS(DD2) EoS at low temperatures, without changing the EoS at beta-equilibrium.

\subsection{Transition between the vdW model and HS(DD2)}

At low densities, such as relevant for example in the crust of the neutron stars, the construction based to the holographic model and vdW gas is too simple to be reliable, so we use instead directly the HS(DD2) EoS.
We implement this by switching from the vdW model of~\eqref{eq:fvdWdef}, which already borrows the $Y_q$ dependence from HS(DD2), smoothly to the exact HS(DD2) EoS at a well chosen transition density. However, before this is possible it is necessary to adjust the HS(DD2) EoS by adding the contribution from the mesons of QCD as indicated in Fig.~\ref{fig:phaseDiagram}.
This contribution is important only in the region of low density and high temperatures (i.e., close to the transition temperature in QCD), which is far from the regime relevant for neutron stars and core-collapse supernovae. 
We however add this contribution since it affects the study of the critical point, which we carry out below.
We write the ``improved'' HS(DD2) EoS as
\begin{eqnarray}
 \hat f_\mathrm{HS(DD2)} (T,n_b,Y_q) &=&   f_\mathrm{HS(DD2)} (T,n_b,Y_q)\nonumber\\&& + \sum_j p_\mathrm{BE}^{(j)}(T,m_j)
\end{eqnarray}
where the sum goes over all mesons from the particle data group~\cite{ParticleDataGroup:2020ssz} with masses below 1~GeV.

After this modification, we define the final nuclear matter EoS as
\begin{eqnarray} \label{eq:DD2vdWm}
    f_\mathrm{NM}(T,n_b,Y_q) &=& [1-w(n_b)] \hat f_\mathrm{HS(DD2)}(T,n_b,Y_q)\nonumber \\
    &&+ w(n_b) f_\mathrm{vdW}(T,n_b,Y_q) 
\end{eqnarray}
where the weight function is
\begin{equation}
 w(n_b) = \frac{1}{2}\left[1+\tanh\left(\frac{\log(n_b/n_0)}{1.75}\right)\right] = \frac{\left(n_b/n_0\right)^{8/7}}{1+\left(n_b/n_0\right)^{8/7}}   
\end{equation}
with $n_0 \approx 0.0694 n_s$. The numerical coefficients were chosen such that the transition from HS(DD2) to vdW is smooth for all temperatures and charge fractions.

\subsection{The mixed phase and the critical point}\label{Dcrit}
The final step in our construction is to combine the NM and QM components into a single EoS.
In order to do this, we first need to adjust the V-QCD QM result: the EoSs constructed in~\cite{Jokela:2018ers} does neither include dependence on the charge fraction nor electron pressure. 
It would be possible to compute the charge fraction dependence from the model directly, but this would require a significant extension of the model, which is beyond the scope of this article in which the focus is on temperature dependence. 
Therefore we resort to approximations. 

There are two simple approximation schemes: First is to assume that the free energy of strongly interacting matter only depends on the total baryon number, given as the sum over the quark number densities as $n_b = (n_u+n_d+n_s)/3$. In this case the free energy of QM arises as the sum over the electromagnetic contribution and the V-QCD pressure:
\begin{eqnarray} \label{eq:fQM1}
 f_\mathrm{QM}(T,n_b,Y_q) &=& f_{e\bar e \gamma}(T,Y_q n_b) \nonumber \\
 && + f_\mathrm{V-QCD}(T,n_b) \ ,
\end{eqnarray}
where the electron density is $n_e = Y_q n_b$ by charge neutrality. The electromagnetic term $f_{e\bar e\gamma}$ is estimated as the sum of the ideal gas free energies of electrons, positrons, and photons at the given electron density.

The second scheme (which was used in~\cite{Chesler:2019osn}) assumes that the free energy arises as a direct sum of the free energies of different quark flavors with equal amount of down and strange quarks. This gives
\begin{eqnarray}
 f_\mathrm{QM}(T,n_b,Y_q) &=& f_{e\bar e \gamma}(T,Y_q n_b) \nonumber \\
 && + \frac{1}{3}f_\mathrm{V-QCD}\left(T,(1+Y_q)n_b\right)\nonumber\\
 && + \frac{2}{3}f_\mathrm{V-QCD}\left(T,(1-Y_q/2)n_b\right) 
\end{eqnarray}
where the second (third) line is the contribution from up (down) type quarks. Notice that this approach includes a simple approximation for the symmetry energy of quark matter, which assumes no interactions between the different quark flavors, and therefore corresponds to the probe limit where the backreaction of the flavors to the gluon dynamics is neglected. However, there is no reason to expect that the effect of the backreaction is small. In this article, we use the simplest approximation of~\eqref{eq:fQM1}.

The final EoS and phase diagram, including the mixed phase between NM and QM, is then found by carrying out a Gibbs construction (see Appendix~\ref{app:3dEoS} for details). As it turns out, two different regimes can be clearly identified from the result. At low temperatures, there is a very strong first order phase transition, which becomes weaker with increasing temperature. At higher temperatures there is a weak first order transition. We interpret this weak transition as the signal of crossover:  continuity over the phases is not possible because this would require a precise match between the EoSs of the meson gas (in the NM phase) with the V-QCD QM EoS. We have not tried to carry out such matching here; this is left for future work. The transition between the two regimes is therefore interpreted as the critical endpoint of the nuclear to quark matter transition line.  We will illustrate the mixed phase and the critical point in detail in Sec.~\ref{sec:Result} and in Appendix~\ref{app:CP}.

\begin{figure*}
    \includegraphics[height=0.32\textwidth]{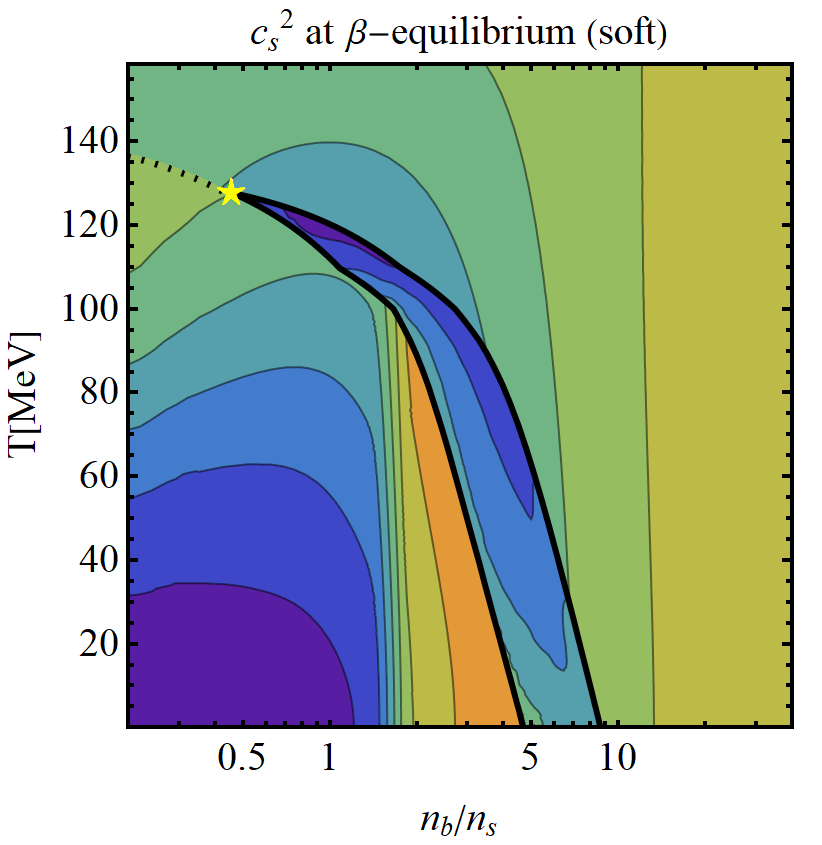}%
     \includegraphics[height=0.32\textwidth]{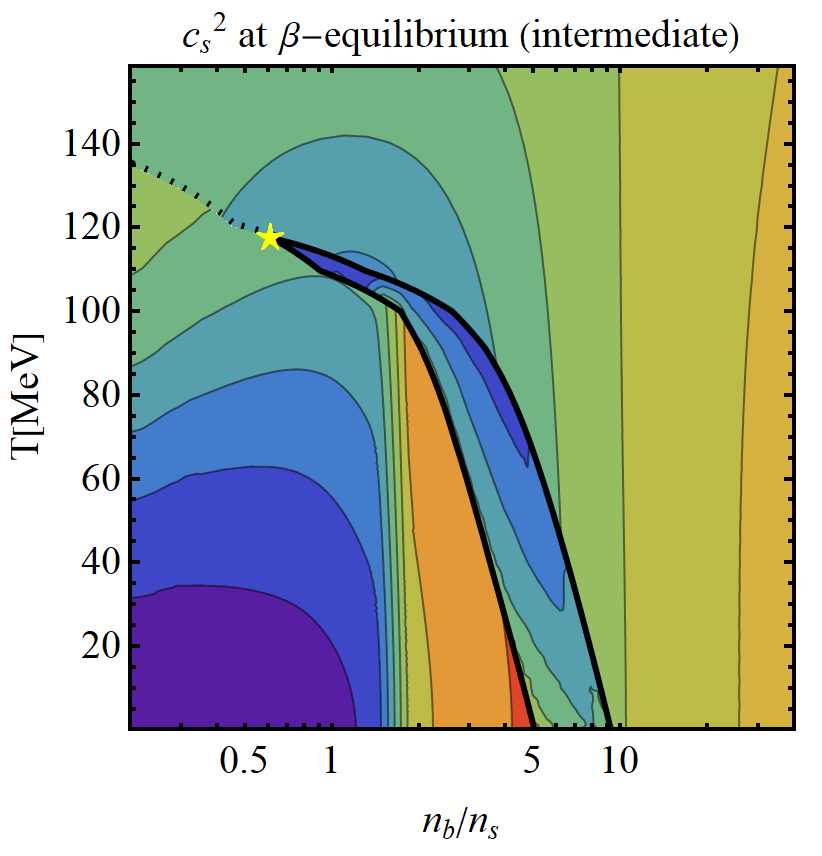}%
    \includegraphics[height=0.33\textwidth,trim=0 6mm 0 0]{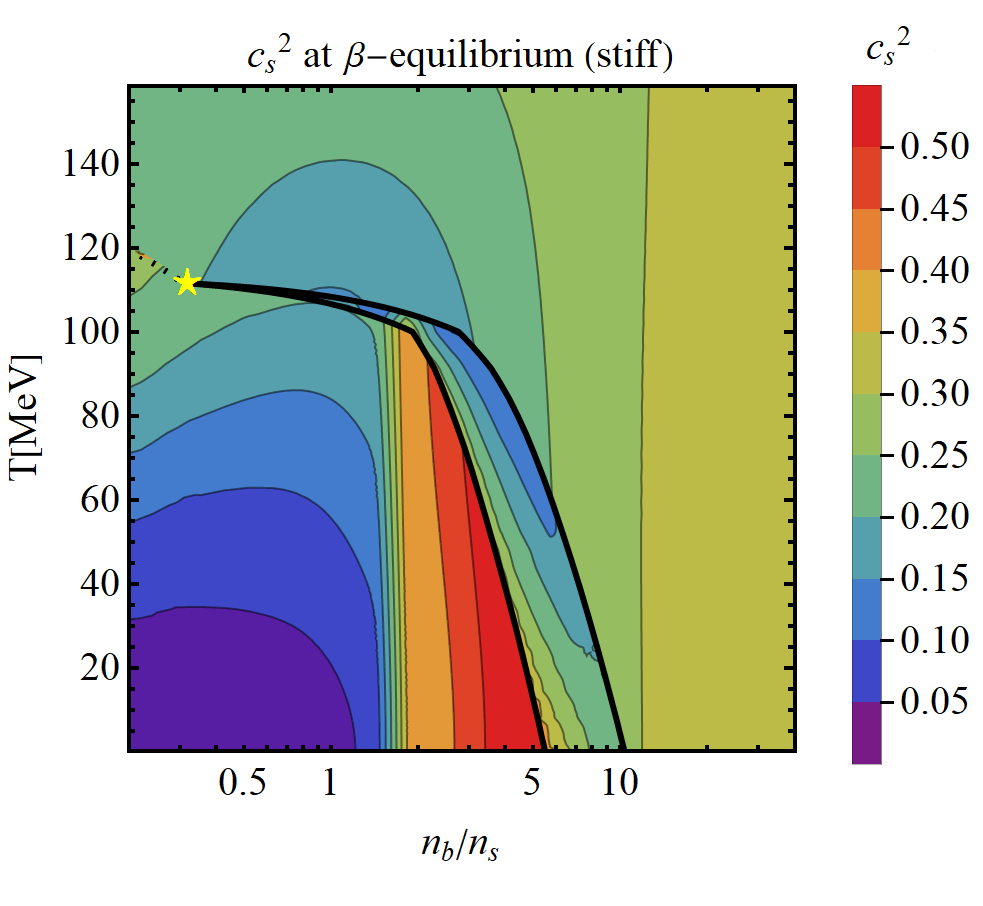}
    \caption{\small Contours of the speed of sound squared in beta-equilibrium for soft (left), intermediate (middle), stiff (right) EoSs.
    Solid black lines separate the mixed phase from NM and QM phases; yellow stars mark the locations of critical points
    whose numerical values are listed in Table~\ref{tab:NSprop}.
    }
    \label{fig:cs2}
\end{figure*}

\section{Results\label{sec:Result}}

We then analyze the thermodynamic properties of the constructed EoSs. We have carried out the steps outlined in Sec.~\ref{sec:comb} for all three versions of the cold hybrid EoSs, leading likewise to three versions of the final equation of state which depends on baryon number density, temperature, and charge fraction. The naming of the models is inherited from the cold EoSs, so that we refer to the three EoSs as ``soft'', ``intermediate'' and ``stiff'', according to the stiffness (i.e., basically the values of the speed of sound) of the EoS in the region of dense nuclear matter.
The stiffness is directly related to the maximal radii and masses of neutron stars described by the corresponding EoS.

Before going to the analysis of our EoSs, let us summarize how they agree with known constraints from various sources. Since the V-QCD QM model was fitted to lattice data, good agreement with lattice results at low density, including the first nontrivial Taylor expansion coefficient in the chemical potential, is guaranteed above the crossover temperature, $T\gtrsim 150$~MeV. 
Because the meson gas contribution is added to HS(DD2) at lower temperatures, agreement with QCD in this opposite region is obtained as well: it is known that even with our simple approximation, good agreement with lattice data, including higher order Taylor coefficients in chemical potential, is found~\cite{Karsch:2003vd,Karsch:2003zq,Vovchenko:2016rkn}. 
Moreover, the QM EoS agrees by construction~\cite{Alho:2012mh,Alho:2013hsa} with leading perturbative QCD results both at high temperatures and high chemical potentials  (see~\cite{Jokela:2018ers,Chesler:2019osn} for explicit comparison); since V-QCD is a strongly coupled model, more detailed agreement with higher order perturbative results is not possible. The hybrid EoSs, which we use for the cold EoSs, agree by construction with models at low density such as chiral effective theory computations. Actually, the cold hybrid EoSs are in excellent agreement (see~\cite{Jokela:2020piw,Jokela:2021vwy}) at all densities with model independent constructions of the EoSs such as polytropic interpolations between the known low and high density limits~\cite{Annala:2017llu,Most:2018hfd,Annala:2019puf,Annala:2021gom} (see also~\cite{Komoltsev:2021jzg}), and consequently also 
agree with constraints from neutron star measurements and from the GW170817 merger event~\cite{TheLIGOScientific:2017qsa,GBM:2017lvd}; see, e.g.,~\cite{Rezzolla:2017aly}.  We will also discuss these constraints explicitly below. Lastly, our model is in good agreement with finite temperature calculations in chiral effective theory as we shall discuss in more detail below.

In Fig.~\ref{fig:cs2} we plot the adiabatic speed of sound squared for our three models in beta equilibrium as function of baryon number density and temperature.
Solid black lines at intermediate densities represent phase boundaries between baryonic, mixed and quark phase, while dashed lines correspond to the artificial phase boundary of our construction at low densities and temperatures close to the crossover between baryonic and quark phase in QCD (see Appendix~\ref{app:CP}).
The speed of sound squared can be expressed in terms of first derivatives as
\begin{equation} \label{eq:cs2formula}
 c_s^2 = \frac{n^2\left[\left(\frac{s}{n_b}-\frac{\partial s}{\partial n_b}\right)^2 + \frac{\partial s}{\partial T}\frac{\partial \hat \mu}{\partial n_b} \right]}{\frac{\partial s}{\partial T}\left(\epsilon +p\right)},
\end{equation}
where we assumed charge neutrality, and all quantities are assumed to be functions of $T$, $n_b$, and $Y_q$. The chemical potential is defined as $\hat{\mu}=\mu_b+Y_q\mu_{le} = \partial f/\partial n_b$, where $\mu_b$ ($\mu_{le}$) is the baryon number (electron lepton number) chemical potential.
In all three models $c_s^2$ exceeds the value of $c_{s,\rm{CFT}}^2=1/3$ in conformal field theory close to the onset of the mixed phase. The maximal values for $c_s^2$  are $0.42$, $0.47$, and $0.59$ 
for  soft, intermediate and stiff variants, respectively. The contours in the mixed phase are also determined by Eq.~\eqref{eq:cs2formula}, but notice that this expression is not the physical speed of sound in this phase.

Yellow stars in Fig.~\ref{fig:cs2} mark the location ($T_c$,$n_{bc}$) of the critical point in the respective models whose precise values are listed in Table~\ref{tab:NSprop}.
The critical point has been analyzed in various models in the literature, and the results for the location vary in a wide range depending on the model~\cite{Stephanov:2004wx,DeWolfe:2010he,Ayala:2021tkm,Aryal:2021ojz}. Recent results in a simpler holographic approach~\cite{DeWolfe:2010he}, which extrapolates results for thermodynamics of QCD from lattice QCD to higher values of baryon chemical potential by using a bottom-up setup, are given by $\{T_c,\mu_{bc}\} = \{112,612\}$~MeV~\cite{Knaute:2017opk} and $\{T_c,\mu_{bc}\} = \{89,724\}$~MeV~\cite{Critelli:2017oub,Grefa:2021qvt}. These numbers are in the same ballpark with ours: we obtain in average slightly higher critical temperatures and lower critical chemical potentials. Notice also that the critical point in all three variants is outside the regime probed by the second phase of the beam energy scan at RHIC~\cite{Yang:2017llt}, but will be probed in future experiments at FAIR and at NICA. Finally, our numbers lie close to the chemical freeze-out curve extracted from heavy-ion experiments (see, e.g.,~\cite{STAR:2017sal,An:2021wof}). Our numbers are actually slightly below the experimental data, but consistent with the curve if the precision of the data and our approach are taken into account.

\begin{figure}
    \includegraphics[height=0.3\textwidth]{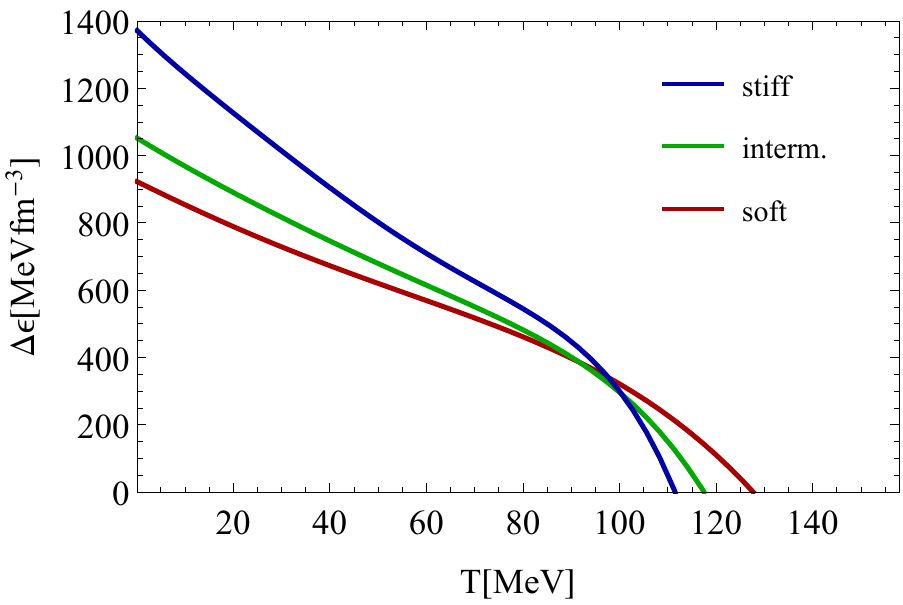}
     \caption{\small Latent heat as function of $T$ are shown for the three variants of EoSs. The location of critical endpoint is determined via the condition  $\Delta\epsilon(T_c,n_{bc})=0$.}
    \label{fig:deltaep}
\end{figure}

In Fig.~\ref{fig:deltaep} we plot the latent heat $\Delta \epsilon=\epsilon_{\text{QM}}-\epsilon_{\text{NM}}$, i.e., the difference between the energy density in the quark phase $\epsilon_{\text{QM}}$ and the nuclear matter phase $\epsilon_{\text{NM}}$, as function of the temperature.
For the three models we have analysed the soft (stiff) model leads to the smallest (largest) latent heat at small temperatures.
Curiously, at $T \approx 100$\,MeV all three models lead to approximately the same value of $\Delta \epsilon$.
The vanishing of the latent heat $\Delta \epsilon=0$ determines the location of the critical point. Notice however that as we pointed out above, the phase transition in our model is always of first order  $\Delta \epsilon$ does not vanish exactly but becomes small above certain temperatures, and the point where $\Delta \epsilon=0$ is obtained via extrapolation (see Appendix~\ref{app:CP} for details).

The temperature $T_c$ of the critical point is correlated with the stiffness of the respective model: larger stiffness results in lower values of $T_c$.
However, there is no clear relation between the critical density and stiffness, as the highest value of $n_{bc}$ is found for the intermediate model. In this case we interpret the variation of $n_{bc}$ as a rough measure of the precision for the value of the critical density. 

\begin{figure}[htb]
    \includegraphics[height=0.33\textwidth]{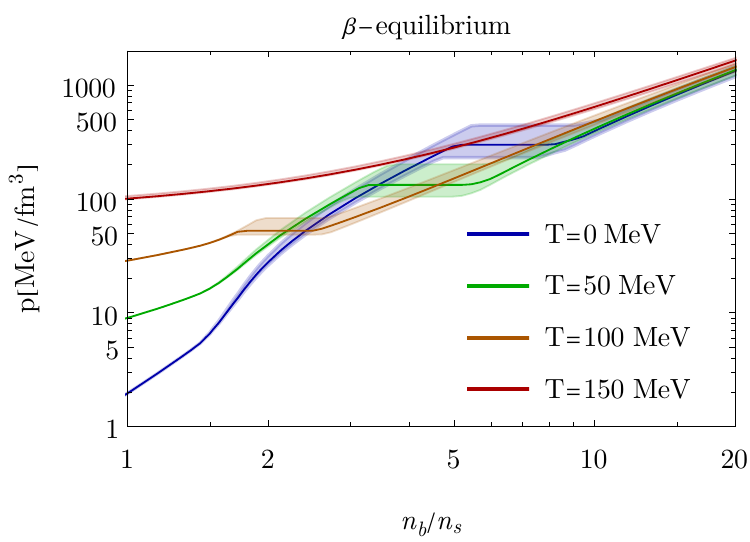}
    \caption{\small
    Pressure a function of the baryon number density in beta-equilibrium for different values of the temperatures.
    The lower (upper) bounds of the coloured bands represent the soft (stiff) model, while central curves correspond to the intermediate version. 
    }
    \label{fig:beta}
\end{figure}

In Fig.~\ref{fig:beta} we plot the pressure as function of the baryon number density at different values of the temperature in beta equilibrium.
The lower (upper) bounds of the coloured bands represent the soft (stiff) model, while the central curves correspond to the intermediate version.
The results in the nuclear matter phase for $n_{b}/n_s<1.4$ are the same for all the three models as there is not input from holography in this region. 
The uppermost curve ($T=150$~MeV) shows the EoS in the quark matter phase entirely described by V-QCD, which is therefore slightly different for all three models even at low $n_b$.
The plateau at intermediate densities is a manifestation of the strong first-order phase transition of the V-QCD model.
The transition density decreases with increasing temperature, and the transition becomes weaker,
as also can be seen from Fig.~\ref{fig:deltaep}.

\begin{figure}[htb]
     \includegraphics[height=0.33\textwidth]{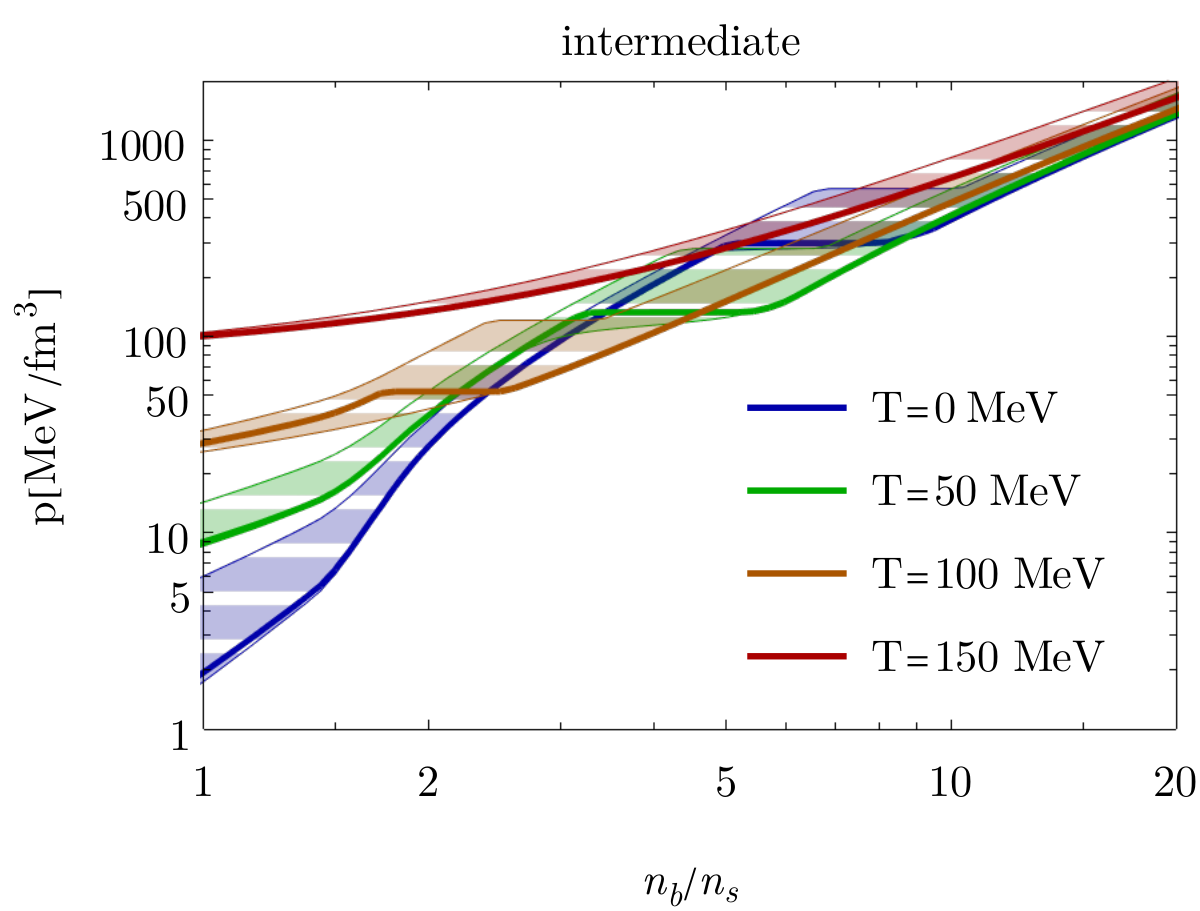}
    \caption{
    \small Pressure for the intermediate model for different values of the temperatures. 
    Solid curves are beta equilibrium, while upper and lower bounds of the hatched coloured bands represent maximum and minimum values of the pressure at given temperature and density.
    }
    \label{fig:pYe}
\end{figure}

To illustrate impact of $Y_q$ outside beta-equilibrium we plot in Fig.~\ref{fig:pYe} the range in pressure that is covered by the intermediate model at various values of the temperature.
Solid curves show the pressure in beta equilibrium, while upper and lower bounds of the hatched coloured bands are maximum and minimum values, respectively at given density and temperature.
Overall the pressure is lowest close to and increases away from beta-equilibrium, except close to the onset of the mixed phase, where the pressure can take smaller values also away from beta-equilibrium.
The variation of the pressure as function of $Y_q$ is largest in the mixed phase.
\begin{figure}
    \includegraphics{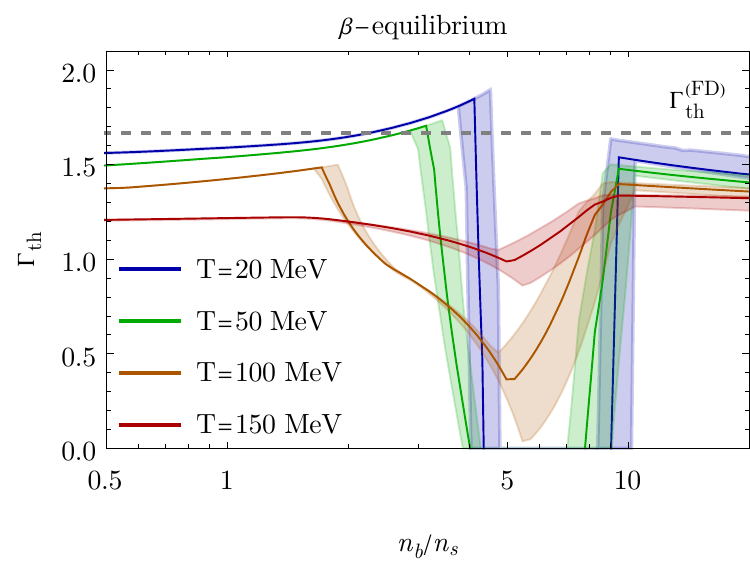}
    \caption{\small Thermal index in beta-equilibrium. Notation for the bands as in Fig.~\protect\ref{fig:beta}.
    }
    \label{fig:Gth}
\end{figure}

A quantity that is useful to explore the thermal contributions to the energy density and the pressure is the thermal index defined as 
\begin{equation}
\Gamma_{\rm{th}}(T,n_b,Y_q)=1+\frac{p(T,n_b,Y_q)-p(0,n_b,Y_q)}{\epsilon(T,n_b,Y_q)-\epsilon(0,n_b,Y_q)}\,.
\end{equation}
In Fig.~\ref{fig:Gth} we show the thermal index on the beta equilibrium slice for various values of the temperature.
As in Fig.~\ref{fig:beta}, the upper (lower) bound of the colored bands represent the stiff (soft) model, while the central curves represent the intermediate case.
The dashed grey line is the thermal index of a free Fermi gas $\Gamma^{(FD)}_{\rm{th}}=5/3$ which is independent of density and temperature.
Any deviations of $\Gamma_{\rm{th}}(T,n_b,Y_q)$ from $\Gamma^{(FD)}_{\rm{th}}$ are due to thermal interaction effects.
Results for the thermal index were computed at low temperatures and up to densities between $1n_s$ and $2n_s$ by using chiral effective theory (CET) in~\cite{Carbone:2019pkr,Keller:2020qhx}. It is important to compare our EoS to these results in particular because the temperature dependence in our setup in this region is based on the vdW setup, which is not guaranteed to be realistic enough to agree with the CET calculations. Interestingly, our result show good overall agreement with the CET predictions, i.e., deviations are mostly below 10\% in the relevant region. Below $\sim 0.5n_s$ the CET predicts values above 1.6 which are slightly higher than our $T=20$~MeV curve. Also the CET analysis predicts mild decrease of the thermal index with increasing density, whereas in our model the index mildly increases with increasing density.
In neutron star simulations a constant thermal index $\Gamma_{\rm{th}}=1-2$ is often assumed to mimic such finite temperature effects \cite{Bauswein:2010dn,DePietri:2019mti,Xie:2020udh}, where $\Gamma_{\rm{th}}\approx 1.7$ has been argued \cite{Figura:2020fkj} to best approximate the dynamical and thermodynamical behavior of neutron star merger simulations with microscopic prescription of finite-temperature effects.
The thermal index in our construction remains also well within these bounds at temperatures relevant in such simulations, except in and close to the mixed phase, where $\Gamma_{\rm{th}}$ can take values smaller smaller than one. 
As expected, thermal interaction effects become more important at higher temperature, where the deviations of the thermal index from $\Gamma^{(FD)}_{\rm{th}}$ are largest.

\begin{figure}
    \includegraphics{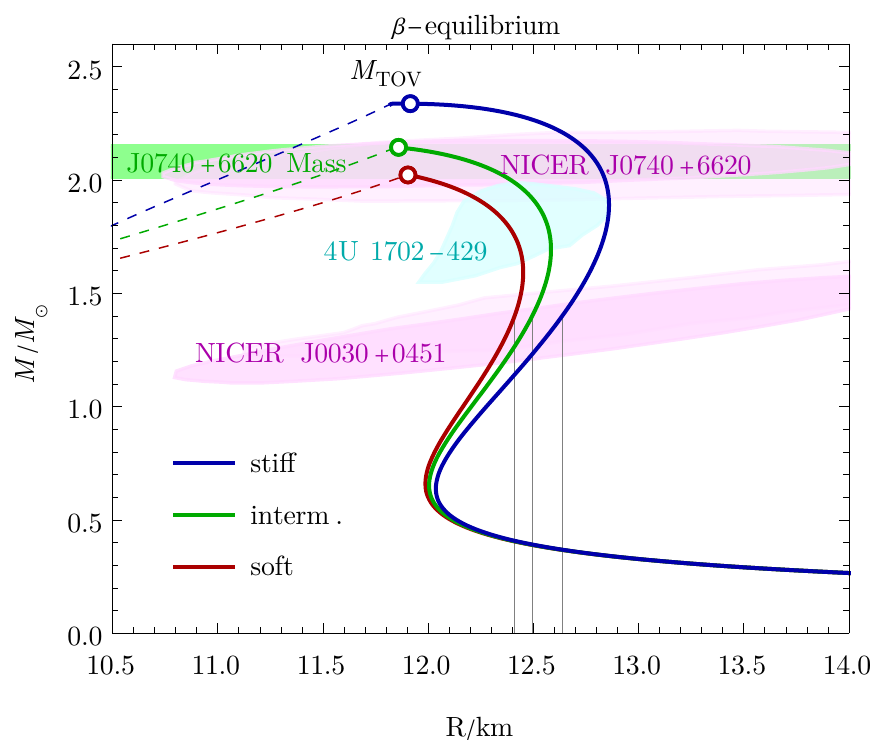}
    \caption{\small Mass-radius relation of non-rotating stars.
    }
    \label{fig:ye}
\end{figure}

Finally, in Fig.~\ref{fig:ye} we show the mass-radius relation of cold isolated non-rotating neutron stars for the soft (red), intermediate (green) and stiff (blue) model, together with the relevant observational constraints.
Solid parts of the curves represent purely baryonic stars, while dashed parts belong to stars with quark matter cores, which in our model turn out to be unstable.
Circles mark the maximum mass of stable non-rotating stars ($M_\mathrm{TOV}$) of the respective model.
For the soft and intermediate model $M_\mathrm{TOV}$ is determined by the onset of the phase transition at which the star becomes unstable to black hole collapse.
For the stiff model the maximum mass is already reached in the baryonic phase and the phase transition is realised only in the unstable branch of the mass-radius sequence.
The green band shows the  
result  
$M_\mathrm{TOV}=2.08 \pm 0.07 M_\odot$ 
from direct mass measurements of the pulsar J0740+6620~\cite{NANOGrav:2019jur,Fonseca:2021wxt}, which sets a lower bound on the maximum mass of nonrotating neutron stars.
Pink ellipses are radius measurements from the NICER experiment for the pulsars J0030+0451~\cite{Riley:2019yda,Miller:2019cac} and J0740+6620~\cite{Miller:2021qha,Riley:2021pdl}, while cyan area is from the measurement of the X-ray binary 4U 1702-429~\cite{Nattila:2017wtj}.
All three examples pass the observational constraints shown in the plot in addition to the constraint on the tidal deformability $\Lambda_{1.4}<580$ of a $M=1.4M_\odot$ star deduced from the analysis of GW170817 by LIGO/Virgo (low-spin prior at 90\% confidence level) \cite{LIGOScientific:2018cki} (see also Table~\ref{tab:NSprop} and the detailed analysis in~\cite{Jokela:2021vwy}).

\begin{table}[h!tb]
 \caption{
 EoS and neutron star properties in beta equilibrium.
 }
 \begin{tabular}{ccccccccc}
 \hline
 \hline
  Model & $\frac{n_{\text{bc}}}{n_{\text{s}}}$ & $\frac{\mu_{bc}}{\text{MeV}}$ & $\frac{T_{\text{c}}}{\text{MeV}}$ & $\frac{M_{\mathrm{TOV}}}{M_\odot}$ &  $\frac{R_{\mathrm{e},1.4}}{\mathrm{km}}$ & $\Lambda_{1.4}$  
  \\
  \hline\hline
  soft & $0.46$  
  & $485$  
  & $128$ 
  & $2.02$ & $12.41$ & $483$  
  \\
  interm. & $0.62$  
  & $575$  
  & $118$ 
  & $2.14$ & $12.50$ & $511$
  \\
  stiff & $0.32$  
  & $565$ 
  & $112$
  & $2.34$ & $12.64$ & $560$ 
  \\
  \hline
  HS(DD2) & -- & -- & -- & $2.45$ & $13.2$ & $686$ 
  \\
  \hline
\end{tabular}
\label{tab:NSprop} 
\end{table}

\section{Conclusion and Outlook\label{sec:Conclusion}}

In this article, we presented a novel framework for the EoS of hot and dense QCD which combines ingredients from various approaches in different regions of the phase diagram including gauge/gravity duality, van der Waals model of nuclear matter, statistical models, and relativistic mean field theory. The aim was to establish an EoS which uses the best available modeling in each of the regimes. An essential new input here was the holographic V-QCD model, which we used to cover the region of intermediate densities where computing theoretical predictions is particularly hard. We presented three versions of the EoS which are in good agreement with QCD data and constraints from measurements of neutron stars and neutron star mergers. Using these models we derived, apart from the properties of the EoS itself, predictions for the location of the critical end point of the nuclear to quark matter transition.

Some details in our approach merit further study. It is expected (see, e.g.,~\cite{Alford:2007xm}) that at low temperatures in the quark matter phase, pairing of quarks takes place, leading to potentially complex phase diagram with various paired, color superconducting phases. Recently, there has been a lot of interest in the analysis of such phases in gauge/gravity duality~\cite{Chen:2009kx,Basu:2011yg,BitaghsirFadafan:2018iqr,Faedo:2018fjw,Ghoroku:2019trx,Henriksson:2019zph}. Future work will study the inclusion of such phases in the V-QCD model and their effect on the EoS. 

While the main focus in this article was the temperature dependence, we also included the dependence on the charge fraction following the HS(DD2) model in the nuclear matter phase, and by using a simple approximation assuming a strongly interacting component of free energy only depending on the total baryon number (and with free electron gas) in the quark matter phase. This latter approach can be improved by including proper flavor dependence in the holographic model, so that one can consider states with unequal amount of different quark flavors. This extension of the model is the topic of ongoing research. 

The inclusion of flavor dependence will also help to generalize the model to analyze transport properties of QCD matter. The strongly interacting components of conductivities and viscosities of QM at high density were solved in \cite{Hoyos:2020hmq,Hoyos:2021njg}. Including flavor dependence will help to properly analyze the correlators of electric and weak currents, which are necessary to estimate the effects of electron and neutrino transport, known to be important, among other things, for neutron star cooling, in core-collapse supernovae, and for the behavior of ejected matter in neutron star mergers (see, e.g.,~\cite{Schmitt:2017efp}). 

Furthermore in this work the treatment of the cross-over at low densities and intermediate temperature was rather simple minded: the EoSs actually have  (very weak) first order phase transition instead of a crossover. It will be interesting to study if the EoS can be improved  in this region by matching the meson gas pressure with the high-temperature holographic quark-gluon plasma pressure more carefully, e.g., by extending the methods of~\cite{Alho:2015zua} to finite density.

We should stress that the framework presented here admits immediate natural generalizations where one replaces some of the building blocks of the model by other approaches. As for the strongly coupled gauge/gravity duality model there is (to our knowledge) currently no alternative to V-QCD that would allow to repeat the analysis as done in this article. This field is however evolving rapidly~\cite{Hoyos:2016zke,Annala:2017tqz,BitaghsirFadafan:2019ofb,Mamani:2020pks,BitaghsirFadafan:2020otb,Ghoroku:2021fos,Hoyos:2021uff}. For example, very recently it was demonstrated that a setup with nuclear matter in the Witten-Sakai-Sugimoto model may lead to realistic neutron stars~\cite{Kovensky:2021kzl}.
For the temperature dependence of dense nuclear matter we used the vdW model for two reasons: it is simple, and agrees reasonably well with the holographic model at zero temperature. A simple model may be the best guess for the temperature dependence at densities around and above the saturation density, where none of the known approaches is reliable. However, also other choices are possible. We have checked that using (for example) the temperature dependence of the HS(DD2) model instead only leads to rather mildly modified EoSs. It would be also interesting to study the Carnahan-Starling generalization of the excluded volume effect in the vdW model, recently considered in~\cite{Fujimoto:2021dvn}, since it may improve the temperature dependence of the EoS at low but nonzero temperatures near the saturation density, where ab initio results are available. Another simple generalization of our approach is to use some other general purpose EoSs for the low density region than the HS(DD2) model.

Apart from varying the building blocks of the model, a future study could further explore the parameter space of the model presented in this article. This would mean, at least, including additional variants of the V-QCD model~\cite{Jokela:2018ers} an studying the effect of varying the excluded volume parameter $v_0$ in the vdW model.

The EoSs constructed in this article will be published in the standard format in the CompOSE database, and can therefore immediately to be used in state-of-the-art simulations of neutron star mergers and core-collapse supernovae. Indeed, there is growing interest in effects arising due to the temperature dependence and the phase transition in merger simulations~\cite{Most:2018eaw,Bauswein:2018bma,Raithel:2021hye,Prakash:2021wpz}. 
Work on applying the EoSs of this article in such simulations is already in progress.

\begin{acknowledgments}
We thank U. G\"ursoy, N. Jokela, E. Kiritsis, L. Rezzolla, W. van der Schee, and A. Vuorinen for useful discussions and comments. 
T.~D. and M.~J. have been supported 
by an appointment to the JRG Program at the APCTP through the Science and Technology Promotion Fund and Lottery Fund of the Korean Government. T.~D. and M.~J. have also been supported by the Korean Local Governments -- Gyeong\-sang\-buk-do Province and Pohang City -- and by the National Research Foundation of Korea (NRF) funded by the Korean government (MSIT) (grant number 2021R1A2C1010834). T. D. would like to thank the Department of Physics of Bo\u{g}azi\c{c}i University for their hospitality during his visit.
C.~E. acknowledges support by the Deutsche Forschungsgemeinschaft (DFG, German Research Foundation) through the CRC-TR 211 'Strong-interaction matter under extreme conditions'-- project number 315477589 -- TRR 211.
\end{acknowledgments}

\appendix

\section{Technical details on the EoS}\label{app:eosdetails}

In this Appendix, we discuss minor technical details which are not important for our conclusions, but need to be discussed for all the results to be fully reproducible.

\subsection{Details of the construction of the cold EoS}\label{app:coldEoS}

We first give more details on the construction of the cold hybrid EoS. Recall that the EoS of cold nuclear matter, as explained in Sec.~\ref{sec:coldEoS}, requires matching between EoSs from a nuclear theory model (in this article, the APR model) at low density with V-QCD nuclear matter at higher density.  As we explained in main text, in order for the cold EoS to be smoother, we choose two matching densities, instead of the single density approach of~\cite{Ecker:2019xrw,Jokela:2020piw}, and connect the speeds of sound from the two approaches though a linear interpolation of the squared speed of sound $c_s^2$ as a function of the baryon number density $n_b$, therefore avoiding a discontinuity in the speed of sound.

To be precise, the matching is carried out as follows. For $n_b<1.4n_s \equiv n_\mathrm{tr}^{(1)}$ we use the APR EoS as such. We then take the speed of sound for $n_\mathrm{tr}^{(1)}<n_b<1.8n_s \equiv n_\mathrm{tr}^{(2)}$ to be
\begin{equation}
    \left(c_s(n_b)\right)^2 = \left(c_s(n_\mathrm{tr}^{(1)})\right)^2 + \kappa \, (n-n_\mathrm{tr}^{(1)})
\end{equation}
where the slope $\kappa$ is a free parameter. The rest of the thermodynamic functions are then obtained  by integration for $n_\mathrm{tr}^{(1)}<n_b< n_\mathrm{tr}^{(2)}$ so that $\kappa$ is the only free parameter. We then require that the pressure, the baryon number chemical potential, and the speed of sound are continuous at $n_b=n_\mathrm{tr}^{(2)}$. These three conditions then determine $\kappa$ as well as the two parameters of V-QCD nuclear matter $c_b$ and $b$ (these parameters appear in the nuclear matter action $S_{nm}$ of~\eqref{action}, see~\cite{Jokela:2020piw} for their definitions).  These latter two parameters were also determined by matching in the simpler approach of~\cite{Ecker:2019xrw,Jokela:2020piw}.

We also modify the cold hybrid EoS in the crust region, $n_b \ll n_s$, before using it to construct the adjusted vdW EoS through~\eqref{eq:Deltaf}, in order to remove some noise from the vdW EoS. This noise would appear because the fine details of the beta-equilibrium, zero temperature APR EoS which turns out to be inconsistent with the $Y_q$-dependence of the HS(DD2) EoS at low density. One might wonder why we need to do this, since we anyhow replace the vdW EoS with the exact HS(DD2) through~\eqref{eq:DD2vdWm} at low densities. However, since we use a smooth weight function instead of an abrupt cutoff, noise in the vdW EoS at low densities would remain in the final matched EoS even if it would be heavily suppressed. We therefore implement the correction by using the HS(DD2) at beta equilibrium for the cold EoS $f_\mathrm{cold}(n_b)$ when $n_b<0.008 n_s$. At this value the pressures of the APR and HS(DD2) EoSs cross, and we further introduce a small shift in the baryon chemical potential for $n_b<0.008 n_s$ in order to make it continuous at $n_b=0.008 n_s$. That is, we in effect introduce a second order phase transition. We stress however that these modifications only remove noise from heavily suppressed terms at low density in our final results.

\subsection{Details of the construction of the three dimensional EoS}\label{app:3dEoS}

We then discuss some technical details on the determination of the final three dimensional EoS and its components.

We carried out some minor but nontrivial modifications in the three dimensional nuclear matter EoS $f_\mathrm{NM}(T,n_b,Y_q)$. The transition from non-uniform to uniform nuclear matter in HS(DD2) proceeds via a first order phase transition at around $0.3 n_s \lesssim n_b \lesssim 0.5 n_s$, at small temperatures $T \le 5$~MeV, and values of $Y_q$ (mostly) far from beta equilibrium~\cite{Hempel:2009mc}. Due to the smooth matching of~\eqref{eq:DD2vdWm}, the mixed phase from this transition causes the resulting nuclear matter to be inconsistent in this narrow range of parameters, so that a thermodynamically unstable region appears. In order to fix this, we carried out a simple one-dimensional Maxwell construction for each value of $T$ and $Y_q$ which removed the inconsistency.

Let us also specify how the EoS for the mixed phase between the NM and QM phases was calculated. It arises from requiring full chemical equilibrium, so that the coexisting phases have the same pressure as well as baryon number and electron lepton number chemical potentials \begin{eqnarray}
 \mu_{le} &=& \frac{1}{n_b}\! \left.\frac{\partial f}{\partial Y_q}\right|_{n_b,T}\  ,\\ \quad \mu_b &=& \left.\frac{\partial f}{\partial n_b}\right|_{n_b Y_q,T} = \left.\frac{\partial f}{\partial n_b}\right|_{Y_q,T} - Y_q  \mu_{le} \ .
\end{eqnarray}

In practice, the mixed phase is found as follows. 
Demanding equilibrium between the phases, we need to solve the following set of equations:
\begin{eqnarray} \label{eq:pmixeq}
 p_\mathrm{NM}(T,n_b^{(1)},Y_q^{(1)}) &=&      p_\mathrm{QM}(T,n_b^{(2)},Y_q^{(2)}) \ , \\
 \mu_{b}^{(\mathrm{NM})}(T,n_b^{(1)},Y_q^{(1)}) &=&      \mu_{b}^{(\mathrm{QM})}(T,n_b^{(2)},Y_q^{(2)}) \ , \\
 \mu_{le}^{(\mathrm{NM})}(T,n_b^{(1)},Y_q^{(1)}) &=&      \mu_{le}^{(\mathrm{QM})}(T,n_b^{(2)},Y_q^{(2)}) \ . \label{eq:mulemixeq}
\end{eqnarray}
We have three conditions and four variables $n_b^{(i)}$, $Y_q^{(i)}$ so the solution will involve one parameter which we call $\gamma$. The solution defines two curves on the $(n_b,Y_q)$-plane, parameterized in terms of $\gamma$, and a mapping between the curves. The mixed phase is found between the curves. The temperature is a ``trivial'' parameter in these equations and we will not denote the dependence on it explicitly below. The construction can be carried out independently for each value of the temperature.

The mixed phase is then a mixture of NM and QM matter in the equilibrium defined by Eqs.~\eqref{eq:pmixeq}--\eqref{eq:mulemixeq}. The thermodynamic functions are most easily written in a parametric representation using the $\gamma$ parameter and the volume fraction $\alpha$ of the NM phase. That is, we may write
\begin{eqnarray}
    n_b(\alpha,\gamma) &=& \alpha  n_b^{(1)}(\gamma ) + (1-\alpha)  n_b^{(2)}(\gamma ) \\
    Y_q(\alpha,\gamma) n_b(\alpha,\gamma) &=& \alpha  Y_q^{(1)}(\gamma )n_b^{(1)}(\gamma )\nonumber\\
    &&\!\!+ (1-\alpha)Y_q^{(2)}(\gamma )  n_b^{(2)}(\gamma ) \\
    f(\alpha,\gamma) &=& \alpha f_\mathrm{NM}(n_b^{(1)}(\gamma),Y_q^{(1)}(\gamma))\nonumber\\
    &&\!\!+  (1-\alpha) f_\mathrm{QM}(n_b^{(2)}(\gamma),Y_q^{(2)}(\gamma)) \ .
\end{eqnarray}
separately for each temperature slice. Note that curves of constant $\gamma$ are straight lines on the $(n_b,Y_q n_b)$-plane. By construction, $p$, $\mu_b$, and $\mu_{le}$ take constant values on these lines.

Notice that in the main text we focused on defining the free energy, which is the natural thermodynamic potential in the canonical ensemble with the parameters $T$, $n_b$, and $Y_q$. Determining first numerically the free energy, and computing the other thermodynamic functions by using it as input, is indeed enough to determine all thermodynamics. However, this procedure requires taking numerical derivatives, which tend to increase numerical noise. When computing the final results for other observables (such as the entropy and the chemical potentials) we have therefore first computed analytically the consequences of the various matching formulas, and computed these quantities directly from the corresponding quantities of the EoSs being matched, avoiding the use of numerical derivatives as much as possible.

\begin{figure*}
    \includegraphics[height=0.31\textwidth]{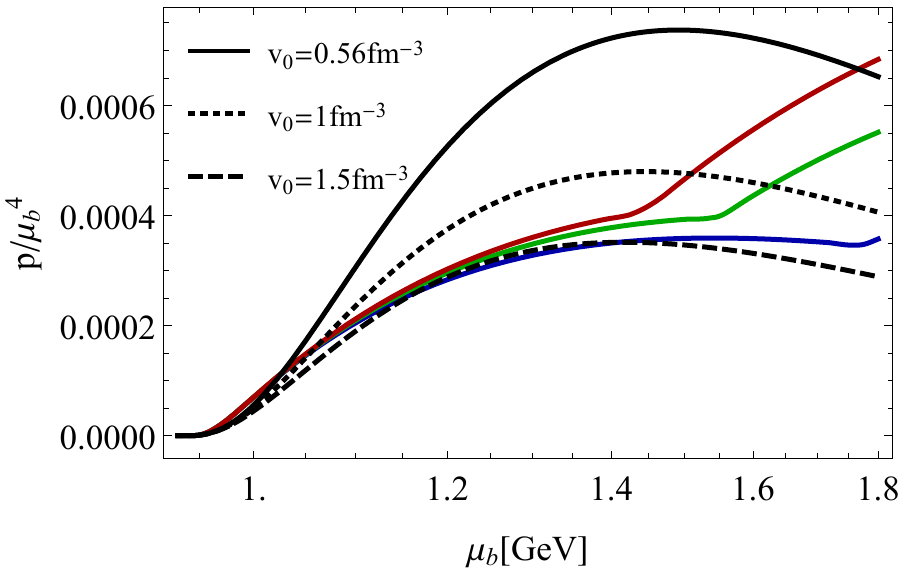}
     \includegraphics[height=0.32\textwidth]{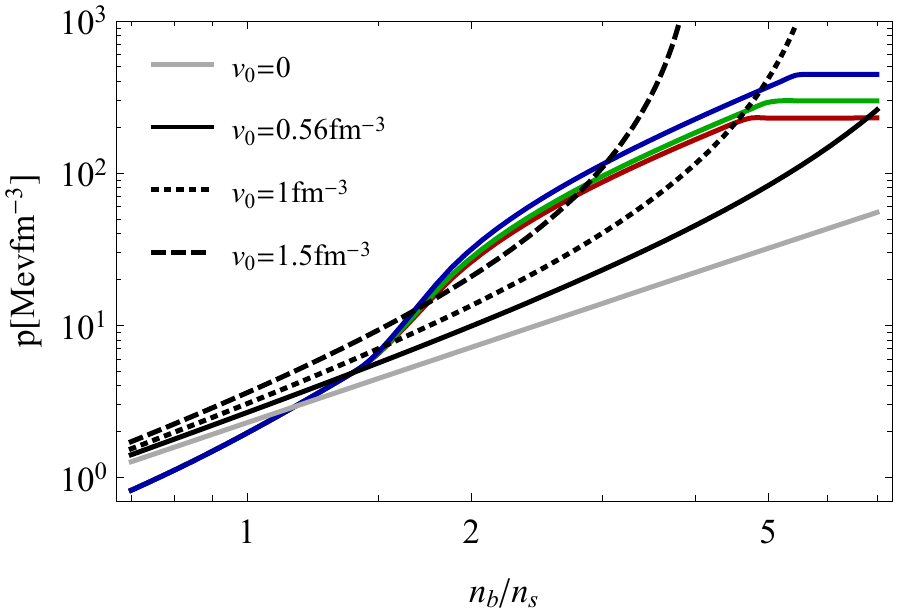}
    \caption{\small Comparison of cold V-QCD EoSs and EoSs of hadron gas with excluded volume correction. Left: normalized pressure ($p/\mu_b^4$) in terms of baryon chemical potential $\mu_b$. Right: pressure in terms of baryon number density $n_b$ in units of $n_s$. In both figures, stiff, intermediate,  soft variants are denoted by blue, green, red curves respectively and hadron gas with excluded volume of $v_0=0.56\mathrm{fm}^{-3}$, $v_0=1\mathrm{fm}^{-3}$and $v_0=1.5\mathrm{fm}^{-3}$ are exhibited by solid, dotted, dashed black curves. In addition, ideal hadron gas result is shown with the gray solid curve in the figure on  the right.}
    \label{fig:v0_comp}
\end{figure*}

\section{Comparison of the vdW EoS to the V-QCD NM EoS}\label{app:vdWcomparison}

A central motivation for the use of the vdW EoS in the article is its agreement with the predicted EoS of cold NM by V-QCD. In this Appendix, we study this by comparing the three different versions of the cold hybrid V-QCD(APR) EoSs to simple vdW EoSs, i.e., those only with electrons, protons, and neutrons with a constant excluded volume correction for the nucleons.

The EoSs are compared in Fig.~\ref{fig:v0_comp}. In both figures, stiff, intermediate, soft variants of V-QCD EoSs up to the onset of the phase transition are showed via blue, green, red curves respectively and simple vdW EoSs with EV parameter $v_0=0.56\mathrm{fm}^{-3}$ (our choice), $v_0=1 \mathrm{fm}^{-3}$, $v_0=1.5 \mathrm{fm}^{-3}$ are denoted by  solid black, dotted, dashed curves. 
The figure on the left shows the baryon chemical potential dependence of the dimensionless  pressure $p/\mu_b^4$, and 
the figure on the right shows the pressure  
as a function of the baryon number density.  
The gray curve in the right-hand side plot 
shows 
the ideal hadron gas result. This curve is not shown in the left plot since it mostly stays outside the plotted range.  It is transparent from the figures that the chemical potential dependence is more sensitive to $v_0$ at small values of the parameter.

We notice that the EV corrected EoS is relatively close to the cold hybrid V-QCD EoSs, in particular when the pressure is plotted as a function of the chemical potential. This happens in part because the left hand plot of Fig.~\ref{fig:v0_comp} ``zooms'' into the region of higher densities where agreement is better. Notice that the range of values of $v_0$ is typical for EV corrections in nuclear matter, see, e.g.,~\cite{Vovchenko:2017zpj}. The best fit between the EV corrected pressure and V-QCD is found around $v_0\approx 1.5 \mathrm{fm}^{-3}$. We however choose a smaller value $v_0\approx 0.56 \mathrm{fm}^{-3}$, because for this value, the potential term of~\eqref{eq:Deltaf} is more natural: this value corresponds to a potential which is attractive (repulsive) at long (short) distances, whereas for larger $v_0$ the potential would be attractive at both short and long distances with an intermediate repulsive range in between. This unnatural behavior is reflected in the nonmonotonic dependence of the pressure difference between the EV corrected and V-QCD pressures as a function of $n_b$ in the left plot of Fig.~\ref{fig:v0_comp}.

\section{Determination of the critical point}\label{app:CP}

As discussed in Sec.\ref{Dcrit}, the EoSs have 
a strong phase transition at low temperature that becomes weaker at higher temperature.
Continuity over the phases is not possible due to limitation of our construction.
Hence, we interpret the weakening in the transition as signal for a crossover at high temperature. 
Using this interpretation, it is possible to calculate the latent heat and obtain an estimate for the location of the critical end point. 
The latent heat is 
\begin{equation}
    \Delta \epsilon(T)=\epsilon_{\text{QM}}(T,n_b^{(2)}\!,Y_q^\mathrm{eq}(n_b^{(2)}))-\epsilon_{\text{NM}}(T,n_b^{(1)}\!,Y_q^\mathrm{eq}(n_b^{(1)})),
\end{equation}
where $Y_q^{eq}$ is the value at beta-equilibrium and $n_b^{(2)}$, $n_b^{(1)}$ are respectively the values for the baryon number density at the onset and the end of the phase transition for a given value of $T$.
Due to the absence of a crossover, the transition lines get close, but do not intersect.
However, the latent heat as function of temperature shows a clean trend. 
By using it, we perform a $5^{th}$-order polynomial fit 
to the data in the range of strong first order transition, i.e., $T \lesssim 115$~MeV.
The fit functions that we obtained are 
\begin{eqnarray}
 \Delta\epsilon_\mathrm{stiff} &=&1.37\times10^{3} - 1.41\times10 T + 
 1.53 \times10^{-1} T^2  \nonumber\\ 
 &&-4.36\times10^{-3} T^3+6.26\times10^{-5} T^4\nonumber\\ 
 && - 3.10\times10^{-7} T^5,\nonumber\\
\Delta\epsilon_\mathrm{interm} &=&1.05\times10^{3} - 8.85 T + 
 5.01\times10^{-2} T^2 \nonumber\\ 
&&-9.04\times10^{-4} T^3+ 1.28\times10^{-5} T^4\nonumber\\ 
 && - 7.53\times10^{-8} T^5,\nonumber\\
\Delta\epsilon_\mathrm{soft}&=&9.23\times10^{2} - 7.07 T +  1.40\times10^{-2} T^2   \nonumber\\ &&-2.90\times10^{-4} T^3-3.40\times 10^{-6} T^4\nonumber\\ 
 &&+ 1.53\times10^{-9} T^5.\nonumber
\end{eqnarray}
The critical temperature is then found by extrapolating to the point where the latent heat vanishes: $\Delta \epsilon(T_c)=0$.
To determine value of the baryon number density at critical end point $n_{bc}$, we use 
the geometric mean 
of the density values on the two transition lines   
at $T_c$. We estimate the uncertainty of this procedure to be below $0.05 n_s$ for the soft EoS and $0.1 n_s$ for the intermediate and stiff EoSs by studying the variation caused by varying the choice of the mean. While this uncertainty is arguably sizeable, it is smaller than the differences between the numbers for the different EoSs in Table~\ref{tab:NSprop}.
Then for the phase diagram, crossover (dashed) lines  are also computed by calculating 
the geometric mean
of $n_b$ on the transition lines at the temperature values above $T_c$.

\bibliography{prx}

\begin{thebibliography}{114}%
\makeatletter
\providecommand \@ifxundefined [1]{%
 \@ifx{#1\undefined}
}%
\providecommand \@ifnum [1]{%
 \ifnum #1\expandafter \@firstoftwo
 \else \expandafter \@secondoftwo
 \fi
}%
\providecommand \@ifx [1]{%
 \ifx #1\expandafter \@firstoftwo
 \else \expandafter \@secondoftwo
 \fi
}%
\providecommand \natexlab [1]{#1}%
\providecommand \enquote  [1]{``#1''}%
\providecommand \bibnamefont  [1]{#1}%
\providecommand \bibfnamefont [1]{#1}%
\providecommand \citenamefont [1]{#1}%
\providecommand \href@noop [0]{\@secondoftwo}%
\providecommand \href [0]{\begingroup \@sanitize@url \@href}%
\providecommand \@href[1]{\@@startlink{#1}\@@href}%
\providecommand \@@href[1]{\endgroup#1\@@endlink}%
\providecommand \@sanitize@url [0]{\catcode `\\12\catcode `\$12\catcode
  `\&12\catcode `\#12\catcode `\^12\catcode `\_12\catcode `\%12\relax}%
\providecommand \@@startlink[1]{}%
\providecommand \@@endlink[0]{}%
\providecommand \url  [0]{\begingroup\@sanitize@url \@url }%
\providecommand \@url [1]{\endgroup\@href {#1}{\urlprefix }}%
\providecommand \urlprefix  [0]{URL }%
\providecommand \Eprint [0]{\href }%
\providecommand \doibase [0]{http://dx.doi.org/}%
\providecommand \selectlanguage [0]{\@gobble}%
\providecommand \bibinfo  [0]{\@secondoftwo}%
\providecommand \bibfield  [0]{\@secondoftwo}%
\providecommand \translation [1]{[#1]}%
\providecommand \BibitemOpen [0]{}%
\providecommand \bibitemStop [0]{}%
\providecommand \bibitemNoStop [0]{.\EOS\space}%
\providecommand \EOS [0]{\spacefactor3000\relax}%
\providecommand \BibitemShut  [1]{\csname bibitem#1\endcsname}%
\let\auto@bib@innerbib\@empty
\bibitem [{\citenamefont {Keller}\ \emph {et~al.}(2021)\citenamefont {Keller},
  \citenamefont {Wellenhofer}, \citenamefont {Hebeler},\ and\ \citenamefont
  {Schwenk}}]{Keller:2020qhx}%
  \BibitemOpen
  \bibfield  {author} {\bibinfo {author} {\bibfnamefont {J.}~\bibnamefont
  {Keller}}, \bibinfo {author} {\bibfnamefont {C.}~\bibnamefont {Wellenhofer}},
  \bibinfo {author} {\bibfnamefont {K.}~\bibnamefont {Hebeler}}, \ and\
  \bibinfo {author} {\bibfnamefont {A.}~\bibnamefont {Schwenk}},\ }\href
  {\doibase 10.1103/PhysRevC.103.055806} {\bibfield  {journal} {\bibinfo
  {journal} {Phys. Rev. C}\ }\textbf {\bibinfo {volume} {103}},\ \bibinfo
  {pages} {055806} (\bibinfo {year} {2021})},\ \Eprint
  {http://arxiv.org/abs/2011.05855} {arXiv:2011.05855 [nucl-th]} \BibitemShut
  {NoStop}%
\bibitem [{\citenamefont {Most}\ \emph {et~al.}(2019)\citenamefont {Most},
  \citenamefont {Papenfort}, \citenamefont {Dexheimer}, \citenamefont
  {Hanauske}, \citenamefont {Schramm}, \citenamefont {St\"ocker},\ and\
  \citenamefont {Rezzolla}}]{Most:2018eaw}%
  \BibitemOpen
  \bibfield  {author} {\bibinfo {author} {\bibfnamefont {E.~R.}\ \bibnamefont
  {Most}}, \bibinfo {author} {\bibfnamefont {L.~J.}\ \bibnamefont {Papenfort}},
  \bibinfo {author} {\bibfnamefont {V.}~\bibnamefont {Dexheimer}}, \bibinfo
  {author} {\bibfnamefont {M.}~\bibnamefont {Hanauske}}, \bibinfo {author}
  {\bibfnamefont {S.}~\bibnamefont {Schramm}}, \bibinfo {author} {\bibfnamefont
  {H.}~\bibnamefont {St\"ocker}}, \ and\ \bibinfo {author} {\bibfnamefont
  {L.}~\bibnamefont {Rezzolla}},\ }\href {\doibase
  10.1103/PhysRevLett.122.061101} {\bibfield  {journal} {\bibinfo  {journal}
  {Phys. Rev. Lett.}\ }\textbf {\bibinfo {volume} {122}},\ \bibinfo {pages}
  {061101} (\bibinfo {year} {2019})},\ \Eprint
  {http://arxiv.org/abs/1807.03684} {arXiv:1807.03684 [astro-ph.HE]}
  \BibitemShut {NoStop}%
\bibitem [{\citenamefont {Perego}\ \emph {et~al.}(2019)\citenamefont {Perego},
  \citenamefont {Bernuzzi},\ and\ \citenamefont {Radice}}]{Perego:2019adq}%
  \BibitemOpen
  \bibfield  {author} {\bibinfo {author} {\bibfnamefont {A.}~\bibnamefont
  {Perego}}, \bibinfo {author} {\bibfnamefont {S.}~\bibnamefont {Bernuzzi}}, \
  and\ \bibinfo {author} {\bibfnamefont {D.}~\bibnamefont {Radice}},\ }\href
  {\doibase 10.1140/epja/i2019-12810-7} {\bibfield  {journal} {\bibinfo
  {journal} {Eur. Phys. J. A}\ }\textbf {\bibinfo {volume} {55}},\ \bibinfo
  {pages} {124} (\bibinfo {year} {2019})},\ \Eprint
  {http://arxiv.org/abs/1903.07898} {arXiv:1903.07898 [gr-qc]} \BibitemShut
  {NoStop}%
\bibitem [{\citenamefont {Endrizzi}\ \emph {et~al.}(2020)\citenamefont
  {Endrizzi}, \citenamefont {Perego}, \citenamefont {Fabbri}, \citenamefont
  {Branca}, \citenamefont {Radice}, \citenamefont {Bernuzzi}, \citenamefont
  {Giacomazzo}, \citenamefont {Pederiva},\ and\ \citenamefont
  {Lovato}}]{Endrizzi:2019trv}%
  \BibitemOpen
  \bibfield  {author} {\bibinfo {author} {\bibfnamefont {A.}~\bibnamefont
  {Endrizzi}}, \bibinfo {author} {\bibfnamefont {A.}~\bibnamefont {Perego}},
  \bibinfo {author} {\bibfnamefont {F.~M.}\ \bibnamefont {Fabbri}}, \bibinfo
  {author} {\bibfnamefont {L.}~\bibnamefont {Branca}}, \bibinfo {author}
  {\bibfnamefont {D.}~\bibnamefont {Radice}}, \bibinfo {author} {\bibfnamefont
  {S.}~\bibnamefont {Bernuzzi}}, \bibinfo {author} {\bibfnamefont
  {B.}~\bibnamefont {Giacomazzo}}, \bibinfo {author} {\bibfnamefont
  {F.}~\bibnamefont {Pederiva}}, \ and\ \bibinfo {author} {\bibfnamefont
  {A.}~\bibnamefont {Lovato}},\ }\href {\doibase
  10.1140/epja/s10050-019-00018-6} {\bibfield  {journal} {\bibinfo  {journal}
  {Eur. Phys. J. A}\ }\textbf {\bibinfo {volume} {56}},\ \bibinfo {pages} {15}
  (\bibinfo {year} {2020})},\ \Eprint {http://arxiv.org/abs/1908.04952}
  {arXiv:1908.04952 [astro-ph.HE]} \BibitemShut {NoStop}%
\bibitem [{\citenamefont {Hammond}\ \emph {et~al.}(2021)\citenamefont
  {Hammond}, \citenamefont {Hawke},\ and\ \citenamefont
  {Andersson}}]{Hammond:2021vtv}%
  \BibitemOpen
  \bibfield  {author} {\bibinfo {author} {\bibfnamefont {P.}~\bibnamefont
  {Hammond}}, \bibinfo {author} {\bibfnamefont {I.}~\bibnamefont {Hawke}}, \
  and\ \bibinfo {author} {\bibfnamefont {N.}~\bibnamefont {Andersson}},\ }\href
  {\doibase 10.1103/PhysRevD.104.103006} {\bibfield  {journal} {\bibinfo
  {journal} {Phys. Rev. D}\ }\textbf {\bibinfo {volume} {104}},\ \bibinfo
  {pages} {103006} (\bibinfo {year} {2021})},\ \Eprint
  {http://arxiv.org/abs/2108.08649} {arXiv:2108.08649 [astro-ph.HE]}
  \BibitemShut {NoStop}%
\bibitem [{\citenamefont {Figura}\ \emph {et~al.}(2021)\citenamefont {Figura},
  \citenamefont {Li}, \citenamefont {Lu}, \citenamefont {Burgio}, \citenamefont
  {Li},\ and\ \citenamefont {Schulze}}]{Figura:2021bcn}%
  \BibitemOpen
  \bibfield  {author} {\bibinfo {author} {\bibfnamefont {A.}~\bibnamefont
  {Figura}}, \bibinfo {author} {\bibfnamefont {F.}~\bibnamefont {Li}}, \bibinfo
  {author} {\bibfnamefont {J.-J.}\ \bibnamefont {Lu}}, \bibinfo {author}
  {\bibfnamefont {G.~F.}\ \bibnamefont {Burgio}}, \bibinfo {author}
  {\bibfnamefont {Z.-H.}\ \bibnamefont {Li}}, \ and\ \bibinfo {author}
  {\bibfnamefont {H.~J.}\ \bibnamefont {Schulze}},\ }\href {\doibase
  10.1103/PhysRevD.103.083012} {\bibfield  {journal} {\bibinfo  {journal}
  {Phys. Rev. D}\ }\textbf {\bibinfo {volume} {103}},\ \bibinfo {pages}
  {083012} (\bibinfo {year} {2021})},\ \Eprint
  {http://arxiv.org/abs/2103.02365} {arXiv:2103.02365 [gr-qc]} \BibitemShut
  {NoStop}%
\bibitem [{\citenamefont {Raithel}\ \emph {et~al.}(2021)\citenamefont
  {Raithel}, \citenamefont {Paschalidis},\ and\ \citenamefont
  {\"Ozel}}]{Raithel:2021hye}%
  \BibitemOpen
  \bibfield  {author} {\bibinfo {author} {\bibfnamefont {C.}~\bibnamefont
  {Raithel}}, \bibinfo {author} {\bibfnamefont {V.}~\bibnamefont
  {Paschalidis}}, \ and\ \bibinfo {author} {\bibfnamefont {F.}~\bibnamefont
  {\"Ozel}},\ }\href {\doibase 10.1103/PhysRevD.104.063016} {\bibfield
  {journal} {\bibinfo  {journal} {Phys. Rev. D}\ }\textbf {\bibinfo {volume}
  {104}},\ \bibinfo {pages} {063016} (\bibinfo {year} {2021})},\ \Eprint
  {http://arxiv.org/abs/2104.07226} {arXiv:2104.07226 [astro-ph.HE]}
  \BibitemShut {NoStop}%
\bibitem [{\citenamefont {Adam}\ \emph {et~al.}(2021)\citenamefont {Adam} \emph
  {et~al.}}]{STAR:2020tga}%
  \BibitemOpen
  \bibfield  {author} {\bibinfo {author} {\bibfnamefont {J.}~\bibnamefont
  {Adam}} \emph {et~al.} (\bibinfo {collaboration} {STAR}),\ }\href {\doibase
  10.1103/PhysRevLett.126.092301} {\bibfield  {journal} {\bibinfo  {journal}
  {Phys. Rev. Lett.}\ }\textbf {\bibinfo {volume} {126}},\ \bibinfo {pages}
  {092301} (\bibinfo {year} {2021})},\ \Eprint
  {http://arxiv.org/abs/2001.02852} {arXiv:2001.02852 [nucl-ex]} \BibitemShut
  {NoStop}%
\bibitem [{\citenamefont {Ablyazimov}\ \emph {et~al.}(2017)\citenamefont
  {Ablyazimov} \emph {et~al.}}]{CBM:2016kpk}%
  \BibitemOpen
  \bibfield  {author} {\bibinfo {author} {\bibfnamefont {T.}~\bibnamefont
  {Ablyazimov}} \emph {et~al.} (\bibinfo {collaboration} {CBM}),\ }\href
  {\doibase 10.1140/epja/i2017-12248-y} {\bibfield  {journal} {\bibinfo
  {journal} {Eur. Phys. J. A}\ }\textbf {\bibinfo {volume} {53}},\ \bibinfo
  {pages} {60} (\bibinfo {year} {2017})},\ \Eprint
  {http://arxiv.org/abs/1607.01487} {arXiv:1607.01487 [nucl-ex]} \BibitemShut
  {NoStop}%
\bibitem [{\citenamefont {Durante}\ \emph {et~al.}(2019)\citenamefont {Durante}
  \emph {et~al.}}]{Durante:2019hzd}%
  \BibitemOpen
  \bibfield  {author} {\bibinfo {author} {\bibfnamefont {M.}~\bibnamefont
  {Durante}} \emph {et~al.},\ }\href {\doibase 10.1088/1402-4896/aaf93f}
  {\bibfield  {journal} {\bibinfo  {journal} {Phys. Scripta}\ }\textbf
  {\bibinfo {volume} {94}},\ \bibinfo {pages} {033001} (\bibinfo {year}
  {2019})},\ \Eprint {http://arxiv.org/abs/1903.05693} {arXiv:1903.05693
  [nucl-th]} \BibitemShut {NoStop}%
\bibitem [{\citenamefont {Sissakian}\ and\ \citenamefont
  {Sorin}(2009)}]{Sissakian:2009zza}%
  \BibitemOpen
  \bibfield  {author} {\bibinfo {author} {\bibfnamefont {A.~N.}\ \bibnamefont
  {Sissakian}}\ and\ \bibinfo {author} {\bibfnamefont {A.~S.}\ \bibnamefont
  {Sorin}} (\bibinfo {collaboration} {NICA}),\ }\href {\doibase
  10.1088/0954-3899/36/6/064069} {\bibfield  {journal} {\bibinfo  {journal} {J.
  Phys. G}\ }\textbf {\bibinfo {volume} {36}},\ \bibinfo {pages} {064069}
  (\bibinfo {year} {2009})}\BibitemShut {NoStop}%
\bibitem [{\citenamefont {J{\"a}rvinen}\ and\ \citenamefont
  {Kiritsis}(2012)}]{Jarvinen:2011qe}%
  \BibitemOpen
  \bibfield  {author} {\bibinfo {author} {\bibfnamefont {M.}~\bibnamefont
  {J{\"a}rvinen}}\ and\ \bibinfo {author} {\bibfnamefont {E.}~\bibnamefont
  {Kiritsis}},\ }\href {\doibase 10.1007/JHEP03(2012)002} {\bibfield  {journal}
  {\bibinfo  {journal} {JHEP}\ }\textbf {\bibinfo {volume} {03}},\ \bibinfo
  {pages} {002} (\bibinfo {year} {2012})},\ \Eprint
  {http://arxiv.org/abs/1112.1261} {arXiv:1112.1261 [hep-ph]} \BibitemShut
  {NoStop}%
\bibitem [{\citenamefont {Ishii}\ \emph {et~al.}(2019)\citenamefont {Ishii},
  \citenamefont {J{\"a}rvinen},\ and\ \citenamefont {Nijs}}]{Ishii:2019gta}%
  \BibitemOpen
  \bibfield  {author} {\bibinfo {author} {\bibfnamefont {T.}~\bibnamefont
  {Ishii}}, \bibinfo {author} {\bibfnamefont {M.}~\bibnamefont {J{\"a}rvinen}},
  \ and\ \bibinfo {author} {\bibfnamefont {G.}~\bibnamefont {Nijs}},\ }\href
  {\doibase 10.1007/JHEP07(2019)003} {\bibfield  {journal} {\bibinfo  {journal}
  {JHEP}\ }\textbf {\bibinfo {volume} {07}},\ \bibinfo {pages} {003} (\bibinfo
  {year} {2019})},\ \Eprint {http://arxiv.org/abs/1903.06169} {arXiv:1903.06169
  [hep-ph]} \BibitemShut {NoStop}%
\bibitem [{\citenamefont {Vovchenko}(2020)}]{Vovchenko:2020lju}%
  \BibitemOpen
  \bibfield  {author} {\bibinfo {author} {\bibfnamefont {V.}~\bibnamefont
  {Vovchenko}},\ }\href {\doibase 10.1142/S0218301320400029} {\bibfield
  {journal} {\bibinfo  {journal} {Int. J. Mod. Phys. E}\ }\textbf {\bibinfo
  {volume} {29}},\ \bibinfo {pages} {2040002} (\bibinfo {year} {2020})},\
  \Eprint {http://arxiv.org/abs/2004.06331} {arXiv:2004.06331 [nucl-th]}
  \BibitemShut {NoStop}%
\bibitem [{\citenamefont {Hempel}\ and\ \citenamefont
  {Schaffner-Bielich}(2010)}]{Hempel:2009mc}%
  \BibitemOpen
  \bibfield  {author} {\bibinfo {author} {\bibfnamefont {M.}~\bibnamefont
  {Hempel}}\ and\ \bibinfo {author} {\bibfnamefont {J.}~\bibnamefont
  {Schaffner-Bielich}},\ }\href {\doibase 10.1016/j.nuclphysa.2010.02.010}
  {\bibfield  {journal} {\bibinfo  {journal} {Nucl. Phys. A}\ }\textbf
  {\bibinfo {volume} {837}},\ \bibinfo {pages} {210} (\bibinfo {year}
  {2010})},\ \Eprint {http://arxiv.org/abs/0911.4073} {arXiv:0911.4073
  [nucl-th]} \BibitemShut {NoStop}%
\bibitem [{\citenamefont {Typel}\ \emph {et~al.}(2010)\citenamefont {Typel},
  \citenamefont {Ropke}, \citenamefont {Klahn}, \citenamefont {Blaschke},\ and\
  \citenamefont {Wolter}}]{Typel:2009sy}%
  \BibitemOpen
  \bibfield  {author} {\bibinfo {author} {\bibfnamefont {S.}~\bibnamefont
  {Typel}}, \bibinfo {author} {\bibfnamefont {G.}~\bibnamefont {Ropke}},
  \bibinfo {author} {\bibfnamefont {T.}~\bibnamefont {Klahn}}, \bibinfo
  {author} {\bibfnamefont {D.}~\bibnamefont {Blaschke}}, \ and\ \bibinfo
  {author} {\bibfnamefont {H.~H.}\ \bibnamefont {Wolter}},\ }\href {\doibase
  10.1103/PhysRevC.81.015803} {\bibfield  {journal} {\bibinfo  {journal} {Phys.
  Rev. C}\ }\textbf {\bibinfo {volume} {81}},\ \bibinfo {pages} {015803}
  (\bibinfo {year} {2010})},\ \Eprint {http://arxiv.org/abs/0908.2344}
  {arXiv:0908.2344 [nucl-th]} \BibitemShut {NoStop}%
\bibitem [{\citenamefont {Akmal}\ \emph
  {et~al.}(1998{\natexlab{a}})\citenamefont {Akmal}, \citenamefont
  {Pandharipande},\ and\ \citenamefont {Ravenhall}}]{Akmal:1998cf}%
  \BibitemOpen
  \bibfield  {author} {\bibinfo {author} {\bibfnamefont {A.}~\bibnamefont
  {Akmal}}, \bibinfo {author} {\bibfnamefont {V.~R.}\ \bibnamefont
  {Pandharipande}}, \ and\ \bibinfo {author} {\bibfnamefont {D.~G.}\
  \bibnamefont {Ravenhall}},\ }\href {\doibase 10.1103/PhysRevC.58.1804}
  {\bibfield  {journal} {\bibinfo  {journal} {Phys. Rev. C}\ }\textbf {\bibinfo
  {volume} {58}},\ \bibinfo {pages} {1804} (\bibinfo {year}
  {1998}{\natexlab{a}})},\ \Eprint {http://arxiv.org/abs/nucl-th/9804027}
  {arXiv:nucl-th/9804027} \BibitemShut {NoStop}%
\bibitem [{Note1()}]{Note1}%
  \BibitemOpen
  \bibinfo {note} {We use APR rather than HS(DD2) for cold nuclear matter near
  the saturation density because if we used HS(DD2) only, the resulting EoSs
  would be so stiff that they would conflict with the LIGO/Virgo measurement of
  the tidal deformability for GW170817.}\BibitemShut {Stop}%
\bibitem [{\citenamefont {Ecker}\ \emph {et~al.}(2020)\citenamefont {Ecker},
  \citenamefont {J{\"a}rvinen}, \citenamefont {Nijs},\ and\ \citenamefont
  {van~der Schee}}]{Ecker:2019xrw}%
  \BibitemOpen
  \bibfield  {author} {\bibinfo {author} {\bibfnamefont {C.}~\bibnamefont
  {Ecker}}, \bibinfo {author} {\bibfnamefont {M.}~\bibnamefont {J{\"a}rvinen}},
  \bibinfo {author} {\bibfnamefont {G.}~\bibnamefont {Nijs}}, \ and\ \bibinfo
  {author} {\bibfnamefont {W.}~\bibnamefont {van~der Schee}},\ }\href {\doibase
  10.1103/PhysRevD.101.103006} {\bibfield  {journal} {\bibinfo  {journal}
  {Phys. Rev. D}\ }\textbf {\bibinfo {volume} {101}},\ \bibinfo {pages}
  {103006} (\bibinfo {year} {2020})},\ \Eprint
  {http://arxiv.org/abs/1908.03213} {arXiv:1908.03213 [astro-ph.HE]}
  \BibitemShut {NoStop}%
\bibitem [{\citenamefont {Jokela}\ \emph
  {et~al.}(2021{\natexlab{a}})\citenamefont {Jokela}, \citenamefont
  {J{\"a}rvinen}, \citenamefont {Nijs},\ and\ \citenamefont
  {Remes}}]{Jokela:2020piw}%
  \BibitemOpen
  \bibfield  {author} {\bibinfo {author} {\bibfnamefont {N.}~\bibnamefont
  {Jokela}}, \bibinfo {author} {\bibfnamefont {M.}~\bibnamefont
  {J{\"a}rvinen}}, \bibinfo {author} {\bibfnamefont {G.}~\bibnamefont {Nijs}},
  \ and\ \bibinfo {author} {\bibfnamefont {J.}~\bibnamefont {Remes}},\ }\href
  {\doibase 10.1103/PhysRevD.103.086004} {\bibfield  {journal} {\bibinfo
  {journal} {Phys. Rev. D}\ }\textbf {\bibinfo {volume} {103}},\ \bibinfo
  {pages} {086004} (\bibinfo {year} {2021}{\natexlab{a}})},\ \Eprint
  {http://arxiv.org/abs/2006.01141} {arXiv:2006.01141 [hep-ph]} \BibitemShut
  {NoStop}%
\bibitem [{\citenamefont {Typel}\ \emph {et~al.}(2015)\citenamefont {Typel},
  \citenamefont {Oertel},\ and\ \citenamefont {Kl\"ahn}}]{Typel:2013rza}%
  \BibitemOpen
  \bibfield  {author} {\bibinfo {author} {\bibfnamefont {S.}~\bibnamefont
  {Typel}}, \bibinfo {author} {\bibfnamefont {M.}~\bibnamefont {Oertel}}, \
  and\ \bibinfo {author} {\bibfnamefont {T.}~\bibnamefont {Kl\"ahn}},\ }\href
  {\doibase 10.1134/S1063779615040061} {\bibfield  {journal} {\bibinfo
  {journal} {Phys. Part. Nucl.}\ }\textbf {\bibinfo {volume} {46}},\ \bibinfo
  {pages} {633} (\bibinfo {year} {2015})},\ \Eprint
  {http://arxiv.org/abs/1307.5715} {arXiv:1307.5715 [astro-ph.SR]} \BibitemShut
  {NoStop}%
\bibitem [{com({\natexlab{a}})}]{composeurl}%
  \BibitemOpen
  \href@noop {} {}\bibinfo {howpublished} {\url{https://compose.obspm.fr}}
  ({\natexlab{a}})\BibitemShut {NoStop}%
\bibitem [{\citenamefont {Demircik}\ \emph {et~al.}(2021)\citenamefont
  {Demircik}, \citenamefont {Ecker},\ and\ \citenamefont
  {J{\"a}rvinen}}]{Demircik:2020jkc}%
  \BibitemOpen
  \bibfield  {author} {\bibinfo {author} {\bibfnamefont {T.}~\bibnamefont
  {Demircik}}, \bibinfo {author} {\bibfnamefont {C.}~\bibnamefont {Ecker}}, \
  and\ \bibinfo {author} {\bibfnamefont {M.}~\bibnamefont {J{\"a}rvinen}},\
  }\href {\doibase 10.3847/2041-8213/abd853} {\bibfield  {journal} {\bibinfo
  {journal} {Astrophys. J. Lett.}\ }\textbf {\bibinfo {volume} {907}},\
  \bibinfo {pages} {L37} (\bibinfo {year} {2021})},\ \Eprint
  {http://arxiv.org/abs/2009.10731} {arXiv:2009.10731 [astro-ph.HE]}
  \BibitemShut {NoStop}%
\bibitem [{\citenamefont {Chesler}\ \emph {et~al.}(2019)\citenamefont
  {Chesler}, \citenamefont {Jokela}, \citenamefont {Loeb},\ and\ \citenamefont
  {Vuorinen}}]{Chesler:2019osn}%
  \BibitemOpen
  \bibfield  {author} {\bibinfo {author} {\bibfnamefont {P.~M.}\ \bibnamefont
  {Chesler}}, \bibinfo {author} {\bibfnamefont {N.}~\bibnamefont {Jokela}},
  \bibinfo {author} {\bibfnamefont {A.}~\bibnamefont {Loeb}}, \ and\ \bibinfo
  {author} {\bibfnamefont {A.}~\bibnamefont {Vuorinen}},\ }\href {\doibase
  10.1103/PhysRevD.100.066027} {\bibfield  {journal} {\bibinfo  {journal}
  {Phys. Rev. D}\ }\textbf {\bibinfo {volume} {100}},\ \bibinfo {pages}
  {066027} (\bibinfo {year} {2019})},\ \Eprint
  {http://arxiv.org/abs/1906.08440} {arXiv:1906.08440 [astro-ph.HE]}
  \BibitemShut {NoStop}%
\bibitem [{\citenamefont {Fujimoto}\ \emph {et~al.}(2021)\citenamefont
  {Fujimoto}, \citenamefont {Fukushima}, \citenamefont {Hidaka}, \citenamefont
  {Hiraguchi},\ and\ \citenamefont {Iida}}]{Fujimoto:2021dvn}%
  \BibitemOpen
  \bibfield  {author} {\bibinfo {author} {\bibfnamefont {Y.}~\bibnamefont
  {Fujimoto}}, \bibinfo {author} {\bibfnamefont {K.}~\bibnamefont {Fukushima}},
  \bibinfo {author} {\bibfnamefont {Y.}~\bibnamefont {Hidaka}}, \bibinfo
  {author} {\bibfnamefont {A.}~\bibnamefont {Hiraguchi}}, \ and\ \bibinfo
  {author} {\bibfnamefont {K.}~\bibnamefont {Iida}},\ }\href@noop {} {\
  (\bibinfo {year} {2021})},\ \Eprint {http://arxiv.org/abs/2109.06799}
  {arXiv:2109.06799 [nucl-th]} \BibitemShut {NoStop}%
\bibitem [{\citenamefont {DeWolfe}\ \emph {et~al.}(2011)\citenamefont
  {DeWolfe}, \citenamefont {Gubser},\ and\ \citenamefont
  {Rosen}}]{DeWolfe:2010he}%
  \BibitemOpen
  \bibfield  {author} {\bibinfo {author} {\bibfnamefont {O.}~\bibnamefont
  {DeWolfe}}, \bibinfo {author} {\bibfnamefont {S.~S.}\ \bibnamefont {Gubser}},
  \ and\ \bibinfo {author} {\bibfnamefont {C.}~\bibnamefont {Rosen}},\ }\href
  {\doibase 10.1103/PhysRevD.83.086005} {\bibfield  {journal} {\bibinfo
  {journal} {Phys. Rev. D}\ }\textbf {\bibinfo {volume} {83}},\ \bibinfo
  {pages} {086005} (\bibinfo {year} {2011})},\ \Eprint
  {http://arxiv.org/abs/1012.1864} {arXiv:1012.1864 [hep-th]} \BibitemShut
  {NoStop}%
\bibitem [{\citenamefont {Knaute}\ \emph {et~al.}(2018)\citenamefont {Knaute},
  \citenamefont {Yaresko},\ and\ \citenamefont {K\"ampfer}}]{Knaute:2017opk}%
  \BibitemOpen
  \bibfield  {author} {\bibinfo {author} {\bibfnamefont {J.}~\bibnamefont
  {Knaute}}, \bibinfo {author} {\bibfnamefont {R.}~\bibnamefont {Yaresko}}, \
  and\ \bibinfo {author} {\bibfnamefont {B.}~\bibnamefont {K\"ampfer}},\ }\href
  {\doibase 10.1016/j.physletb.2018.01.053} {\bibfield  {journal} {\bibinfo
  {journal} {Phys. Lett. B}\ }\textbf {\bibinfo {volume} {778}},\ \bibinfo
  {pages} {419} (\bibinfo {year} {2018})},\ \Eprint
  {http://arxiv.org/abs/1702.06731} {arXiv:1702.06731 [hep-ph]} \BibitemShut
  {NoStop}%
\bibitem [{\citenamefont {Critelli}\ \emph {et~al.}(2017)\citenamefont
  {Critelli}, \citenamefont {Noronha}, \citenamefont {Noronha-Hostler},
  \citenamefont {Portillo}, \citenamefont {Ratti},\ and\ \citenamefont
  {Rougemont}}]{Critelli:2017oub}%
  \BibitemOpen
  \bibfield  {author} {\bibinfo {author} {\bibfnamefont {R.}~\bibnamefont
  {Critelli}}, \bibinfo {author} {\bibfnamefont {J.}~\bibnamefont {Noronha}},
  \bibinfo {author} {\bibfnamefont {J.}~\bibnamefont {Noronha-Hostler}},
  \bibinfo {author} {\bibfnamefont {I.}~\bibnamefont {Portillo}}, \bibinfo
  {author} {\bibfnamefont {C.}~\bibnamefont {Ratti}}, \ and\ \bibinfo {author}
  {\bibfnamefont {R.}~\bibnamefont {Rougemont}},\ }\href {\doibase
  10.1103/PhysRevD.96.096026} {\bibfield  {journal} {\bibinfo  {journal} {Phys.
  Rev. D}\ }\textbf {\bibinfo {volume} {96}},\ \bibinfo {pages} {096026}
  (\bibinfo {year} {2017})},\ \Eprint {http://arxiv.org/abs/1706.00455}
  {arXiv:1706.00455 [nucl-th]} \BibitemShut {NoStop}%
\bibitem [{\citenamefont {Grefa}\ \emph {et~al.}(2021)\citenamefont {Grefa},
  \citenamefont {Noronha}, \citenamefont {Noronha-Hostler}, \citenamefont
  {Portillo}, \citenamefont {Ratti},\ and\ \citenamefont
  {Rougemont}}]{Grefa:2021qvt}%
  \BibitemOpen
  \bibfield  {author} {\bibinfo {author} {\bibfnamefont {J.}~\bibnamefont
  {Grefa}}, \bibinfo {author} {\bibfnamefont {J.}~\bibnamefont {Noronha}},
  \bibinfo {author} {\bibfnamefont {J.}~\bibnamefont {Noronha-Hostler}},
  \bibinfo {author} {\bibfnamefont {I.}~\bibnamefont {Portillo}}, \bibinfo
  {author} {\bibfnamefont {C.}~\bibnamefont {Ratti}}, \ and\ \bibinfo {author}
  {\bibfnamefont {R.}~\bibnamefont {Rougemont}},\ }\href {\doibase
  10.1103/PhysRevD.104.034002} {\bibfield  {journal} {\bibinfo  {journal}
  {Phys. Rev. D}\ }\textbf {\bibinfo {volume} {104}},\ \bibinfo {pages}
  {034002} (\bibinfo {year} {2021})},\ \Eprint
  {http://arxiv.org/abs/2102.12042} {arXiv:2102.12042 [nucl-th]} \BibitemShut
  {NoStop}%
\bibitem [{\citenamefont {Jokela}\ \emph {et~al.}(2019)\citenamefont {Jokela},
  \citenamefont {J{\"a}rvinen},\ and\ \citenamefont {Remes}}]{Jokela:2018ers}%
  \BibitemOpen
  \bibfield  {author} {\bibinfo {author} {\bibfnamefont {N.}~\bibnamefont
  {Jokela}}, \bibinfo {author} {\bibfnamefont {M.}~\bibnamefont
  {J{\"a}rvinen}}, \ and\ \bibinfo {author} {\bibfnamefont {J.}~\bibnamefont
  {Remes}},\ }\href {\doibase 10.1007/JHEP03(2019)041} {\bibfield  {journal}
  {\bibinfo  {journal} {JHEP}\ }\textbf {\bibinfo {volume} {03}},\ \bibinfo
  {pages} {041} (\bibinfo {year} {2019})},\ \Eprint
  {http://arxiv.org/abs/1809.07770} {arXiv:1809.07770 [hep-ph]} \BibitemShut
  {NoStop}%
\bibitem [{\citenamefont {J{\"a}rvinen}(2021)}]{Jarvinen:2021jbd}%
  \BibitemOpen
  \bibfield  {author} {\bibinfo {author} {\bibfnamefont {M.}~\bibnamefont
  {J{\"a}rvinen}},\ }\href@noop {} {\  (\bibinfo {year} {2021})},\ \Eprint
  {http://arxiv.org/abs/2110.08281} {arXiv:2110.08281 [hep-ph]} \BibitemShut
  {NoStop}%
\bibitem [{\citenamefont {Gursoy}\ and\ \citenamefont
  {Kiritsis}(2008)}]{Gursoy:2007cb}%
  \BibitemOpen
  \bibfield  {author} {\bibinfo {author} {\bibfnamefont {U.}~\bibnamefont
  {Gursoy}}\ and\ \bibinfo {author} {\bibfnamefont {E.}~\bibnamefont
  {Kiritsis}},\ }\href {\doibase 10.1088/1126-6708/2008/02/032} {\bibfield
  {journal} {\bibinfo  {journal} {JHEP}\ }\textbf {\bibinfo {volume} {02}},\
  \bibinfo {pages} {032} (\bibinfo {year} {2008})},\ \Eprint
  {http://arxiv.org/abs/0707.1324} {arXiv:0707.1324 [hep-th]} \BibitemShut
  {NoStop}%
\bibitem [{\citenamefont {Gursoy}\ \emph
  {et~al.}(2008{\natexlab{a}})\citenamefont {Gursoy}, \citenamefont
  {Kiritsis},\ and\ \citenamefont {Nitti}}]{Gursoy:2007er}%
  \BibitemOpen
  \bibfield  {author} {\bibinfo {author} {\bibfnamefont {U.}~\bibnamefont
  {Gursoy}}, \bibinfo {author} {\bibfnamefont {E.}~\bibnamefont {Kiritsis}}, \
  and\ \bibinfo {author} {\bibfnamefont {F.}~\bibnamefont {Nitti}},\ }\href
  {\doibase 10.1088/1126-6708/2008/02/019} {\bibfield  {journal} {\bibinfo
  {journal} {JHEP}\ }\textbf {\bibinfo {volume} {02}},\ \bibinfo {pages} {019}
  (\bibinfo {year} {2008}{\natexlab{a}})},\ \Eprint
  {http://arxiv.org/abs/0707.1349} {arXiv:0707.1349 [hep-th]} \BibitemShut
  {NoStop}%
\bibitem [{\citenamefont {Bigazzi}\ \emph {et~al.}(2005)\citenamefont
  {Bigazzi}, \citenamefont {Casero}, \citenamefont {Cotrone}, \citenamefont
  {Kiritsis},\ and\ \citenamefont {Paredes}}]{Bigazzi:2005md}%
  \BibitemOpen
  \bibfield  {author} {\bibinfo {author} {\bibfnamefont {F.}~\bibnamefont
  {Bigazzi}}, \bibinfo {author} {\bibfnamefont {R.}~\bibnamefont {Casero}},
  \bibinfo {author} {\bibfnamefont {A.~L.}\ \bibnamefont {Cotrone}}, \bibinfo
  {author} {\bibfnamefont {E.}~\bibnamefont {Kiritsis}}, \ and\ \bibinfo
  {author} {\bibfnamefont {A.}~\bibnamefont {Paredes}},\ }\href {\doibase
  10.1088/1126-6708/2005/10/012} {\bibfield  {journal} {\bibinfo  {journal}
  {JHEP}\ }\textbf {\bibinfo {volume} {10}},\ \bibinfo {pages} {012} (\bibinfo
  {year} {2005})},\ \Eprint {http://arxiv.org/abs/hep-th/0505140}
  {arXiv:hep-th/0505140} \BibitemShut {NoStop}%
\bibitem [{\citenamefont {Casero}\ \emph {et~al.}(2007)\citenamefont {Casero},
  \citenamefont {Kiritsis},\ and\ \citenamefont {Paredes}}]{Casero:2007ae}%
  \BibitemOpen
  \bibfield  {author} {\bibinfo {author} {\bibfnamefont {R.}~\bibnamefont
  {Casero}}, \bibinfo {author} {\bibfnamefont {E.}~\bibnamefont {Kiritsis}}, \
  and\ \bibinfo {author} {\bibfnamefont {A.}~\bibnamefont {Paredes}},\ }\href
  {\doibase 10.1016/j.nuclphysb.2007.07.009} {\bibfield  {journal} {\bibinfo
  {journal} {Nucl. Phys. B}\ }\textbf {\bibinfo {volume} {787}},\ \bibinfo
  {pages} {98} (\bibinfo {year} {2007})},\ \Eprint
  {http://arxiv.org/abs/hep-th/0702155} {arXiv:hep-th/0702155} \BibitemShut
  {NoStop}%
\bibitem [{\citenamefont {Veneziano}(1979)}]{Veneziano:1979ec}%
  \BibitemOpen
  \bibfield  {author} {\bibinfo {author} {\bibfnamefont {G.}~\bibnamefont
  {Veneziano}},\ }\href {\doibase 10.1016/0550-3213(79)90332-8} {\bibfield
  {journal} {\bibinfo  {journal} {Nucl. Phys. B}\ }\textbf {\bibinfo {volume}
  {159}},\ \bibinfo {pages} {213} (\bibinfo {year} {1979})}\BibitemShut
  {NoStop}%
\bibitem [{\citenamefont {Gubser}(2000)}]{Gubser:2000nd}%
  \BibitemOpen
  \bibfield  {author} {\bibinfo {author} {\bibfnamefont {S.~S.}\ \bibnamefont
  {Gubser}},\ }\href {\doibase 10.4310/ATMP.2000.v4.n3.a6} {\bibfield
  {journal} {\bibinfo  {journal} {Adv. Theor. Math. Phys.}\ }\textbf {\bibinfo
  {volume} {4}},\ \bibinfo {pages} {679} (\bibinfo {year} {2000})},\ \Eprint
  {http://arxiv.org/abs/hep-th/0002160} {arXiv:hep-th/0002160} \BibitemShut
  {NoStop}%
\bibitem [{\citenamefont {Alho}\ \emph {et~al.}(2013)\citenamefont {Alho},
  \citenamefont {J{\"a}rvinen}, \citenamefont {Kajantie}, \citenamefont
  {Kiritsis},\ and\ \citenamefont {Tuominen}}]{Alho:2012mh}%
  \BibitemOpen
  \bibfield  {author} {\bibinfo {author} {\bibfnamefont {T.}~\bibnamefont
  {Alho}}, \bibinfo {author} {\bibfnamefont {M.}~\bibnamefont {J{\"a}rvinen}},
  \bibinfo {author} {\bibfnamefont {K.}~\bibnamefont {Kajantie}}, \bibinfo
  {author} {\bibfnamefont {E.}~\bibnamefont {Kiritsis}}, \ and\ \bibinfo
  {author} {\bibfnamefont {K.}~\bibnamefont {Tuominen}},\ }\href {\doibase
  10.1007/JHEP01(2013)093} {\bibfield  {journal} {\bibinfo  {journal} {JHEP}\
  }\textbf {\bibinfo {volume} {01}},\ \bibinfo {pages} {093} (\bibinfo {year}
  {2013})},\ \Eprint {http://arxiv.org/abs/1210.4516} {arXiv:1210.4516
  [hep-ph]} \BibitemShut {NoStop}%
\bibitem [{\citenamefont {Alho}\ \emph {et~al.}(2014)\citenamefont {Alho},
  \citenamefont {J{\"a}rvinen}, \citenamefont {Kajantie}, \citenamefont
  {Kiritsis}, \citenamefont {Rosen},\ and\ \citenamefont
  {Tuominen}}]{Alho:2013hsa}%
  \BibitemOpen
  \bibfield  {author} {\bibinfo {author} {\bibfnamefont {T.}~\bibnamefont
  {Alho}}, \bibinfo {author} {\bibfnamefont {M.}~\bibnamefont {J{\"a}rvinen}},
  \bibinfo {author} {\bibfnamefont {K.}~\bibnamefont {Kajantie}}, \bibinfo
  {author} {\bibfnamefont {E.}~\bibnamefont {Kiritsis}}, \bibinfo {author}
  {\bibfnamefont {C.}~\bibnamefont {Rosen}}, \ and\ \bibinfo {author}
  {\bibfnamefont {K.}~\bibnamefont {Tuominen}},\ }\href {\doibase
  10.1007/JHEP04(2014)124} {\bibfield  {journal} {\bibinfo  {journal} {JHEP}\
  }\textbf {\bibinfo {volume} {04}},\ \bibinfo {pages} {124} (\bibinfo {year}
  {2014})},\ \bibinfo {note} {[Erratum: JHEP 02, 033 (2015)]},\ \Eprint
  {http://arxiv.org/abs/1312.5199} {arXiv:1312.5199 [hep-ph]} \BibitemShut
  {NoStop}%
\bibitem [{\citenamefont {Gursoy}\ \emph
  {et~al.}(2008{\natexlab{b}})\citenamefont {Gursoy}, \citenamefont {Kiritsis},
  \citenamefont {Mazzanti},\ and\ \citenamefont {Nitti}}]{Gursoy:2008bu}%
  \BibitemOpen
  \bibfield  {author} {\bibinfo {author} {\bibfnamefont {U.}~\bibnamefont
  {Gursoy}}, \bibinfo {author} {\bibfnamefont {E.}~\bibnamefont {Kiritsis}},
  \bibinfo {author} {\bibfnamefont {L.}~\bibnamefont {Mazzanti}}, \ and\
  \bibinfo {author} {\bibfnamefont {F.}~\bibnamefont {Nitti}},\ }\href
  {\doibase 10.1103/PhysRevLett.101.181601} {\bibfield  {journal} {\bibinfo
  {journal} {Phys. Rev. Lett.}\ }\textbf {\bibinfo {volume} {101}},\ \bibinfo
  {pages} {181601} (\bibinfo {year} {2008}{\natexlab{b}})},\ \Eprint
  {http://arxiv.org/abs/0804.0899} {arXiv:0804.0899 [hep-th]} \BibitemShut
  {NoStop}%
\bibitem [{\citenamefont {Gursoy}\ \emph {et~al.}(2009)\citenamefont {Gursoy},
  \citenamefont {Kiritsis}, \citenamefont {Mazzanti},\ and\ \citenamefont
  {Nitti}}]{Gursoy:2008za}%
  \BibitemOpen
  \bibfield  {author} {\bibinfo {author} {\bibfnamefont {U.}~\bibnamefont
  {Gursoy}}, \bibinfo {author} {\bibfnamefont {E.}~\bibnamefont {Kiritsis}},
  \bibinfo {author} {\bibfnamefont {L.}~\bibnamefont {Mazzanti}}, \ and\
  \bibinfo {author} {\bibfnamefont {F.}~\bibnamefont {Nitti}},\ }\href
  {\doibase 10.1088/1126-6708/2009/05/033} {\bibfield  {journal} {\bibinfo
  {journal} {JHEP}\ }\textbf {\bibinfo {volume} {05}},\ \bibinfo {pages} {033}
  (\bibinfo {year} {2009})},\ \Eprint {http://arxiv.org/abs/0812.0792}
  {arXiv:0812.0792 [hep-th]} \BibitemShut {NoStop}%
\bibitem [{\citenamefont {Alho}\ \emph {et~al.}(2015)\citenamefont {Alho},
  \citenamefont {J{\"a}rvinen}, \citenamefont {Kajantie}, \citenamefont
  {Kiritsis},\ and\ \citenamefont {Tuominen}}]{Alho:2015zua}%
  \BibitemOpen
  \bibfield  {author} {\bibinfo {author} {\bibfnamefont {T.}~\bibnamefont
  {Alho}}, \bibinfo {author} {\bibfnamefont {M.}~\bibnamefont {J{\"a}rvinen}},
  \bibinfo {author} {\bibfnamefont {K.}~\bibnamefont {Kajantie}}, \bibinfo
  {author} {\bibfnamefont {E.}~\bibnamefont {Kiritsis}}, \ and\ \bibinfo
  {author} {\bibfnamefont {K.}~\bibnamefont {Tuominen}},\ }\href {\doibase
  10.1103/PhysRevD.91.055017} {\bibfield  {journal} {\bibinfo  {journal} {Phys.
  Rev. D}\ }\textbf {\bibinfo {volume} {91}},\ \bibinfo {pages} {055017}
  (\bibinfo {year} {2015})},\ \Eprint {http://arxiv.org/abs/1501.06379}
  {arXiv:1501.06379 [hep-ph]} \BibitemShut {NoStop}%
\bibitem [{\citenamefont {Panero}(2009)}]{Panero:2009tv}%
  \BibitemOpen
  \bibfield  {author} {\bibinfo {author} {\bibfnamefont {M.}~\bibnamefont
  {Panero}},\ }\href {\doibase 10.1103/PhysRevLett.103.232001} {\bibfield
  {journal} {\bibinfo  {journal} {Phys. Rev. Lett.}\ }\textbf {\bibinfo
  {volume} {103}},\ \bibinfo {pages} {232001} (\bibinfo {year} {2009})},\
  \Eprint {http://arxiv.org/abs/0907.3719} {arXiv:0907.3719 [hep-lat]}
  \BibitemShut {NoStop}%
\bibitem [{\citenamefont {Borsanyi}\ \emph {et~al.}(2014)\citenamefont
  {Borsanyi}, \citenamefont {Fodor}, \citenamefont {Hoelbling}, \citenamefont
  {Katz}, \citenamefont {Krieg},\ and\ \citenamefont
  {Szabo}}]{Borsanyi:2013bia}%
  \BibitemOpen
  \bibfield  {author} {\bibinfo {author} {\bibfnamefont {S.}~\bibnamefont
  {Borsanyi}}, \bibinfo {author} {\bibfnamefont {Z.}~\bibnamefont {Fodor}},
  \bibinfo {author} {\bibfnamefont {C.}~\bibnamefont {Hoelbling}}, \bibinfo
  {author} {\bibfnamefont {S.~D.}\ \bibnamefont {Katz}}, \bibinfo {author}
  {\bibfnamefont {S.}~\bibnamefont {Krieg}}, \ and\ \bibinfo {author}
  {\bibfnamefont {K.~K.}\ \bibnamefont {Szabo}},\ }\href {\doibase
  10.1016/j.physletb.2014.01.007} {\bibfield  {journal} {\bibinfo  {journal}
  {Phys. Lett. B}\ }\textbf {\bibinfo {volume} {730}},\ \bibinfo {pages} {99}
  (\bibinfo {year} {2014})},\ \Eprint {http://arxiv.org/abs/1309.5258}
  {arXiv:1309.5258 [hep-lat]} \BibitemShut {NoStop}%
\bibitem [{\citenamefont {Borsanyi}\ \emph {et~al.}(2012)\citenamefont
  {Borsanyi}, \citenamefont {Fodor}, \citenamefont {Katz}, \citenamefont
  {Krieg}, \citenamefont {Ratti},\ and\ \citenamefont
  {Szabo}}]{Borsanyi:2011sw}%
  \BibitemOpen
  \bibfield  {author} {\bibinfo {author} {\bibfnamefont {S.}~\bibnamefont
  {Borsanyi}}, \bibinfo {author} {\bibfnamefont {Z.}~\bibnamefont {Fodor}},
  \bibinfo {author} {\bibfnamefont {S.~D.}\ \bibnamefont {Katz}}, \bibinfo
  {author} {\bibfnamefont {S.}~\bibnamefont {Krieg}}, \bibinfo {author}
  {\bibfnamefont {C.}~\bibnamefont {Ratti}}, \ and\ \bibinfo {author}
  {\bibfnamefont {K.}~\bibnamefont {Szabo}},\ }\href {\doibase
  10.1007/JHEP01(2012)138} {\bibfield  {journal} {\bibinfo  {journal} {JHEP}\
  }\textbf {\bibinfo {volume} {01}},\ \bibinfo {pages} {138} (\bibinfo {year}
  {2012})},\ \Eprint {http://arxiv.org/abs/1112.4416} {arXiv:1112.4416
  [hep-lat]} \BibitemShut {NoStop}%
\bibitem [{\citenamefont {Jokela}\ \emph
  {et~al.}(2021{\natexlab{b}})\citenamefont {Jokela}, \citenamefont
  {J{\"a}rvinen},\ and\ \citenamefont {Remes}}]{Jokela:2021vwy}%
  \BibitemOpen
  \bibfield  {author} {\bibinfo {author} {\bibfnamefont {N.}~\bibnamefont
  {Jokela}}, \bibinfo {author} {\bibfnamefont {M.}~\bibnamefont
  {J{\"a}rvinen}}, \ and\ \bibinfo {author} {\bibfnamefont {J.}~\bibnamefont
  {Remes}},\ }\href@noop {} {\  (\bibinfo {year} {2021}{\natexlab{b}})},\
  \Eprint {http://arxiv.org/abs/2111.12101} {arXiv:2111.12101 [hep-ph]}
  \BibitemShut {NoStop}%
\bibitem [{Note2()}]{Note2}%
  \BibitemOpen
  \bibinfo {note} {In principle working in the Veneziano limit, where $N_f \sim
  N_c$ should enhance the pressure of the confined matter. However it turns out
  that such contributions are not captured only by backreacting the flavor
  action to the geometry, but one should include nontrivial string loop
  diagrams~\cite {Alho:2015zua}.}\BibitemShut {Stop}%
\bibitem [{\citenamefont {Zyla}\ \emph {et~al.}(2020)\citenamefont {Zyla} \emph
  {et~al.}}]{ParticleDataGroup:2020ssz}%
  \BibitemOpen
  \bibfield  {author} {\bibinfo {author} {\bibfnamefont {P.~A.}\ \bibnamefont
  {Zyla}} \emph {et~al.} (\bibinfo {collaboration} {Particle Data Group}),\
  }\href {\doibase 10.1093/ptep/ptaa104} {\bibfield  {journal} {\bibinfo
  {journal} {PTEP}\ }\textbf {\bibinfo {volume} {2020}},\ \bibinfo {pages}
  {083C01} (\bibinfo {year} {2020})}\BibitemShut {NoStop}%
\bibitem [{\citenamefont {Rischke}\ \emph {et~al.}(1991)\citenamefont
  {Rischke}, \citenamefont {Gorenstein}, \citenamefont {Stoecker},\ and\
  \citenamefont {Greiner}}]{Rischke:1991ke}%
  \BibitemOpen
  \bibfield  {author} {\bibinfo {author} {\bibfnamefont {D.~H.}\ \bibnamefont
  {Rischke}}, \bibinfo {author} {\bibfnamefont {M.~I.}\ \bibnamefont
  {Gorenstein}}, \bibinfo {author} {\bibfnamefont {H.}~\bibnamefont
  {Stoecker}}, \ and\ \bibinfo {author} {\bibfnamefont {W.}~\bibnamefont
  {Greiner}},\ }\href {\doibase 10.1007/BF01548574} {\bibfield  {journal}
  {\bibinfo  {journal} {Z. Phys. C}\ }\textbf {\bibinfo {volume} {51}},\
  \bibinfo {pages} {485} (\bibinfo {year} {1991})}\BibitemShut {NoStop}%
\bibitem [{\citenamefont {Hagedorn}\ and\ \citenamefont
  {Rafelski}(1980)}]{Hagedorn:1980kb}%
  \BibitemOpen
  \bibfield  {author} {\bibinfo {author} {\bibfnamefont {R.}~\bibnamefont
  {Hagedorn}}\ and\ \bibinfo {author} {\bibfnamefont {J.}~\bibnamefont
  {Rafelski}},\ }\href {\doibase 10.1016/0370-2693(80)90566-3} {\bibfield
  {journal} {\bibinfo  {journal} {Phys. Lett. B}\ }\textbf {\bibinfo {volume}
  {97}},\ \bibinfo {pages} {136} (\bibinfo {year} {1980})}\BibitemShut
  {NoStop}%
\bibitem [{\citenamefont {Vovchenko}\ \emph
  {et~al.}(2017{\natexlab{a}})\citenamefont {Vovchenko}, \citenamefont
  {Gorenstein},\ and\ \citenamefont {Stoecker}}]{Vovchenko:2016rkn}%
  \BibitemOpen
  \bibfield  {author} {\bibinfo {author} {\bibfnamefont {V.}~\bibnamefont
  {Vovchenko}}, \bibinfo {author} {\bibfnamefont {M.~I.}\ \bibnamefont
  {Gorenstein}}, \ and\ \bibinfo {author} {\bibfnamefont {H.}~\bibnamefont
  {Stoecker}},\ }\href {\doibase 10.1103/PhysRevLett.118.182301} {\bibfield
  {journal} {\bibinfo  {journal} {Phys. Rev. Lett.}\ }\textbf {\bibinfo
  {volume} {118}},\ \bibinfo {pages} {182301} (\bibinfo {year}
  {2017}{\natexlab{a}})},\ \Eprint {http://arxiv.org/abs/1609.03975}
  {arXiv:1609.03975 [hep-ph]} \BibitemShut {NoStop}%
\bibitem [{\citenamefont {Vovchenko}\ \emph
  {et~al.}(2017{\natexlab{b}})\citenamefont {Vovchenko}, \citenamefont
  {Motornenko}, \citenamefont {Alba}, \citenamefont {Gorenstein}, \citenamefont
  {Satarov},\ and\ \citenamefont {Stoecker}}]{Vovchenko:2017zpj}%
  \BibitemOpen
  \bibfield  {author} {\bibinfo {author} {\bibfnamefont {V.}~\bibnamefont
  {Vovchenko}}, \bibinfo {author} {\bibfnamefont {A.}~\bibnamefont
  {Motornenko}}, \bibinfo {author} {\bibfnamefont {P.}~\bibnamefont {Alba}},
  \bibinfo {author} {\bibfnamefont {M.~I.}\ \bibnamefont {Gorenstein}},
  \bibinfo {author} {\bibfnamefont {L.~M.}\ \bibnamefont {Satarov}}, \ and\
  \bibinfo {author} {\bibfnamefont {H.}~\bibnamefont {Stoecker}},\ }\href
  {\doibase 10.1103/PhysRevC.96.045202} {\bibfield  {journal} {\bibinfo
  {journal} {Phys. Rev. C}\ }\textbf {\bibinfo {volume} {96}},\ \bibinfo
  {pages} {045202} (\bibinfo {year} {2017}{\natexlab{b}})},\ \Eprint
  {http://arxiv.org/abs/1707.09215} {arXiv:1707.09215 [nucl-th]} \BibitemShut
  {NoStop}%
\bibitem [{\citenamefont {Vovchenko}\ \emph {et~al.}(2015)\citenamefont
  {Vovchenko}, \citenamefont {Anchishkin}, \citenamefont {Gorenstein},\ and\
  \citenamefont {Poberezhnyuk}}]{Vovchenko:2015pya}%
  \BibitemOpen
  \bibfield  {author} {\bibinfo {author} {\bibfnamefont {V.}~\bibnamefont
  {Vovchenko}}, \bibinfo {author} {\bibfnamefont {D.~V.}\ \bibnamefont
  {Anchishkin}}, \bibinfo {author} {\bibfnamefont {M.~I.}\ \bibnamefont
  {Gorenstein}}, \ and\ \bibinfo {author} {\bibfnamefont {R.~V.}\ \bibnamefont
  {Poberezhnyuk}},\ }\href {\doibase 10.1103/PhysRevC.92.054901} {\bibfield
  {journal} {\bibinfo  {journal} {Phys. Rev. C}\ }\textbf {\bibinfo {volume}
  {92}},\ \bibinfo {pages} {054901} (\bibinfo {year} {2015})},\ \Eprint
  {http://arxiv.org/abs/1506.05763} {arXiv:1506.05763 [nucl-th]} \BibitemShut
  {NoStop}%
\bibitem [{\citenamefont {Redlich}\ and\ \citenamefont
  {Zalewski}(2016)}]{Redlich:2016dpb}%
  \BibitemOpen
  \bibfield  {author} {\bibinfo {author} {\bibfnamefont {K.}~\bibnamefont
  {Redlich}}\ and\ \bibinfo {author} {\bibfnamefont {K.}~\bibnamefont
  {Zalewski}},\ }\href {\doibase 10.5506/APhysPolB.47.1943} {\bibfield
  {journal} {\bibinfo  {journal} {Acta Phys. Polon. B}\ }\textbf {\bibinfo
  {volume} {47}},\ \bibinfo {pages} {1943} (\bibinfo {year} {2016})},\ \Eprint
  {http://arxiv.org/abs/1605.09686} {arXiv:1605.09686 [cond-mat.quant-gas]}
  \BibitemShut {NoStop}%
\bibitem [{\citenamefont {Toki}\ \emph {et~al.}(1995)\citenamefont {Toki},
  \citenamefont {Hirata}, \citenamefont {Sugahara}, \citenamefont {Sumiyoshi},\
  and\ \citenamefont {Tanihata}}]{TOKI1995c357}%
  \BibitemOpen
  \bibfield  {author} {\bibinfo {author} {\bibfnamefont {H.}~\bibnamefont
  {Toki}}, \bibinfo {author} {\bibfnamefont {D.}~\bibnamefont {Hirata}},
  \bibinfo {author} {\bibfnamefont {Y.}~\bibnamefont {Sugahara}}, \bibinfo
  {author} {\bibfnamefont {K.}~\bibnamefont {Sumiyoshi}}, \ and\ \bibinfo
  {author} {\bibfnamefont {I.}~\bibnamefont {Tanihata}},\ }\href {\doibase
  https://doi.org/10.1016/0375-9474(95)00161-S} {\bibfield  {journal} {\bibinfo
   {journal} {Nuclear Physics A}\ }\textbf {\bibinfo {volume} {588}},\ \bibinfo
  {pages} {c357} (\bibinfo {year} {1995})},\ \bibinfo {note} {proceedings of
  the Fifth International Symposium on Physics of Unstable Nuclei}\BibitemShut
  {NoStop}%
\bibitem [{\citenamefont {Audi}\ \emph {et~al.}(2003)\citenamefont {Audi},
  \citenamefont {Wapstra},\ and\ \citenamefont {Thibault}}]{AUDI2003337}%
  \BibitemOpen
  \bibfield  {author} {\bibinfo {author} {\bibfnamefont {G.}~\bibnamefont
  {Audi}}, \bibinfo {author} {\bibfnamefont {A.}~\bibnamefont {Wapstra}}, \
  and\ \bibinfo {author} {\bibfnamefont {C.}~\bibnamefont {Thibault}},\ }\href
  {\doibase https://doi.org/10.1016/j.nuclphysa.2003.11.003} {\bibfield
  {journal} {\bibinfo  {journal} {Nuclear Physics A}\ }\textbf {\bibinfo
  {volume} {729}},\ \bibinfo {pages} {337} (\bibinfo {year} {2003})},\ \bibinfo
  {note} {the 2003 NUBASE and Atomic Mass Evaluations}\BibitemShut {NoStop}%
\bibitem [{\citenamefont {Geng}\ \emph {et~al.}(2005)\citenamefont {Geng},
  \citenamefont {Toki},\ and\ \citenamefont {Meng}}]{Geng:2005yu}%
  \BibitemOpen
  \bibfield  {author} {\bibinfo {author} {\bibfnamefont {L.-S.}\ \bibnamefont
  {Geng}}, \bibinfo {author} {\bibfnamefont {H.}~\bibnamefont {Toki}}, \ and\
  \bibinfo {author} {\bibfnamefont {J.}~\bibnamefont {Meng}},\ }\href {\doibase
  10.1143/PTP.113.785} {\bibfield  {journal} {\bibinfo  {journal} {Prog. Theor.
  Phys.}\ }\textbf {\bibinfo {volume} {113}},\ \bibinfo {pages} {785} (\bibinfo
  {year} {2005})},\ \Eprint {http://arxiv.org/abs/nucl-th/0503086}
  {arXiv:nucl-th/0503086} \BibitemShut {NoStop}%
\bibitem [{\citenamefont {Hempel}\ \emph {et~al.}(2012)\citenamefont {Hempel},
  \citenamefont {Fischer}, \citenamefont {Schaffner-Bielich},\ and\
  \citenamefont {Liebendorfer}}]{Hempel:2011mk}%
  \BibitemOpen
  \bibfield  {author} {\bibinfo {author} {\bibfnamefont {M.}~\bibnamefont
  {Hempel}}, \bibinfo {author} {\bibfnamefont {T.}~\bibnamefont {Fischer}},
  \bibinfo {author} {\bibfnamefont {J.}~\bibnamefont {Schaffner-Bielich}}, \
  and\ \bibinfo {author} {\bibfnamefont {M.}~\bibnamefont {Liebendorfer}},\
  }\href {\doibase 10.1088/0004-637X/748/1/70} {\bibfield  {journal} {\bibinfo
  {journal} {Astrophys. J.}\ }\textbf {\bibinfo {volume} {748}},\ \bibinfo
  {pages} {70} (\bibinfo {year} {2012})},\ \Eprint
  {http://arxiv.org/abs/1108.0848} {arXiv:1108.0848 [astro-ph.HE]} \BibitemShut
  {NoStop}%
\bibitem [{\citenamefont {Steiner}\ \emph {et~al.}(2013)\citenamefont
  {Steiner}, \citenamefont {Hempel},\ and\ \citenamefont
  {Fischer}}]{Steiner:2012rk}%
  \BibitemOpen
  \bibfield  {author} {\bibinfo {author} {\bibfnamefont {A.~W.}\ \bibnamefont
  {Steiner}}, \bibinfo {author} {\bibfnamefont {M.}~\bibnamefont {Hempel}}, \
  and\ \bibinfo {author} {\bibfnamefont {T.}~\bibnamefont {Fischer}},\ }\href
  {\doibase 10.1088/0004-637X/774/1/17} {\bibfield  {journal} {\bibinfo
  {journal} {Astrophys. J.}\ }\textbf {\bibinfo {volume} {774}},\ \bibinfo
  {pages} {17} (\bibinfo {year} {2013})},\ \Eprint
  {http://arxiv.org/abs/1207.2184} {arXiv:1207.2184 [astro-ph.SR]} \BibitemShut
  {NoStop}%
\bibitem [{\citenamefont {Akmal}\ \emph
  {et~al.}(1998{\natexlab{b}})\citenamefont {Akmal}, \citenamefont
  {Pandharipande},\ and\ \citenamefont {Ravenhall}}]{PhysRevC.58.1804}%
  \BibitemOpen
  \bibfield  {author} {\bibinfo {author} {\bibfnamefont {A.}~\bibnamefont
  {Akmal}}, \bibinfo {author} {\bibfnamefont {V.~R.}\ \bibnamefont
  {Pandharipande}}, \ and\ \bibinfo {author} {\bibfnamefont {D.~G.}\
  \bibnamefont {Ravenhall}},\ }\href {\doibase 10.1103/PhysRevC.58.1804}
  {\bibfield  {journal} {\bibinfo  {journal} {Phys. Rev. C}\ }\textbf {\bibinfo
  {volume} {58}},\ \bibinfo {pages} {1804} (\bibinfo {year}
  {1998}{\natexlab{b}})}\BibitemShut {NoStop}%
\bibitem [{com({\natexlab{b}})}]{composeurlVQCDsoft}%
  \BibitemOpen
  \href@noop {} {}\bibinfo {howpublished}
  {\url{https://compose.obspm.fr/eos/197}} ({\natexlab{b}})\BibitemShut
  {NoStop}%
\bibitem [{com({\natexlab{c}})}]{composeurlVQCDinter}%
  \BibitemOpen
  \href@noop {} {}\bibinfo {howpublished}
  {\url{https://compose.obspm.fr/eos/198}} ({\natexlab{c}})\BibitemShut
  {NoStop}%
\bibitem [{com({\natexlab{d}})}]{composeurlVQCDstiff}%
  \BibitemOpen
  \href@noop {} {}\bibinfo {howpublished}
  {\url{https://compose.obspm.fr/eos/199}} ({\natexlab{d}})\BibitemShut
  {NoStop}%
\bibitem [{\citenamefont {Baldo}\ and\ \citenamefont
  {Burgio}(2016)}]{Baldo:2016jhp}%
  \BibitemOpen
  \bibfield  {author} {\bibinfo {author} {\bibfnamefont {M.}~\bibnamefont
  {Baldo}}\ and\ \bibinfo {author} {\bibfnamefont {G.~F.}\ \bibnamefont
  {Burgio}},\ }\href {\doibase 10.1016/j.ppnp.2016.06.006} {\bibfield
  {journal} {\bibinfo  {journal} {Prog. Part. Nucl. Phys.}\ }\textbf {\bibinfo
  {volume} {91}},\ \bibinfo {pages} {203} (\bibinfo {year} {2016})},\ \Eprint
  {http://arxiv.org/abs/1606.08838} {arXiv:1606.08838 [nucl-th]} \BibitemShut
  {NoStop}%
\bibitem [{Note3()}]{Note3}%
  \BibitemOpen
  \bibinfo {note} {We use $T_\protect \mathrm {min}=0.1$~MeV instead of $T=0$
  when evaluating these definitions numerically because this value is the
  lowest available in the data grid of the HS(DD2) EoS.}\BibitemShut {Stop}%
\bibitem [{\citenamefont {Karsch}\ \emph
  {et~al.}(2003{\natexlab{a}})\citenamefont {Karsch}, \citenamefont {Redlich},\
  and\ \citenamefont {Tawfik}}]{Karsch:2003vd}%
  \BibitemOpen
  \bibfield  {author} {\bibinfo {author} {\bibfnamefont {F.}~\bibnamefont
  {Karsch}}, \bibinfo {author} {\bibfnamefont {K.}~\bibnamefont {Redlich}}, \
  and\ \bibinfo {author} {\bibfnamefont {A.}~\bibnamefont {Tawfik}},\ }\href
  {\doibase 10.1140/epjc/s2003-01228-y} {\bibfield  {journal} {\bibinfo
  {journal} {Eur. Phys. J. C}\ }\textbf {\bibinfo {volume} {29}},\ \bibinfo
  {pages} {549} (\bibinfo {year} {2003}{\natexlab{a}})},\ \Eprint
  {http://arxiv.org/abs/hep-ph/0303108} {arXiv:hep-ph/0303108} \BibitemShut
  {NoStop}%
\bibitem [{\citenamefont {Karsch}\ \emph
  {et~al.}(2003{\natexlab{b}})\citenamefont {Karsch}, \citenamefont {Redlich},\
  and\ \citenamefont {Tawfik}}]{Karsch:2003zq}%
  \BibitemOpen
  \bibfield  {author} {\bibinfo {author} {\bibfnamefont {F.}~\bibnamefont
  {Karsch}}, \bibinfo {author} {\bibfnamefont {K.}~\bibnamefont {Redlich}}, \
  and\ \bibinfo {author} {\bibfnamefont {A.}~\bibnamefont {Tawfik}},\ }\href
  {\doibase 10.1016/j.physletb.2003.08.001} {\bibfield  {journal} {\bibinfo
  {journal} {Phys. Lett. B}\ }\textbf {\bibinfo {volume} {571}},\ \bibinfo
  {pages} {67} (\bibinfo {year} {2003}{\natexlab{b}})},\ \Eprint
  {http://arxiv.org/abs/hep-ph/0306208} {arXiv:hep-ph/0306208} \BibitemShut
  {NoStop}%
\bibitem [{\citenamefont {Annala}\ \emph
  {et~al.}(2018{\natexlab{a}})\citenamefont {Annala}, \citenamefont {Gorda},
  \citenamefont {Kurkela},\ and\ \citenamefont {Vuorinen}}]{Annala:2017llu}%
  \BibitemOpen
  \bibfield  {author} {\bibinfo {author} {\bibfnamefont {E.}~\bibnamefont
  {Annala}}, \bibinfo {author} {\bibfnamefont {T.}~\bibnamefont {Gorda}},
  \bibinfo {author} {\bibfnamefont {A.}~\bibnamefont {Kurkela}}, \ and\
  \bibinfo {author} {\bibfnamefont {A.}~\bibnamefont {Vuorinen}},\ }\href
  {\doibase 10.1103/PhysRevLett.120.172703} {\bibfield  {journal} {\bibinfo
  {journal} {Phys. Rev. Lett.}\ }\textbf {\bibinfo {volume} {120}},\ \bibinfo
  {pages} {172703} (\bibinfo {year} {2018}{\natexlab{a}})},\ \Eprint
  {http://arxiv.org/abs/1711.02644} {arXiv:1711.02644 [astro-ph.HE]}
  \BibitemShut {NoStop}%
\bibitem [{\citenamefont {Most}\ \emph {et~al.}(2018)\citenamefont {Most},
  \citenamefont {Weih}, \citenamefont {Rezzolla},\ and\ \citenamefont
  {Schaffner-Bielich}}]{Most:2018hfd}%
  \BibitemOpen
  \bibfield  {author} {\bibinfo {author} {\bibfnamefont {E.~R.}\ \bibnamefont
  {Most}}, \bibinfo {author} {\bibfnamefont {L.~R.}\ \bibnamefont {Weih}},
  \bibinfo {author} {\bibfnamefont {L.}~\bibnamefont {Rezzolla}}, \ and\
  \bibinfo {author} {\bibfnamefont {J.}~\bibnamefont {Schaffner-Bielich}},\
  }\href {\doibase 10.1103/PhysRevLett.120.261103} {\bibfield  {journal}
  {\bibinfo  {journal} {Phys. Rev. Lett.}\ }\textbf {\bibinfo {volume} {120}},\
  \bibinfo {pages} {261103} (\bibinfo {year} {2018})},\ \Eprint
  {http://arxiv.org/abs/1803.00549} {arXiv:1803.00549 [gr-qc]} \BibitemShut
  {NoStop}%
\bibitem [{\citenamefont {Annala}\ \emph {et~al.}(2020)\citenamefont {Annala},
  \citenamefont {Gorda}, \citenamefont {Kurkela}, \citenamefont {N\"attil\"a},\
  and\ \citenamefont {Vuorinen}}]{Annala:2019puf}%
  \BibitemOpen
  \bibfield  {author} {\bibinfo {author} {\bibfnamefont {E.}~\bibnamefont
  {Annala}}, \bibinfo {author} {\bibfnamefont {T.}~\bibnamefont {Gorda}},
  \bibinfo {author} {\bibfnamefont {A.}~\bibnamefont {Kurkela}}, \bibinfo
  {author} {\bibfnamefont {J.}~\bibnamefont {N\"attil\"a}}, \ and\ \bibinfo
  {author} {\bibfnamefont {A.}~\bibnamefont {Vuorinen}},\ }\href {\doibase
  10.1038/s41567-020-0914-9} {\bibfield  {journal} {\bibinfo  {journal} {Nature
  Phys.}\ }\textbf {\bibinfo {volume} {16}},\ \bibinfo {pages} {907} (\bibinfo
  {year} {2020})},\ \Eprint {http://arxiv.org/abs/1903.09121} {arXiv:1903.09121
  [astro-ph.HE]} \BibitemShut {NoStop}%
\bibitem [{\citenamefont {Annala}\ \emph {et~al.}(2021)\citenamefont {Annala},
  \citenamefont {Gorda}, \citenamefont {Katerini}, \citenamefont {Kurkela},
  \citenamefont {N\"attil\"a}, \citenamefont {Paschalidis},\ and\ \citenamefont
  {Vuorinen}}]{Annala:2021gom}%
  \BibitemOpen
  \bibfield  {author} {\bibinfo {author} {\bibfnamefont {E.}~\bibnamefont
  {Annala}}, \bibinfo {author} {\bibfnamefont {T.}~\bibnamefont {Gorda}},
  \bibinfo {author} {\bibfnamefont {E.}~\bibnamefont {Katerini}}, \bibinfo
  {author} {\bibfnamefont {A.}~\bibnamefont {Kurkela}}, \bibinfo {author}
  {\bibfnamefont {J.}~\bibnamefont {N\"attil\"a}}, \bibinfo {author}
  {\bibfnamefont {V.}~\bibnamefont {Paschalidis}}, \ and\ \bibinfo {author}
  {\bibfnamefont {A.}~\bibnamefont {Vuorinen}},\ }\href@noop {} {\  (\bibinfo
  {year} {2021})},\ \Eprint {http://arxiv.org/abs/2105.05132} {arXiv:2105.05132
  [astro-ph.HE]} \BibitemShut {NoStop}%
\bibitem [{\citenamefont {Komoltsev}\ and\ \citenamefont
  {Kurkela}(2021)}]{Komoltsev:2021jzg}%
  \BibitemOpen
  \bibfield  {author} {\bibinfo {author} {\bibfnamefont {O.}~\bibnamefont
  {Komoltsev}}\ and\ \bibinfo {author} {\bibfnamefont {A.}~\bibnamefont
  {Kurkela}},\ }\href@noop {} {\  (\bibinfo {year} {2021})},\ \Eprint
  {http://arxiv.org/abs/2111.05350} {arXiv:2111.05350 [nucl-th]} \BibitemShut
  {NoStop}%
\bibitem [{\citenamefont {Abbott}\ \emph
  {et~al.}(2017{\natexlab{a}})\citenamefont {Abbott} \emph
  {et~al.}}]{TheLIGOScientific:2017qsa}%
  \BibitemOpen
  \bibfield  {author} {\bibinfo {author} {\bibfnamefont {B.~P.}\ \bibnamefont
  {Abbott}} \emph {et~al.} (\bibinfo {collaboration} {LIGO Scientific,
  Virgo}),\ }\href {\doibase 10.1103/PhysRevLett.119.161101} {\bibfield
  {journal} {\bibinfo  {journal} {Phys. Rev. Lett.}\ }\textbf {\bibinfo
  {volume} {119}},\ \bibinfo {pages} {161101} (\bibinfo {year}
  {2017}{\natexlab{a}})},\ \Eprint {http://arxiv.org/abs/1710.05832}
  {arXiv:1710.05832 [gr-qc]} \BibitemShut {NoStop}%
\bibitem [{\citenamefont {Abbott}\ \emph
  {et~al.}(2017{\natexlab{b}})\citenamefont {Abbott} \emph
  {et~al.}}]{GBM:2017lvd}%
  \BibitemOpen
  \bibfield  {author} {\bibinfo {author} {\bibfnamefont {B.~P.}\ \bibnamefont
  {Abbott}} \emph {et~al.} (\bibinfo {collaboration} {LIGO Scientific, Virgo,
  Fermi GBM, INTEGRAL, IceCube, AstroSat Cadmium Zinc Telluride Imager Team,
  IPN, Insight-Hxmt, ANTARES, Swift, AGILE Team, 1M2H Team, Dark Energy Camera
  GW-EM, DES, DLT40, GRAWITA, Fermi-LAT, ATCA, ASKAP, Las Cumbres Observatory
  Group, OzGrav, DWF (Deeper Wider Faster Program), AST3, CAASTRO, VINROUGE,
  MASTER, J-GEM, GROWTH, JAGWAR, CaltechNRAO, TTU-NRAO, NuSTAR, Pan-STARRS,
  MAXI Team, TZAC Consortium, KU, Nordic Optical Telescope, ePESSTO, GROND,
  Texas Tech University, SALT Group, TOROS, BOOTES, MWA, CALET, IKI-GW
  Follow-up, H.E.S.S., LOFAR, LWA, HAWC, Pierre Auger, ALMA, Euro VLBI Team, Pi
  of Sky, Chandra Team at McGill University, DFN, ATLAS Telescopes, High Time
  Resolution Universe Survey, RIMAS, RATIR, SKA South Africa/MeerKAT}),\ }\href
  {\doibase 10.3847/2041-8213/aa91c9} {\bibfield  {journal} {\bibinfo
  {journal} {Astrophys. J. Lett.}\ }\textbf {\bibinfo {volume} {848}},\
  \bibinfo {pages} {L12} (\bibinfo {year} {2017}{\natexlab{b}})},\ \Eprint
  {http://arxiv.org/abs/1710.05833} {arXiv:1710.05833 [astro-ph.HE]}
  \BibitemShut {NoStop}%
\bibitem [{\citenamefont {Rezzolla}\ \emph {et~al.}(2018)\citenamefont
  {Rezzolla}, \citenamefont {Most},\ and\ \citenamefont
  {Weih}}]{Rezzolla:2017aly}%
  \BibitemOpen
  \bibfield  {author} {\bibinfo {author} {\bibfnamefont {L.}~\bibnamefont
  {Rezzolla}}, \bibinfo {author} {\bibfnamefont {E.~R.}\ \bibnamefont {Most}},
  \ and\ \bibinfo {author} {\bibfnamefont {L.~R.}\ \bibnamefont {Weih}},\
  }\href {\doibase 10.3847/2041-8213/aaa401} {\bibfield  {journal} {\bibinfo
  {journal} {Astrophys. J. Lett.}\ }\textbf {\bibinfo {volume} {852}},\
  \bibinfo {pages} {L25} (\bibinfo {year} {2018})},\ \Eprint
  {http://arxiv.org/abs/1711.00314} {arXiv:1711.00314 [astro-ph.HE]}
  \BibitemShut {NoStop}%
\bibitem [{\citenamefont {Stephanov}(2004)}]{Stephanov:2004wx}%
  \BibitemOpen
  \bibfield  {author} {\bibinfo {author} {\bibfnamefont {M.~A.}\ \bibnamefont
  {Stephanov}},\ }\href {\doibase 10.1142/S0217751X05027965} {\bibfield
  {journal} {\bibinfo  {journal} {Prog. Theor. Phys. Suppl.}\ }\textbf
  {\bibinfo {volume} {153}},\ \bibinfo {pages} {139} (\bibinfo {year}
  {2004})},\ \Eprint {http://arxiv.org/abs/hep-ph/0402115}
  {arXiv:hep-ph/0402115} \BibitemShut {NoStop}%
\bibitem [{\citenamefont {Ayala}\ \emph {et~al.}(2021)\citenamefont {Ayala},
  \citenamefont {Zamora}, \citenamefont {Cobos-Mart\'\i{}nez}, \citenamefont
  {Hern\'andez-Ortiz}, \citenamefont {Hern\'andez}, \citenamefont {Raya},\ and\
  \citenamefont {Tejeda-Yeomans}}]{Ayala:2021tkm}%
  \BibitemOpen
  \bibfield  {author} {\bibinfo {author} {\bibfnamefont {A.}~\bibnamefont
  {Ayala}}, \bibinfo {author} {\bibfnamefont {B.~A.}\ \bibnamefont {Zamora}},
  \bibinfo {author} {\bibfnamefont {J.~J.}\ \bibnamefont
  {Cobos-Mart\'\i{}nez}}, \bibinfo {author} {\bibfnamefont {S.}~\bibnamefont
  {Hern\'andez-Ortiz}}, \bibinfo {author} {\bibfnamefont {L.~A.}\ \bibnamefont
  {Hern\'andez}}, \bibinfo {author} {\bibfnamefont {A.}~\bibnamefont {Raya}}, \
  and\ \bibinfo {author} {\bibfnamefont {M.~E.}\ \bibnamefont
  {Tejeda-Yeomans}},\ }\href@noop {} {\  (\bibinfo {year} {2021})},\ \Eprint
  {http://arxiv.org/abs/2108.02362} {arXiv:2108.02362 [hep-ph]} \BibitemShut
  {NoStop}%
\bibitem [{\citenamefont {Aryal}\ \emph {et~al.}(2021)\citenamefont {Aryal},
  \citenamefont {Constantinou}, \citenamefont {Farias},\ and\ \citenamefont
  {Dexheimer}}]{Aryal:2021ojz}%
  \BibitemOpen
  \bibfield  {author} {\bibinfo {author} {\bibfnamefont {K.}~\bibnamefont
  {Aryal}}, \bibinfo {author} {\bibfnamefont {C.}~\bibnamefont {Constantinou}},
  \bibinfo {author} {\bibfnamefont {R.~L.~S.}\ \bibnamefont {Farias}}, \ and\
  \bibinfo {author} {\bibfnamefont {V.}~\bibnamefont {Dexheimer}},\ }\href
  {\doibase 10.3390/universe7110454} {\bibfield  {journal} {\bibinfo  {journal}
  {Universe}\ }\textbf {\bibinfo {volume} {7}},\ \bibinfo {pages} {454}
  (\bibinfo {year} {2021})},\ \Eprint {http://arxiv.org/abs/2109.14787}
  {arXiv:2109.14787 [nucl-th]} \BibitemShut {NoStop}%
\bibitem [{\citenamefont {Yang}(2017)}]{Yang:2017llt}%
  \BibitemOpen
  \bibfield  {author} {\bibinfo {author} {\bibfnamefont {C.}~\bibnamefont
  {Yang}} (\bibinfo {collaboration} {STAR}),\ }\href {\doibase
  10.1016/j.nuclphysa.2017.05.042} {\bibfield  {journal} {\bibinfo  {journal}
  {Nucl. Phys. A}\ }\textbf {\bibinfo {volume} {967}},\ \bibinfo {pages} {800}
  (\bibinfo {year} {2017})}\BibitemShut {NoStop}%
\bibitem [{\citenamefont {Adamczyk}\ \emph {et~al.}(2017)\citenamefont
  {Adamczyk} \emph {et~al.}}]{STAR:2017sal}%
  \BibitemOpen
  \bibfield  {author} {\bibinfo {author} {\bibfnamefont {L.}~\bibnamefont
  {Adamczyk}} \emph {et~al.} (\bibinfo {collaboration} {STAR}),\ }\href
  {\doibase 10.1103/PhysRevC.96.044904} {\bibfield  {journal} {\bibinfo
  {journal} {Phys. Rev. C}\ }\textbf {\bibinfo {volume} {96}},\ \bibinfo
  {pages} {044904} (\bibinfo {year} {2017})},\ \Eprint
  {http://arxiv.org/abs/1701.07065} {arXiv:1701.07065 [nucl-ex]} \BibitemShut
  {NoStop}%
\bibitem [{\citenamefont {An}\ \emph {et~al.}(2022)\citenamefont {An} \emph
  {et~al.}}]{An:2021wof}%
  \BibitemOpen
  \bibfield  {author} {\bibinfo {author} {\bibfnamefont {X.}~\bibnamefont {An}}
  \emph {et~al.},\ }\href {\doibase 10.1016/j.nuclphysa.2021.122343} {\bibfield
   {journal} {\bibinfo  {journal} {Nucl. Phys. A}\ }\textbf {\bibinfo {volume}
  {1017}},\ \bibinfo {pages} {122343} (\bibinfo {year} {2022})},\ \Eprint
  {http://arxiv.org/abs/2108.13867} {arXiv:2108.13867 [nucl-th]} \BibitemShut
  {NoStop}%
\bibitem [{\citenamefont {Carbone}\ and\ \citenamefont
  {Schwenk}(2019)}]{Carbone:2019pkr}%
  \BibitemOpen
  \bibfield  {author} {\bibinfo {author} {\bibfnamefont {A.}~\bibnamefont
  {Carbone}}\ and\ \bibinfo {author} {\bibfnamefont {A.}~\bibnamefont
  {Schwenk}},\ }\href {\doibase 10.1103/PhysRevC.100.025805} {\bibfield
  {journal} {\bibinfo  {journal} {Phys. Rev. C}\ }\textbf {\bibinfo {volume}
  {100}},\ \bibinfo {pages} {025805} (\bibinfo {year} {2019})},\ \Eprint
  {http://arxiv.org/abs/1904.00924} {arXiv:1904.00924 [nucl-th]} \BibitemShut
  {NoStop}%
\bibitem [{\citenamefont {Bauswein}\ \emph {et~al.}(2010)\citenamefont
  {Bauswein}, \citenamefont {Janka},\ and\ \citenamefont
  {Oechslin}}]{Bauswein:2010dn}%
  \BibitemOpen
  \bibfield  {author} {\bibinfo {author} {\bibfnamefont {A.}~\bibnamefont
  {Bauswein}}, \bibinfo {author} {\bibfnamefont {H.~T.}\ \bibnamefont {Janka}},
  \ and\ \bibinfo {author} {\bibfnamefont {R.}~\bibnamefont {Oechslin}},\
  }\href {\doibase 10.1103/PhysRevD.82.084043} {\bibfield  {journal} {\bibinfo
  {journal} {Phys. Rev. D}\ }\textbf {\bibinfo {volume} {82}},\ \bibinfo
  {pages} {084043} (\bibinfo {year} {2010})},\ \Eprint
  {http://arxiv.org/abs/1006.3315} {arXiv:1006.3315 [astro-ph.SR]} \BibitemShut
  {NoStop}%
\bibitem [{\citenamefont {De~Pietri}\ \emph {et~al.}(2020)\citenamefont
  {De~Pietri}, \citenamefont {Feo}, \citenamefont {Font}, \citenamefont
  {L\"offler}, \citenamefont {Pasquali},\ and\ \citenamefont
  {Stergioulas}}]{DePietri:2019mti}%
  \BibitemOpen
  \bibfield  {author} {\bibinfo {author} {\bibfnamefont {R.}~\bibnamefont
  {De~Pietri}}, \bibinfo {author} {\bibfnamefont {A.}~\bibnamefont {Feo}},
  \bibinfo {author} {\bibfnamefont {J.~A.}\ \bibnamefont {Font}}, \bibinfo
  {author} {\bibfnamefont {F.}~\bibnamefont {L\"offler}}, \bibinfo {author}
  {\bibfnamefont {M.}~\bibnamefont {Pasquali}}, \ and\ \bibinfo {author}
  {\bibfnamefont {N.}~\bibnamefont {Stergioulas}},\ }\href {\doibase
  10.1103/PhysRevD.101.064052} {\bibfield  {journal} {\bibinfo  {journal}
  {Phys. Rev. D}\ }\textbf {\bibinfo {volume} {101}},\ \bibinfo {pages}
  {064052} (\bibinfo {year} {2020})},\ \Eprint
  {http://arxiv.org/abs/1910.04036} {arXiv:1910.04036 [gr-qc]} \BibitemShut
  {NoStop}%
\bibitem [{\citenamefont {Xie}\ \emph {et~al.}(2020)\citenamefont {Xie},
  \citenamefont {Hawke}, \citenamefont {Passamonti},\ and\ \citenamefont
  {Andersson}}]{Xie:2020udh}%
  \BibitemOpen
  \bibfield  {author} {\bibinfo {author} {\bibfnamefont {X.}~\bibnamefont
  {Xie}}, \bibinfo {author} {\bibfnamefont {I.}~\bibnamefont {Hawke}}, \bibinfo
  {author} {\bibfnamefont {A.}~\bibnamefont {Passamonti}}, \ and\ \bibinfo
  {author} {\bibfnamefont {N.}~\bibnamefont {Andersson}},\ }\href {\doibase
  10.1103/PhysRevD.102.044040} {\bibfield  {journal} {\bibinfo  {journal}
  {Phys. Rev. D}\ }\textbf {\bibinfo {volume} {102}},\ \bibinfo {pages}
  {044040} (\bibinfo {year} {2020})},\ \Eprint
  {http://arxiv.org/abs/2005.13696} {arXiv:2005.13696 [astro-ph.HE]}
  \BibitemShut {NoStop}%
\bibitem [{\citenamefont {Figura}\ \emph {et~al.}(2020)\citenamefont {Figura},
  \citenamefont {Lu}, \citenamefont {Burgio}, \citenamefont {Li},\ and\
  \citenamefont {Schulze}}]{Figura:2020fkj}%
  \BibitemOpen
  \bibfield  {author} {\bibinfo {author} {\bibfnamefont {A.}~\bibnamefont
  {Figura}}, \bibinfo {author} {\bibfnamefont {J.~J.}\ \bibnamefont {Lu}},
  \bibinfo {author} {\bibfnamefont {G.~F.}\ \bibnamefont {Burgio}}, \bibinfo
  {author} {\bibfnamefont {Z.~H.}\ \bibnamefont {Li}}, \ and\ \bibinfo {author}
  {\bibfnamefont {H.~J.}\ \bibnamefont {Schulze}},\ }\href {\doibase
  10.1103/PhysRevD.102.043006} {\bibfield  {journal} {\bibinfo  {journal}
  {Phys. Rev. D}\ }\textbf {\bibinfo {volume} {102}},\ \bibinfo {pages}
  {043006} (\bibinfo {year} {2020})},\ \Eprint
  {http://arxiv.org/abs/2005.08691} {arXiv:2005.08691 [gr-qc]} \BibitemShut
  {NoStop}%
\bibitem [{\citenamefont {Cromartie}\ \emph {et~al.}(2019)\citenamefont
  {Cromartie} \emph {et~al.}}]{NANOGrav:2019jur}%
  \BibitemOpen
  \bibfield  {author} {\bibinfo {author} {\bibfnamefont {H.~T.}\ \bibnamefont
  {Cromartie}} \emph {et~al.} (\bibinfo {collaboration} {NANOGrav}),\ }\href
  {\doibase 10.1038/s41550-019-0880-2} {\bibfield  {journal} {\bibinfo
  {journal} {Nature Astron.}\ }\textbf {\bibinfo {volume} {4}},\ \bibinfo
  {pages} {72} (\bibinfo {year} {2019})},\ \Eprint
  {http://arxiv.org/abs/1904.06759} {arXiv:1904.06759 [astro-ph.HE]}
  \BibitemShut {NoStop}%
\bibitem [{\citenamefont {Fonseca}\ \emph {et~al.}(2021)\citenamefont {Fonseca}
  \emph {et~al.}}]{Fonseca:2021wxt}%
  \BibitemOpen
  \bibfield  {author} {\bibinfo {author} {\bibfnamefont {E.}~\bibnamefont
  {Fonseca}} \emph {et~al.},\ }\href {\doibase 10.3847/2041-8213/ac03b8}
  {\bibfield  {journal} {\bibinfo  {journal} {Astrophys. J. Lett.}\ }\textbf
  {\bibinfo {volume} {915}},\ \bibinfo {pages} {L12} (\bibinfo {year}
  {2021})},\ \Eprint {http://arxiv.org/abs/2104.00880} {arXiv:2104.00880
  [astro-ph.HE]} \BibitemShut {NoStop}%
\bibitem [{\citenamefont {Riley}\ \emph {et~al.}(2019)\citenamefont {Riley}
  \emph {et~al.}}]{Riley:2019yda}%
  \BibitemOpen
  \bibfield  {author} {\bibinfo {author} {\bibfnamefont {T.~E.}\ \bibnamefont
  {Riley}} \emph {et~al.},\ }\href {\doibase 10.3847/2041-8213/ab481c}
  {\bibfield  {journal} {\bibinfo  {journal} {Astrophys. J. Lett.}\ }\textbf
  {\bibinfo {volume} {887}},\ \bibinfo {pages} {L21} (\bibinfo {year}
  {2019})},\ \Eprint {http://arxiv.org/abs/1912.05702} {arXiv:1912.05702
  [astro-ph.HE]} \BibitemShut {NoStop}%
\bibitem [{\citenamefont {Miller}\ \emph {et~al.}(2019)\citenamefont {Miller}
  \emph {et~al.}}]{Miller:2019cac}%
  \BibitemOpen
  \bibfield  {author} {\bibinfo {author} {\bibfnamefont {M.}~\bibnamefont
  {Miller}} \emph {et~al.},\ }\href {\doibase 10.3847/2041-8213/ab50c5}
  {\bibfield  {journal} {\bibinfo  {journal} {Astrophys. J. Lett.}\ }\textbf
  {\bibinfo {volume} {887}},\ \bibinfo {pages} {L24} (\bibinfo {year}
  {2019})},\ \Eprint {http://arxiv.org/abs/1912.05705} {arXiv:1912.05705
  [astro-ph.HE]} \BibitemShut {NoStop}%
\bibitem [{\citenamefont {Miller}\ \emph {et~al.}(2021)\citenamefont {Miller}
  \emph {et~al.}}]{Miller:2021qha}%
  \BibitemOpen
  \bibfield  {author} {\bibinfo {author} {\bibfnamefont {M.~C.}\ \bibnamefont
  {Miller}} \emph {et~al.},\ }\href {\doibase 10.3847/2041-8213/ac089b}
  {\bibfield  {journal} {\bibinfo  {journal} {Astrophys. J. Lett.}\ }\textbf
  {\bibinfo {volume} {918}},\ \bibinfo {pages} {L28} (\bibinfo {year}
  {2021})},\ \Eprint {http://arxiv.org/abs/2105.06979} {arXiv:2105.06979
  [astro-ph.HE]} \BibitemShut {NoStop}%
\bibitem [{\citenamefont {Riley}\ \emph {et~al.}(2021)\citenamefont {Riley}
  \emph {et~al.}}]{Riley:2021pdl}%
  \BibitemOpen
  \bibfield  {author} {\bibinfo {author} {\bibfnamefont {T.~E.}\ \bibnamefont
  {Riley}} \emph {et~al.},\ }\href {\doibase 10.3847/2041-8213/ac0a81}
  {\bibfield  {journal} {\bibinfo  {journal} {Astrophys. J. Lett.}\ }\textbf
  {\bibinfo {volume} {918}},\ \bibinfo {pages} {L27} (\bibinfo {year}
  {2021})},\ \Eprint {http://arxiv.org/abs/2105.06980} {arXiv:2105.06980
  [astro-ph.HE]} \BibitemShut {NoStop}%
\bibitem [{\citenamefont {N\"attil\"a}\ \emph {et~al.}(2017)\citenamefont
  {N\"attil\"a}, \citenamefont {Miller}, \citenamefont {Steiner}, \citenamefont
  {Kajava}, \citenamefont {Suleimanov},\ and\ \citenamefont
  {Poutanen}}]{Nattila:2017wtj}%
  \BibitemOpen
  \bibfield  {author} {\bibinfo {author} {\bibfnamefont {J.}~\bibnamefont
  {N\"attil\"a}}, \bibinfo {author} {\bibfnamefont {M.~C.}\ \bibnamefont
  {Miller}}, \bibinfo {author} {\bibfnamefont {A.~W.}\ \bibnamefont {Steiner}},
  \bibinfo {author} {\bibfnamefont {J.~J.~E.}\ \bibnamefont {Kajava}}, \bibinfo
  {author} {\bibfnamefont {V.~F.}\ \bibnamefont {Suleimanov}}, \ and\ \bibinfo
  {author} {\bibfnamefont {J.}~\bibnamefont {Poutanen}},\ }\href {\doibase
  10.1051/0004-6361/201731082} {\bibfield  {journal} {\bibinfo  {journal}
  {Astron. Astrophys.}\ }\textbf {\bibinfo {volume} {608}},\ \bibinfo {pages}
  {A31} (\bibinfo {year} {2017})},\ \Eprint {http://arxiv.org/abs/1709.09120}
  {arXiv:1709.09120 [astro-ph.HE]} \BibitemShut {NoStop}%
\bibitem [{\citenamefont {Abbott}\ \emph {et~al.}(2018)\citenamefont {Abbott}
  \emph {et~al.}}]{LIGOScientific:2018cki}%
  \BibitemOpen
  \bibfield  {author} {\bibinfo {author} {\bibfnamefont {B.~P.}\ \bibnamefont
  {Abbott}} \emph {et~al.} (\bibinfo {collaboration} {LIGO Scientific,
  Virgo}),\ }\href {\doibase 10.1103/PhysRevLett.121.161101} {\bibfield
  {journal} {\bibinfo  {journal} {Phys. Rev. Lett.}\ }\textbf {\bibinfo
  {volume} {121}},\ \bibinfo {pages} {161101} (\bibinfo {year} {2018})},\
  \Eprint {http://arxiv.org/abs/1805.11581} {arXiv:1805.11581 [gr-qc]}
  \BibitemShut {NoStop}%
\bibitem [{\citenamefont {Alford}\ \emph {et~al.}(2008)\citenamefont {Alford},
  \citenamefont {Schmitt}, \citenamefont {Rajagopal},\ and\ \citenamefont
  {Sch\"afer}}]{Alford:2007xm}%
  \BibitemOpen
  \bibfield  {author} {\bibinfo {author} {\bibfnamefont {M.~G.}\ \bibnamefont
  {Alford}}, \bibinfo {author} {\bibfnamefont {A.}~\bibnamefont {Schmitt}},
  \bibinfo {author} {\bibfnamefont {K.}~\bibnamefont {Rajagopal}}, \ and\
  \bibinfo {author} {\bibfnamefont {T.}~\bibnamefont {Sch\"afer}},\ }\href
  {\doibase 10.1103/RevModPhys.80.1455} {\bibfield  {journal} {\bibinfo
  {journal} {Rev. Mod. Phys.}\ }\textbf {\bibinfo {volume} {80}},\ \bibinfo
  {pages} {1455} (\bibinfo {year} {2008})},\ \Eprint
  {http://arxiv.org/abs/0709.4635} {arXiv:0709.4635 [hep-ph]} \BibitemShut
  {NoStop}%
\bibitem [{\citenamefont {Chen}\ \emph {et~al.}(2010)\citenamefont {Chen},
  \citenamefont {Hashimoto},\ and\ \citenamefont {Matsuura}}]{Chen:2009kx}%
  \BibitemOpen
  \bibfield  {author} {\bibinfo {author} {\bibfnamefont {H.-Y.}\ \bibnamefont
  {Chen}}, \bibinfo {author} {\bibfnamefont {K.}~\bibnamefont {Hashimoto}}, \
  and\ \bibinfo {author} {\bibfnamefont {S.}~\bibnamefont {Matsuura}},\ }\href
  {\doibase 10.1007/JHEP02(2010)104} {\bibfield  {journal} {\bibinfo  {journal}
  {JHEP}\ }\textbf {\bibinfo {volume} {02}},\ \bibinfo {pages} {104} (\bibinfo
  {year} {2010})},\ \Eprint {http://arxiv.org/abs/0909.1296} {arXiv:0909.1296
  [hep-th]} \BibitemShut {NoStop}%
\bibitem [{\citenamefont {Basu}\ \emph {et~al.}(2011)\citenamefont {Basu},
  \citenamefont {Nogueira}, \citenamefont {Rozali}, \citenamefont {Stang},\
  and\ \citenamefont {Van~Raamsdonk}}]{Basu:2011yg}%
  \BibitemOpen
  \bibfield  {author} {\bibinfo {author} {\bibfnamefont {P.}~\bibnamefont
  {Basu}}, \bibinfo {author} {\bibfnamefont {F.}~\bibnamefont {Nogueira}},
  \bibinfo {author} {\bibfnamefont {M.}~\bibnamefont {Rozali}}, \bibinfo
  {author} {\bibfnamefont {J.~B.}\ \bibnamefont {Stang}}, \ and\ \bibinfo
  {author} {\bibfnamefont {M.}~\bibnamefont {Van~Raamsdonk}},\ }\href {\doibase
  10.1088/1367-2630/13/5/055001} {\bibfield  {journal} {\bibinfo  {journal}
  {New J. Phys.}\ }\textbf {\bibinfo {volume} {13}},\ \bibinfo {pages} {055001}
  (\bibinfo {year} {2011})},\ \Eprint {http://arxiv.org/abs/1101.4042}
  {arXiv:1101.4042 [hep-th]} \BibitemShut {NoStop}%
\bibitem [{\citenamefont {Bitaghsir~Fadafan}\ \emph {et~al.}(2018)\citenamefont
  {Bitaghsir~Fadafan}, \citenamefont {Cruz~Rojas},\ and\ \citenamefont
  {Evans}}]{BitaghsirFadafan:2018iqr}%
  \BibitemOpen
  \bibfield  {author} {\bibinfo {author} {\bibfnamefont {K.}~\bibnamefont
  {Bitaghsir~Fadafan}}, \bibinfo {author} {\bibfnamefont {J.}~\bibnamefont
  {Cruz~Rojas}}, \ and\ \bibinfo {author} {\bibfnamefont {N.}~\bibnamefont
  {Evans}},\ }\href {\doibase 10.1103/PhysRevD.98.066010} {\bibfield  {journal}
  {\bibinfo  {journal} {Phys. Rev. D}\ }\textbf {\bibinfo {volume} {98}},\
  \bibinfo {pages} {066010} (\bibinfo {year} {2018})},\ \Eprint
  {http://arxiv.org/abs/1803.03107} {arXiv:1803.03107 [hep-ph]} \BibitemShut
  {NoStop}%
\bibitem [{\citenamefont {Faedo}\ \emph {et~al.}(2019)\citenamefont {Faedo},
  \citenamefont {Mateos}, \citenamefont {Pantelidou},\ and\ \citenamefont
  {Tarr\'\i{}o}}]{Faedo:2018fjw}%
  \BibitemOpen
  \bibfield  {author} {\bibinfo {author} {\bibfnamefont {A.~F.}\ \bibnamefont
  {Faedo}}, \bibinfo {author} {\bibfnamefont {D.}~\bibnamefont {Mateos}},
  \bibinfo {author} {\bibfnamefont {C.}~\bibnamefont {Pantelidou}}, \ and\
  \bibinfo {author} {\bibfnamefont {J.}~\bibnamefont {Tarr\'\i{}o}},\ }\href
  {\doibase 10.1007/JHEP05(2019)106} {\bibfield  {journal} {\bibinfo  {journal}
  {JHEP}\ }\textbf {\bibinfo {volume} {05}},\ \bibinfo {pages} {106} (\bibinfo
  {year} {2019})},\ \Eprint {http://arxiv.org/abs/1807.09712} {arXiv:1807.09712
  [hep-th]} \BibitemShut {NoStop}%
\bibitem [{\citenamefont {Ghoroku}\ \emph {et~al.}(2019)\citenamefont
  {Ghoroku}, \citenamefont {Kashiwa}, \citenamefont {Nakano}, \citenamefont
  {Tachibana},\ and\ \citenamefont {Toyoda}}]{Ghoroku:2019trx}%
  \BibitemOpen
  \bibfield  {author} {\bibinfo {author} {\bibfnamefont {K.}~\bibnamefont
  {Ghoroku}}, \bibinfo {author} {\bibfnamefont {K.}~\bibnamefont {Kashiwa}},
  \bibinfo {author} {\bibfnamefont {Y.}~\bibnamefont {Nakano}}, \bibinfo
  {author} {\bibfnamefont {M.}~\bibnamefont {Tachibana}}, \ and\ \bibinfo
  {author} {\bibfnamefont {F.}~\bibnamefont {Toyoda}},\ }\href {\doibase
  10.1103/PhysRevD.99.106011} {\bibfield  {journal} {\bibinfo  {journal} {Phys.
  Rev. D}\ }\textbf {\bibinfo {volume} {99}},\ \bibinfo {pages} {106011}
  (\bibinfo {year} {2019})},\ \Eprint {http://arxiv.org/abs/1902.01093}
  {arXiv:1902.01093 [hep-th]} \BibitemShut {NoStop}%
\bibitem [{\citenamefont {Henriksson}\ \emph {et~al.}(2019)\citenamefont
  {Henriksson}, \citenamefont {Hoyos},\ and\ \citenamefont
  {Jokela}}]{Henriksson:2019zph}%
  \BibitemOpen
  \bibfield  {author} {\bibinfo {author} {\bibfnamefont {O.}~\bibnamefont
  {Henriksson}}, \bibinfo {author} {\bibfnamefont {C.}~\bibnamefont {Hoyos}}, \
  and\ \bibinfo {author} {\bibfnamefont {N.}~\bibnamefont {Jokela}},\ }\href
  {\doibase 10.1007/JHEP09(2019)088} {\bibfield  {journal} {\bibinfo  {journal}
  {JHEP}\ }\textbf {\bibinfo {volume} {09}},\ \bibinfo {pages} {088} (\bibinfo
  {year} {2019})},\ \Eprint {http://arxiv.org/abs/1907.01562} {arXiv:1907.01562
  [hep-th]} \BibitemShut {NoStop}%
\bibitem [{\citenamefont {Hoyos}\ \emph {et~al.}(2020)\citenamefont {Hoyos},
  \citenamefont {Jokela}, \citenamefont {J{\"a}rvinen}, \citenamefont {Subils},
  \citenamefont {Tarrio},\ and\ \citenamefont {Vuorinen}}]{Hoyos:2020hmq}%
  \BibitemOpen
  \bibfield  {author} {\bibinfo {author} {\bibfnamefont {C.}~\bibnamefont
  {Hoyos}}, \bibinfo {author} {\bibfnamefont {N.}~\bibnamefont {Jokela}},
  \bibinfo {author} {\bibfnamefont {M.}~\bibnamefont {J{\"a}rvinen}}, \bibinfo
  {author} {\bibfnamefont {J.~G.}\ \bibnamefont {Subils}}, \bibinfo {author}
  {\bibfnamefont {J.}~\bibnamefont {Tarrio}}, \ and\ \bibinfo {author}
  {\bibfnamefont {A.}~\bibnamefont {Vuorinen}},\ }\href {\doibase
  10.1103/PhysRevLett.125.241601} {\bibfield  {journal} {\bibinfo  {journal}
  {Phys. Rev. Lett.}\ }\textbf {\bibinfo {volume} {125}},\ \bibinfo {pages}
  {241601} (\bibinfo {year} {2020})},\ \Eprint
  {http://arxiv.org/abs/2005.14205} {arXiv:2005.14205 [hep-th]} \BibitemShut
  {NoStop}%
\bibitem [{\citenamefont {Hoyos}\ \emph
  {et~al.}(2021{\natexlab{a}})\citenamefont {Hoyos}, \citenamefont {Jokela},
  \citenamefont {J{\"a}rvinen}, \citenamefont {Subils}, \citenamefont
  {Tarrio},\ and\ \citenamefont {Vuorinen}}]{Hoyos:2021njg}%
  \BibitemOpen
  \bibfield  {author} {\bibinfo {author} {\bibfnamefont {C.}~\bibnamefont
  {Hoyos}}, \bibinfo {author} {\bibfnamefont {N.}~\bibnamefont {Jokela}},
  \bibinfo {author} {\bibfnamefont {M.}~\bibnamefont {J{\"a}rvinen}}, \bibinfo
  {author} {\bibfnamefont {J.~G.}\ \bibnamefont {Subils}}, \bibinfo {author}
  {\bibfnamefont {J.}~\bibnamefont {Tarrio}}, \ and\ \bibinfo {author}
  {\bibfnamefont {A.}~\bibnamefont {Vuorinen}},\ }\href@noop {} {\  (\bibinfo
  {year} {2021}{\natexlab{a}})},\ \Eprint {http://arxiv.org/abs/2109.12122}
  {arXiv:2109.12122 [hep-th]} \BibitemShut {NoStop}%
\bibitem [{\citenamefont {Schmitt}\ and\ \citenamefont
  {Shternin}(2018)}]{Schmitt:2017efp}%
  \BibitemOpen
  \bibfield  {author} {\bibinfo {author} {\bibfnamefont {A.}~\bibnamefont
  {Schmitt}}\ and\ \bibinfo {author} {\bibfnamefont {P.}~\bibnamefont
  {Shternin}},\ }\href {\doibase 10.1007/978-3-319-97616-7_9} {\bibfield
  {journal} {\bibinfo  {journal} {Astrophys. Space Sci. Libr.}\ }\textbf
  {\bibinfo {volume} {457}},\ \bibinfo {pages} {455} (\bibinfo {year}
  {2018})},\ \Eprint {http://arxiv.org/abs/1711.06520} {arXiv:1711.06520
  [astro-ph.HE]} \BibitemShut {NoStop}%
\bibitem [{\citenamefont {Hoyos}\ \emph {et~al.}(2016)\citenamefont {Hoyos},
  \citenamefont {Rodr\'\i{}guez~Fern\'andez}, \citenamefont {Jokela},\ and\
  \citenamefont {Vuorinen}}]{Hoyos:2016zke}%
  \BibitemOpen
  \bibfield  {author} {\bibinfo {author} {\bibfnamefont {C.}~\bibnamefont
  {Hoyos}}, \bibinfo {author} {\bibfnamefont {D.}~\bibnamefont
  {Rodr\'\i{}guez~Fern\'andez}}, \bibinfo {author} {\bibfnamefont
  {N.}~\bibnamefont {Jokela}}, \ and\ \bibinfo {author} {\bibfnamefont
  {A.}~\bibnamefont {Vuorinen}},\ }\href {\doibase
  10.1103/PhysRevLett.117.032501} {\bibfield  {journal} {\bibinfo  {journal}
  {Phys. Rev. Lett.}\ }\textbf {\bibinfo {volume} {117}},\ \bibinfo {pages}
  {032501} (\bibinfo {year} {2016})},\ \Eprint
  {http://arxiv.org/abs/1603.02943} {arXiv:1603.02943 [hep-ph]} \BibitemShut
  {NoStop}%
\bibitem [{\citenamefont {Annala}\ \emph
  {et~al.}(2018{\natexlab{b}})\citenamefont {Annala}, \citenamefont {Ecker},
  \citenamefont {Hoyos}, \citenamefont {Jokela}, \citenamefont
  {Rodr\'\i{}guez~Fern\'andez},\ and\ \citenamefont
  {Vuorinen}}]{Annala:2017tqz}%
  \BibitemOpen
  \bibfield  {author} {\bibinfo {author} {\bibfnamefont {E.}~\bibnamefont
  {Annala}}, \bibinfo {author} {\bibfnamefont {C.}~\bibnamefont {Ecker}},
  \bibinfo {author} {\bibfnamefont {C.}~\bibnamefont {Hoyos}}, \bibinfo
  {author} {\bibfnamefont {N.}~\bibnamefont {Jokela}}, \bibinfo {author}
  {\bibfnamefont {D.}~\bibnamefont {Rodr\'\i{}guez~Fern\'andez}}, \ and\
  \bibinfo {author} {\bibfnamefont {A.}~\bibnamefont {Vuorinen}},\ }\href
  {\doibase 10.1007/JHEP12(2018)078} {\bibfield  {journal} {\bibinfo  {journal}
  {JHEP}\ }\textbf {\bibinfo {volume} {12}},\ \bibinfo {pages} {078} (\bibinfo
  {year} {2018}{\natexlab{b}})},\ \Eprint {http://arxiv.org/abs/1711.06244}
  {arXiv:1711.06244 [astro-ph.HE]} \BibitemShut {NoStop}%
\bibitem [{\citenamefont {Bitaghsir~Fadafan}\ \emph {et~al.}(2020)\citenamefont
  {Bitaghsir~Fadafan}, \citenamefont {Cruz~Rojas},\ and\ \citenamefont
  {Evans}}]{BitaghsirFadafan:2019ofb}%
  \BibitemOpen
  \bibfield  {author} {\bibinfo {author} {\bibfnamefont {K.}~\bibnamefont
  {Bitaghsir~Fadafan}}, \bibinfo {author} {\bibfnamefont {J.}~\bibnamefont
  {Cruz~Rojas}}, \ and\ \bibinfo {author} {\bibfnamefont {N.}~\bibnamefont
  {Evans}},\ }\href {\doibase 10.1103/PhysRevD.101.126005} {\bibfield
  {journal} {\bibinfo  {journal} {Phys. Rev. D}\ }\textbf {\bibinfo {volume}
  {101}},\ \bibinfo {pages} {126005} (\bibinfo {year} {2020})},\ \Eprint
  {http://arxiv.org/abs/1911.12705} {arXiv:1911.12705 [hep-ph]} \BibitemShut
  {NoStop}%
\bibitem [{\citenamefont {Mamani}\ \emph {et~al.}(2020)\citenamefont {Mamani},
  \citenamefont {Flores},\ and\ \citenamefont {Zanchin}}]{Mamani:2020pks}%
  \BibitemOpen
  \bibfield  {author} {\bibinfo {author} {\bibfnamefont {L.~A.~H.}\
  \bibnamefont {Mamani}}, \bibinfo {author} {\bibfnamefont {C.~V.}\
  \bibnamefont {Flores}}, \ and\ \bibinfo {author} {\bibfnamefont {V.~T.}\
  \bibnamefont {Zanchin}},\ }\href {\doibase 10.1103/PhysRevD.102.066006}
  {\bibfield  {journal} {\bibinfo  {journal} {Phys. Rev. D}\ }\textbf {\bibinfo
  {volume} {102}},\ \bibinfo {pages} {066006} (\bibinfo {year} {2020})},\
  \Eprint {http://arxiv.org/abs/2006.09401} {arXiv:2006.09401 [hep-th]}
  \BibitemShut {NoStop}%
\bibitem [{\citenamefont {Bitaghsir~Fadafan}\ \emph {et~al.}(2021)\citenamefont
  {Bitaghsir~Fadafan}, \citenamefont {Cruz~Rojas},\ and\ \citenamefont
  {Evans}}]{BitaghsirFadafan:2020otb}%
  \BibitemOpen
  \bibfield  {author} {\bibinfo {author} {\bibfnamefont {K.}~\bibnamefont
  {Bitaghsir~Fadafan}}, \bibinfo {author} {\bibfnamefont {J.}~\bibnamefont
  {Cruz~Rojas}}, \ and\ \bibinfo {author} {\bibfnamefont {N.}~\bibnamefont
  {Evans}},\ }\href {\doibase 10.1103/PhysRevD.103.026012} {\bibfield
  {journal} {\bibinfo  {journal} {Phys. Rev. D}\ }\textbf {\bibinfo {volume}
  {103}},\ \bibinfo {pages} {026012} (\bibinfo {year} {2021})},\ \Eprint
  {http://arxiv.org/abs/2009.14079} {arXiv:2009.14079 [hep-ph]} \BibitemShut
  {NoStop}%
\bibitem [{\citenamefont {Ghoroku}\ \emph {et~al.}(2021)\citenamefont
  {Ghoroku}, \citenamefont {Kashiwa}, \citenamefont {Nakano}, \citenamefont
  {Tachibana},\ and\ \citenamefont {Toyoda}}]{Ghoroku:2021fos}%
  \BibitemOpen
  \bibfield  {author} {\bibinfo {author} {\bibfnamefont {K.}~\bibnamefont
  {Ghoroku}}, \bibinfo {author} {\bibfnamefont {K.}~\bibnamefont {Kashiwa}},
  \bibinfo {author} {\bibfnamefont {Y.}~\bibnamefont {Nakano}}, \bibinfo
  {author} {\bibfnamefont {M.}~\bibnamefont {Tachibana}}, \ and\ \bibinfo
  {author} {\bibfnamefont {F.}~\bibnamefont {Toyoda}},\ }\href {\doibase
  10.1103/PhysRevD.104.126002} {\bibfield  {journal} {\bibinfo  {journal}
  {Phys. Rev. D}\ }\textbf {\bibinfo {volume} {104}},\ \bibinfo {pages}
  {126002} (\bibinfo {year} {2021})},\ \Eprint
  {http://arxiv.org/abs/2107.14450} {arXiv:2107.14450 [hep-th]} \BibitemShut
  {NoStop}%
\bibitem [{\citenamefont {Hoyos}\ \emph
  {et~al.}(2021{\natexlab{b}})\citenamefont {Hoyos}, \citenamefont {Jokela},\
  and\ \citenamefont {Vuorinen}}]{Hoyos:2021uff}%
  \BibitemOpen
  \bibfield  {author} {\bibinfo {author} {\bibfnamefont {C.}~\bibnamefont
  {Hoyos}}, \bibinfo {author} {\bibfnamefont {N.}~\bibnamefont {Jokela}}, \
  and\ \bibinfo {author} {\bibfnamefont {A.}~\bibnamefont {Vuorinen}},\
  }\href@noop {} {\  (\bibinfo {year} {2021}{\natexlab{b}})},\ \Eprint
  {http://arxiv.org/abs/2112.08422} {arXiv:2112.08422 [hep-th]} \BibitemShut
  {NoStop}%
\bibitem [{\citenamefont {Kovensky}\ \emph {et~al.}(2021)\citenamefont
  {Kovensky}, \citenamefont {Poole},\ and\ \citenamefont
  {Schmitt}}]{Kovensky:2021kzl}%
  \BibitemOpen
  \bibfield  {author} {\bibinfo {author} {\bibfnamefont {N.}~\bibnamefont
  {Kovensky}}, \bibinfo {author} {\bibfnamefont {A.}~\bibnamefont {Poole}}, \
  and\ \bibinfo {author} {\bibfnamefont {A.}~\bibnamefont {Schmitt}},\
  }\href@noop {} {\  (\bibinfo {year} {2021})},\ \Eprint
  {http://arxiv.org/abs/2111.03374} {arXiv:2111.03374 [hep-ph]} \BibitemShut
  {NoStop}%
\bibitem [{\citenamefont {Bauswein}\ \emph {et~al.}(2019)\citenamefont
  {Bauswein}, \citenamefont {Bastian}, \citenamefont {Blaschke}, \citenamefont
  {Chatziioannou}, \citenamefont {Clark}, \citenamefont {Fischer},\ and\
  \citenamefont {Oertel}}]{Bauswein:2018bma}%
  \BibitemOpen
  \bibfield  {author} {\bibinfo {author} {\bibfnamefont {A.}~\bibnamefont
  {Bauswein}}, \bibinfo {author} {\bibfnamefont {N.-U.~F.}\ \bibnamefont
  {Bastian}}, \bibinfo {author} {\bibfnamefont {D.~B.}\ \bibnamefont
  {Blaschke}}, \bibinfo {author} {\bibfnamefont {K.}~\bibnamefont
  {Chatziioannou}}, \bibinfo {author} {\bibfnamefont {J.~A.}\ \bibnamefont
  {Clark}}, \bibinfo {author} {\bibfnamefont {T.}~\bibnamefont {Fischer}}, \
  and\ \bibinfo {author} {\bibfnamefont {M.}~\bibnamefont {Oertel}},\ }\href
  {\doibase 10.1103/PhysRevLett.122.061102} {\bibfield  {journal} {\bibinfo
  {journal} {Phys. Rev. Lett.}\ }\textbf {\bibinfo {volume} {122}},\ \bibinfo
  {pages} {061102} (\bibinfo {year} {2019})},\ \Eprint
  {http://arxiv.org/abs/1809.01116} {arXiv:1809.01116 [astro-ph.HE]}
  \BibitemShut {NoStop}%
\bibitem [{\citenamefont {Prakash}\ \emph {et~al.}(2021)\citenamefont
  {Prakash}, \citenamefont {Radice}, \citenamefont {Logoteta}, \citenamefont
  {Perego}, \citenamefont {Nedora}, \citenamefont {Bombaci}, \citenamefont
  {Kashyap}, \citenamefont {Bernuzzi},\ and\ \citenamefont
  {Endrizzi}}]{Prakash:2021wpz}%
  \BibitemOpen
  \bibfield  {author} {\bibinfo {author} {\bibfnamefont {A.}~\bibnamefont
  {Prakash}}, \bibinfo {author} {\bibfnamefont {D.}~\bibnamefont {Radice}},
  \bibinfo {author} {\bibfnamefont {D.}~\bibnamefont {Logoteta}}, \bibinfo
  {author} {\bibfnamefont {A.}~\bibnamefont {Perego}}, \bibinfo {author}
  {\bibfnamefont {V.}~\bibnamefont {Nedora}}, \bibinfo {author} {\bibfnamefont
  {I.}~\bibnamefont {Bombaci}}, \bibinfo {author} {\bibfnamefont
  {R.}~\bibnamefont {Kashyap}}, \bibinfo {author} {\bibfnamefont
  {S.}~\bibnamefont {Bernuzzi}}, \ and\ \bibinfo {author} {\bibfnamefont
  {A.}~\bibnamefont {Endrizzi}},\ }\href {\doibase 10.1103/PhysRevD.104.083029}
  {\bibfield  {journal} {\bibinfo  {journal} {Phys. Rev. D}\ }\textbf {\bibinfo
  {volume} {104}},\ \bibinfo {pages} {083029} (\bibinfo {year} {2021})},\
  \Eprint {http://arxiv.org/abs/2106.07885} {arXiv:2106.07885 [astro-ph.HE]}
  \BibitemShut {NoStop}%
\end{thebibliography}%

\end{document}